\documentclass[aps,prx,twocolumn,superscriptaddress,floatfix,longbibliography,10pt]{revtex4-2}

\usepackage[T1]{fontenc}
\usepackage[utf8]{inputenc}

\usepackage{amsmath,amsfonts,amssymb,amsthm,bm,mathtools}

\usepackage{graphicx}
\usepackage{booktabs}

\usepackage[stretch=10,shrink=10]{microtype}
\usepackage[mathlines]{lineno}
\modulolinenumbers[1]
\makeatletter
\newcommand*\patchAmsMathEnvironmentForLineno[1]{%
  \expandafter\let\csname old#1\expandafter\endcsname\csname #1\endcsname
  \expandafter\let\csname oldend#1\expandafter\endcsname\csname end#1\endcsname
  \renewenvironment{#1}%
     {\linenomath\csname old#1\endcsname}%
     {\csname oldend#1\endcsname\endlinenomath}%
}
\newcommand*\patchBothAmsMathEnvironmentsForLineno[1]{%
  \patchAmsMathEnvironmentForLineno{#1}%
  \patchAmsMathEnvironmentForLineno{#1*}%
}
\AtBeginDocument{%
  \patchBothAmsMathEnvironmentsForLineno{equation}%
  \patchBothAmsMathEnvironmentsForLineno{align}%
  \patchBothAmsMathEnvironmentsForLineno{flalign}%
  \patchBothAmsMathEnvironmentsForLineno{alignat}%
  \patchBothAmsMathEnvironmentsForLineno{gather}%
  \patchBothAmsMathEnvironmentsForLineno{multline}%
}
\makeatother

\usepackage[dvipsnames]{xcolor}
\definecolor{linkblue}{RGB}{0,75,135}
\definecolor{citeteal}{RGB}{0,110,95}

\usepackage{tcolorbox}
\tcbuselibrary{skins}

\tcbset{
    prxmathbox/.style={
        enhanced,
        colback=black!2,
        colframe=black!70,
        coltitle=black,
        fonttitle=\bfseries\small,
        boxrule=0.45pt,
        arc=2pt,
        boxsep=1pt,
        left=6pt,
        right=6pt,
        top=8pt,
        bottom=3pt,
        width=\columnwidth,
        center,
        before skip=6pt,
        after skip=6pt,
        attach boxed title to top left={xshift=6pt,yshift=-1.5mm},
        boxed title style={
            colback=white,
            colframe=black!70,
            boxrule=0.4pt,
            arc=2pt,
            left=3pt,
            right=3pt,
            top=0pt,
            bottom=0pt
        },
        before upper={\raggedright\setlength{\abovedisplayskip}{1pt}%
            \setlength{\belowdisplayskip}{1pt}%
            \setlength{\abovedisplayshortskip}{0pt}%
            \setlength{\belowdisplayshortskip}{1pt}}
    }
}

\usepackage{hyperref}
\hypersetup{
    colorlinks=true,
    linkcolor=linkblue,
    citecolor=citeteal,
    urlcolor=citeteal,
    pdftitle={Linear equivalence of nonlinear recurrent neural networks},
    pdfauthor={David G. Clark}
}

\newcommand{\tavg}[1]{\left\langle #1 \right\rangle}

\newcommand{\tavgshort}[1]{\langle #1 \rangle}

\newcommand{\order}[1]{\mathcal{O}\!(#1)}

\newcommand{\dd}{\mathrm{d}}
\newcommand{\Tr}{\mathrm{Tr}}
\newcommand{\Sstar}{S^\phi_\star}
\newcommand{\Cstar}{C^\phi_\star}
\newcommand{\CstarDelta}{C^\Delta_\star}
\newcommand{\etac}{\eta^{\mathrm{c}}}
\newcommand{\free}{\mathrm{c}}
\newcommand{\sstavg}[1]{\left\langle #1 \right\rangle_\star}

\makeatletter
\newif\if@apptoc \@apptocfalse
\newcommand{\apptocstart}{\global\@apptoctrue}
\AtBeginDocument{%
  \let\apt@lsec\l@section \let\apt@lsub\l@subsection
  \renewcommand*{\l@section}[2]{\if@apptoc\apt@lsec{#1}{#2}\fi}%
  \renewcommand*{\l@subsection}[2]{\if@apptoc\apt@lsub{#1}{#2}\fi}%
  \renewcommand*{\l@subsubsection}[2]{}%
  \providecommand*{\l@title}[2]{}\renewcommand*{\l@title}[2]{}%
  \providecommand*{\l@abstract}[2]{}\renewcommand*{\l@abstract}[2]{}%
}
\newcommand{\appendixcontents}{%
  \begingroup\let\addcontentsline\@gobblethree\let\appendix\relax
  \section*{Contents of the appendices}\small\@starttoc{toc}\endgroup}
\makeatother
\makeatletter
\def\frontmatter@RRAP@format{\small\centering}
\makeatother

\begin{document}

\title{Linear equivalence of nonlinear recurrent neural networks}
\author{David G. Clark}
\email{dclark@flatironinstitute.org}
\altaffiliation[Initial stages of this work were performed at ]{the Kempner Institute for the Study of Natural and Artificial Intelligence, Harvard University, Cambridge, MA 02138, USA}
\affiliation{Center for Computational Neuroscience, Flatiron Institute, New York, NY 10010, USA}
\date[]{November 9, 2026}

\begin{abstract}
The covariance matrix of a large nonlinear recurrent neural network is shown to match that of a linear network with the same couplings, driven by independent noise, for typical random couplings with independent, identically distributed elements. This linear equivalence, previously proposed as an ansatz, is derived using a two-site cavity method and confirmed in simulations. With zero mean activities, and with sign-flip symmetry when units repeat, weak correlations between units reduce even-order moments at leading order to products of single-site and two-site moments, recovering results first obtained diagrammatically. In models with nonzero mean activities, including a recurrent network with Hebbian plasticity and a saturating Lotka--Volterra model, the temporal means obey a self-consistent mean-field equation, and linear equivalence holds for the connected covariance matrix. Parts of this work were done in collaboration with AI models.
\end{abstract}

\maketitle

\section{Introduction}
\label{sec:introduction}

Nonlinear recurrent neural networks are used widely in neuroscience and machine learning. In both fields, networks with random couplings play a central theoretical role. As models of cortex, such networks generate chaotic activity reminiscent of spontaneous cortical activity observed \emph{in vivo}~\cite{arieli1996dynamics}. To make such a network perform a task, its couplings can be held fixed while a linear readout of its activity is trained, the idea underlying reservoir computing~\cite{jaeger2004harnessing, takasu2025neuronal}. Alternatively, the couplings themselves can be trained~\cite{sussillo2015neural}; in this case, training reshapes the activity of the initial random network, so understanding this activity is a prerequisite for understanding that of the trained network~\cite{clark2026structure}. More broadly, nonlinear recurrent neural networks with random couplings are among the many high-dimensional disordered dynamical systems studied in statistical physics, ecology, and other fields.

Beyond single-unit statistics, network activity has \emph{collective} structure, reflected in how the activities of different units co-fluctuate. In neuroscience, this collective structure is of great interest, and large-scale recording technologies~\cite{trautmann2025large, manley2024simultaneous} now permit its detailed characterization, revealing distributed computations invisible in single-neuron responses~\cite{chung2021neural}. At the level of second-order statistics, the collective structure is captured by the $N \times N$ covariance matrix of the activities. This matrix underlies principal component analysis and other dimensionality-reduction methods~\cite{cunningham2014dimensionality, gao2015simplicity}, and its eigenvalue spectrum has been studied directly in cortical recordings~\cite{stringer2019high, pospisil2025revisiting, pachitariu2026critical}. This covariance matrix is the focus of the present paper.

For noise-driven linear networks, the covariance matrix has an exact closed-form expression, and its eigenvalue spectrum follows from random matrix theory~\cite{hu2022spectrum}. For nonlinear networks, whether noise-driven or autonomous, single-site dynamical mean-field theory (DMFT)~\cite{sompolinsky1988chaos} provides order parameters for the statistics of individual units at large $N$. Later work computed the disorder-averaged second moment of the cross covariances between units~\cite{clark2023dimension, clark2025connectivity, mato2026statistics}. However, apart from approximations for networks of stochastic neurons that linearize the response of each neuron to its recurrent input~\cite{ginzburg1994theory, trousdale2012impact, grytskyy2013unified, dahmen2019second} (Sec.~\ref{sec:discussion}), the full $N \times N$ matrix for fixed couplings $\bm{J}$ had not been characterized analytically, leaving a theoretical gap between linear and nonlinear networks.

\begin{figure}
    \centering
    \includegraphics[width=\columnwidth]{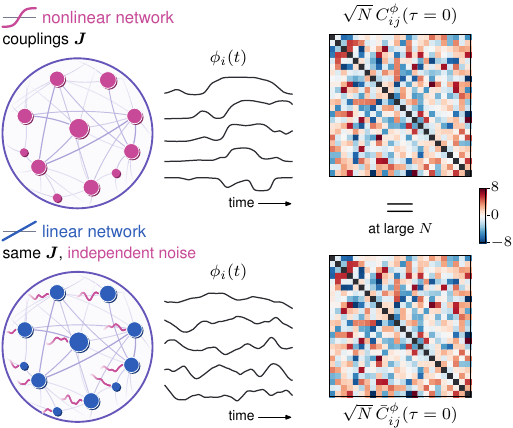}
    \caption{%
    Linear equivalence of a nonlinear recurrent network's covariance matrix. Top, a nonlinear recurrent network with sigmoidal units and couplings $\bm{J}$, without external drive, generates chaotic activities $\phi_i(t)$. Bottom, a linear network with the same $\bm{J}$ is driven by independent noise at each unit. At large $N$ and for typical quenched $\bm{J}$, the two networks have the same covariance matrix. Right, equal-time cross covariances $\sqrt{N}\,C^\phi_{ij}(\tau = 0)$ of the nonlinear network in simulation (top) and the ansatz $\sqrt{N}\,\bar{C}^\phi_{ij}(\tau = 0)$ (bottom) for $24$ units of the Sompolinsky--Crisanti--Sommers (SCS) model at $N = 10000$ and $g = 2.5$ (Sec.~\ref{sec:numerics}). Traces are illustrative.}
    \label{fig:schematic}
\end{figure}

\citet{shen2025covariance} proposed an ansatz that would close this gap. According to this ansatz, at large $N$ and for typical quenched $\bm{J}$, the covariance matrix of a nonlinear recurrent network takes the same form as that of a linear network with the same couplings, driven by independent noise, with DMFT order parameters setting the transfer function of the linear units and the power spectrum of the noise (Fig.~\ref{fig:schematic}). If correct, this \emph{linear-equivalent ansatz} would bring the full machinery of random matrix theory to bear on strongly nonlinear, potentially chaotic recurrent networks. \citet{shen2025covariance} supported the ansatz with simulations and, in a later version, a heuristic two-site cavity argument. In work concurrent with ours, \citet{wakhloo2026solution} derived it with diagrammatic techniques for the rate model of \citet{sompolinsky1988chaos} (Sec.~\ref{sec:discussion}). Here, we derive it with the two-site cavity method for general single-unit dynamics and explain why it holds.

The structure of this paper is as follows. In the remainder of \textbf{Sec.~\ref{sec:introduction}}, we introduce the model and dynamics, define the covariance and response matrices, review single-site DMFT via a single-site cavity method, and state the ansatz. In \textbf{Sec.~\ref{sec:cavity}}, we review the two-site cavity method~\cite{clark2023dimension}, which gives access to joint statistics of activities at a pair of sites. This method rests on two properties of the cavity quantities that hold under the time average at typical quenched $\bm{J}$, namely, joint Gaussianity of the cavity fields (Property~1) and decoupling of an interaction kernel from cavity-field functionals (Property~2), which we derive in Appendix~\ref{app:cavity_properties}.

In \textbf{Sec.~\ref{sec:method1}}, following \citet{shen2025covariance}, we decompose the activity of each unit into a linear response to its recurrent input and a residual. We then show that residual cross covariances at distinct sites are suppressed below the scale of the activity cross covariances, so that the residuals act as independent noise driving a linear network. In \textbf{Sec.~\ref{sec:method2}}, we give an alternative derivation based on a self-consistent equation for the covariance matrix. Cross terms that, in a linear network, would be generated by an external drive are missing under a naive Gaussian closure, but are recovered through non-Gaussian contributions revealed by the cavity method.

In \textbf{Sec.~\ref{sec:moment_decomp}}, we decompose even-order cross-site moments at leading order into products of pairwise covariances, given by the ansatz, or, for repeated sites, into products of single-site moments and two-site moments of odd powers, the latter given by an extension of the ansatz. The same decompositions were obtained by \citet{wakhloo2026solution} with diagrammatic techniques. Here, they follow directly from weak correlations between units, quantified by the scaling of cumulants with the number of distinct sites, together with zero mean activities when all sites are distinct, and sign-flip symmetry when sites repeat. In \textbf{Sec.~\ref{sec:numerics}}, we confirm the predicted scalings in simulations of a chaotic rate model across a range of network sizes. In \textbf{Sec.~\ref{sec:other_models}}, we extend the theory to networks with nonzero mean activities and test it in a network of sigmoid units with nonnegative activities, a network with Hebbian plasticity, and a saturating Lotka--Volterra model. Finally, in \textbf{Sec.~\ref{sec:discussion}}, we discuss implications of the results. Longer derivations and numerical details are given in the appendices.

Over the course of this work, the author increasingly used AI models, for components of the proofs and for the code, figures, and writing, as well as to audit all results (details in Appendix~\ref{app:ai}).

\subsection{Model and dynamics}
\label{sec:model}

We study nonlinear recurrent networks of $N$ units, indexed by $i \in \{1, \ldots, N\}$, with activities $\phi_i(t)$, external drives $\xi_i(t)$, and couplings $J_{ij}$. We collect the activities and external drives into vectors $\bm{\phi}(t)$ and $\bm{\xi}(t)$, and the couplings into the matrix $\bm{J}$. The single-unit dynamics are defined by a causal functional $\mathcal{T}[h](t)$,
\begin{align}
    \phi_i(t) &= \mathcal{T}[h_i](t), \nonumber \\
    h_i(t) &= \eta_i(t) + \xi_i(t), \label{eq:dynamics} \\
    \eta_i(t) &= \sum_{j=1}^N J_{ij}\, \phi_j(t),
    \label{eq:local_field}
\end{align}
where $\eta_i(t)$ is the local field at site $i$ and $h_i(t)$ is the total input, including the external drive $\xi_i(t)$.

Our focus is the rate model of Sompolinsky, Crisanti, and Sommers~\cite{sompolinsky1988chaos}, the \emph{SCS model}, for which $\mathcal{T}[h](t)$ is defined implicitly by
\begin{align}
    \phi(t) &= f(x(t)), \nonumber \\
    \partial_t x(t) &= -x(t) + h(t),
    \label{eq:sompolinsky}
\end{align}
where $x(t)$ is the preactivation and $f(x)$ is an odd-symmetric sigmoidal nonlinearity with $f'(0) = 1$. For $g > 1$ (the coupling strength of Eq.~\eqref{eq:coupling_moments}) and no external drive, this model exhibits a phase of deterministic chaos~\cite{sompolinsky1988chaos}.

A variety of other models in neuroscience and ecology fit into this framework, and the theory applies to them (Sec.~\ref{sec:other_models}). Structurally similar equations also describe spin-glass dynamics (which are distinguished by having \emph{symmetric} couplings), including effects such as aging~\cite{sompolinsky1981dynamic, cugliandolo1993analytical}; gradient-flow and stochastic-gradient-descent training on high-dimensional random data~\cite{mignacco2020dynamical}; and agent-based models of collective decision making or financial markets~\cite{garnier2024unlearnable}. In all of these settings, the couplings $\bm{J}$ (or, for training on random data, the data) play the role of quenched disorder.

We make the following assumptions throughout.
\begin{itemize}
    \item The couplings are i.i.d., the simplest instantiation of quenched disorder, with first and second moments
    \begin{align}
        \tavg{J_{ij}}_{\bm{J}} &= 0, \nonumber \\
        \tavg{J_{ij}^2}_{\bm{J}} &= \frac{g^2}{N},
    \label{eq:coupling_moments}
\end{align}
    where $g$ is the coupling strength, and higher moments scale as $\tavgshort{J_{ij}^m}_{\bm{J}} = \order{N^{-m/2}}$. The diagonal elements $J_{ii}$ can be drawn in the same way or set to zero, with the same results at leading order. Our simulations set $J_{ii} = 0$.
    \item The external drives $\xi_i(t)$ are stationary processes drawn i.i.d.\ across units from a distribution $\mathcal{P}_\xi(\xi)$, not necessarily Gaussian.
    \item The initial conditions are drawn i.i.d.\ across units.
    \item The activities reach a stationary, ergodic state, so that averages over external drives and initial conditions equal time averages, denoted $\tavgshort{\cdots}_t$.
\end{itemize}
Two further assumptions on the stationary state apply only in parts of the paper.
\begin{itemize}
    \item In Secs.~\ref{sec:introduction}--\ref{sec:numerics}, we make the \emph{zero-mean assumption} that $\tavgshort{\phi_i(t)}_t = 0$ for all $i$ and every $N$. We lift this assumption for the covariance matrix in Sec.~\ref{sec:other_models}.
    \item For moments with repeated sites (Sec.~\ref{sec:moment_decomp} and Appendix~\ref{app:pairwise_moments}), we make the stronger \emph{sign-flip symmetry assumption} that, for every $N$, the stationary state is invariant under flipping the signs of all activities, $\phi_i(t) \to -\phi_i(t)$ for all $i$. Consequently, the time average of any product of an odd number of activities vanishes. In the SCS model, the dynamics are invariant under this flip for odd $f(x)$ and $\mathcal{P}_\xi(\xi) = \mathcal{P}_\xi(-\xi)$, so, with ergodicity, the stationary state has this symmetry.
\end{itemize}
Uses of these two assumptions are noted where they occur.

\subsection{Covariance and response matrices, single-site cavity method}
\label{sec:dmft}

The $N \times N$ time-lagged covariance matrix has elements
\begin{equation}
    C^\phi_{ij}(\tau) = \tavg{\phi_i(t+\tau)\, \phi_j(t)}_t,
    \label{eq:cov_def}
\end{equation}
where the average is over time at fixed $\bm{J}$. We also define the instantaneous functional derivatives and their time averages, which comprise the $N \times N$ response matrix,
\begin{align}
    S^\phi_{ij}(t,t') &= \frac{\delta \phi_i(t)}{\delta \xi_j(t')}, \nonumber \\
    S^\phi_{ij}(\tau) &= \tavg{S^\phi_{ij}(t+\tau, t)}_t.
    \label{eq:resp_def}
\end{align}
We further define the \emph{local response}
\begin{equation}
    R_{i}(t,t') = \left. \frac{\delta\mathcal{T}[h](t)}{\delta h(t')}\right|_{h = \eta_i + \xi_i},
    \label{eq:Ri_def}
\end{equation}
which depends only on the local field and external drive at site $i$, not accounting for propagation of perturbations through the couplings. We collect the covariance and response elements into matrices $\bm{C}^\phi(\tau)$ and $\bm{S}^\phi(\tau)$, with Fourier transforms $\bm{C}^\phi(\omega)$ and $\bm{S}^\phi(\omega)$~\footnote{We adopt the Fourier-transform convention $a(\omega) = \int_{-\infty}^{\infty} \dd\tau\, e^{-i\omega\tau}\, a(\tau)$ and $a(\tau) = \frac{1}{2\pi}\int_{-\infty}^{\infty} \dd\omega\, e^{i\omega\tau}\, a(\omega)$.}. For stationary signals $a(t)$ and $b(t)$, we write $\tavgshort{a(\omega)\,b(\omega)^*}_t$ for their cross-spectrum, the Fourier transform in $\tau$ of $\tavgshort{a(t+\tau)\,b(t)}_t$, and averages of products of Fourier-space quantities under the time average $\tavgshort{\cdots}_t$ are meant in this sense. In a chaotic network, the response $S^\phi_{ij}(t,t')$ along a trajectory, before time averaging, grows exponentially in $t - t'$. We assume that its time average, $S^\phi_{ij}(\tau)$, nonetheless decays integrably in $\tau$, so that $\bm{S}^\phi(\omega)$ exists.

The matrices $\bm{C}^\phi(\tau)$ and $\bm{S}^\phi(\tau)$ contain two qualitatively different types of information. The first type is contained in the self-averaging diagonal elements of the covariance and response matrices, $C^\phi_{ii}(\tau)$ and $S^\phi_{ii}(\tau)$. Under the zero-mean assumption, these elements concentrate around single-site order parameters at large $N$,
\begin{align}
    C^\phi_{ii}(\tau) &= \Cstar(\tau) + \order{N^{-1/2}}, \label{eq:C_diagonal_concentration} \\
    S^\phi_{ii}(\tau) &= \Sstar(\tau) + \order{N^{-1/2}}. \label{eq:S_diagonal_concentration}
\end{align}
The limiting values $\Cstar(\tau) = \lim_{N \to \infty} C^\phi_{ii}(\tau)$ and $\Sstar(\tau) = \lim_{N \to \infty} S^\phi_{ii}(\tau)$ are determined via DMFT~\cite{sompolinsky1988chaos}, as we now review using a cavity method.

The cavity method singles out a small number of units, called the \emph{cavity units}, for analysis; the remaining units form a \emph{reservoir}. For the single-site DMFT, we introduce one cavity unit, labeled $0$, and define the \emph{cavity field}
\begin{equation}
    \etac_0(t) = \sum_{i=1}^N J_{0i}\,\tilde{\phi}_i(t),
    \label{eq:cavity_field_single_site}
\end{equation}
the input that site $0$ would receive from a reservoir evolving without it, with the tilde indicating this absence. The full local field $\eta_0(t)$ differs from $\etac_0(t)$ through two mechanisms, namely, a direct self-coupling via the autapse $J_{00}$ and an indirect self-coupling in which a perturbation from the cavity unit propagates through the reservoir and back to its source. Both mechanisms contribute at $\order{N^{-1/2}}$ and can be neglected at the level of single-site DMFT.

Due to independence of the couplings $J_{0i}$ and the unperturbed reservoir activities $\tilde{\phi}_i(t)$, the cavity field $\etac_0(t)$ is a Gaussian process at large $N$. Conventionally, this Gaussianity is meant in a disorder-averaged sense~\cite{sompolinsky1988chaos, crisanti2018path}, with no time average performed; in that sense, the mean vanishes because $\tavgshort{J_{0i}}_{\bm{J}} = 0$, and the covariance is $g^2\,\Cstar(\tau)$. In this paper, we instead mean it in a time-averaged sense, for typical quenched $\bm{J}$. Gaussianity holds in this sense as well (Property~1, stated in Sec.~\ref{sec:cavity} and derived in Appendix~\ref{app:property_1}). The temporal mean $\sum_{i=1}^N J_{0i}\,\tavgshort{\tilde{\phi}_i(t)}_t$ vanishes (zero-mean assumption), and the covariance is the same $g^2\,\Cstar(\tau)$. The single-site DMFT therefore proceeds exactly as in the conventional disorder-averaged treatment. We denote by $\sstavg{\cdots}$ the $N \to \infty$ measure on $(\etac_0, \xi_0)$ under which $\etac_0 \sim \mathcal{GP}(0, g^2\Cstar)$ and $\xi_0 \sim \mathcal{P}_\xi$ independently. Averages of products of Fourier-space quantities under $\sstavg{\cdots}$ are meant in the same sense as under $\tavgshort{\cdots}_t$.

In terms of the cavity-field-dependent quantities
\begin{align}
    \phi^\free_0(t) &= \mathcal{T}[\etac_0 + \xi_0](t), \nonumber \\
    R^\free_0(t,t') &= \left. \frac{\delta \mathcal{T}[h](t)}{\delta h(t')} \right|_{h = \etac_0 + \xi_0},
    \label{eq:cavity_quantities}
\end{align}
the order parameters satisfy
\begin{align}
    \Cstar(\tau) &= \sstavg{\phi^\free_0(t+\tau)\,\phi^\free_0(t)}, \label{eq:dmft_C_sc}\\
    \Sstar(\tau) &= \sstavg{R^\free_0(t+\tau,t)}. \label{eq:dmft_S_sc}
\end{align}
Due to the dependence of the measure $\sstavg{\cdots}$ on $\Cstar(\tau)$, Eq.~\eqref{eq:dmft_C_sc} constitutes a self-consistent equation for $\Cstar(\tau)$. Once it is solved, $\Sstar(\tau)$ is given by Eq.~\eqref{eq:dmft_S_sc}. Under the zero-mean assumption, units are statistically identical at large $N$, so every diagonal element concentrates around the same value, with $\order{N^{-1/2}}$ site-to-site fluctuations (Eqs.~\eqref{eq:C_diagonal_concentration} and~\eqref{eq:S_diagonal_concentration}). Without the zero-mean assumption, the cavity field has an $\order{1}$, site-specific temporal mean, and each diagonal element instead concentrates around a value set by the temporal mean input to that site (Sec.~\ref{sec:other_models}).

The second type of information is contained in the non-self-averaging off-diagonal elements of the covariance and response matrices, $C^\phi_{ij}(\tau)$ and $S^\phi_{ij}(\tau)$ ($i \neq j$). To study cross covariances, \citet{clark2023dimension} used a two-site cavity method followed by a disorder average to obtain, for $N \to \infty$, a low-dimensional summary statistic that encodes the second moment of the cross covariances, namely,
\begin{equation}
    \psi^\phi(\tau_1, \tau_2) = \frac{1}{N} \sum_{\substack{i,j=1 \\ i \neq j}}^N C^\phi_{ij}(\tau_1)\,C^\phi_{ij}(\tau_2).
    \label{eq:psi_def}
\end{equation}
Together with $\Cstar(\omega)$, the Fourier transform of $\psi^\phi(\tau_1, \tau_2)$ gives the second moment of the eigenvalue spectrum of $\bm{C}^\phi(\omega)$ and determines the participation-ratio dimension of activity, as a fraction of $N$~\cite{gao2017theory, litwin2017optimal}.

\subsection{Linear-equivalent ansatz}
\label{sec:ansatz}

In contrast to the aforementioned prior work on low-dimensional summary statistics, here we consider the full matrix $\bm{C}^\phi(\omega)$. \citet{shen2025covariance} proposed that this matrix takes the same form as that of a particular linear network. To make this precise, consider a noise-driven linear network with dynamics
\begin{equation}
    \phi_i(\omega) = \mathcal{T}(\omega)\,\bigl(\eta_i(\omega) + \xi_i(\omega)\bigr),
    \label{eq:linear_dynamics}
\end{equation}
where $\mathcal{T}(\omega)$ is the transfer function of each unit and the external drives $\xi_i(t)$ are i.i.d.\ stationary noise with power spectrum $\sigma^2_\xi(\omega)$. Single-site DMFT gives the order parameters at large $N$,
\begin{align}
    \Cstar(\omega)
    &= \frac{|\mathcal{T}(\omega)|^2\,\sigma^2_\xi(\omega)}
    {1 - g^2\,|\mathcal{T}(\omega)|^2}
    \quad \text{(linear network)},
    \label{eq:linear_op_cov} \\
    \Sstar(\omega)
    &= \mathcal{T}(\omega)
    \hspace{5.08em} \text{(linear network)}.
    \label{eq:linear_op_resp}
\end{align}
Beyond these order parameters, Eq.~\eqref{eq:linear_dynamics} has the closed-form solution
\begin{equation}
    \bm{\phi}(\omega) = \bigl(\bm{I}_N - \mathcal{T}(\omega)\,\bm{J}\bigr)^{-1}\,\mathcal{T}(\omega)\,\bm{\xi}(\omega).
    \label{eq:linear_solution}
\end{equation}
Taking the outer product of this solution with its conjugate transpose and averaging over time yields
\begin{align}
    &\bm{C}^\phi(\omega) \nonumber \\
    &= |\mathcal{T}(\omega)|^2\sigma^2_\xi(\omega)
    \bigl(\bm{I}_N - \mathcal{T}(\omega)\bm{J}\bigr)^{-1}
    \bigl(\bm{I}_N - \mathcal{T}(\omega)\bm{J}\bigr)^{-\dagger},
    \label{eq:linear_cov}
\end{align}
where $\bm{A}^{-\dagger} = (\bm{A}^{-1})^\dagger$.

The ansatz states that the covariance matrix of the nonlinear network takes the same form as Eq.~\eqref{eq:linear_cov} under two replacements. First, the transfer function $\mathcal{T}(\omega)$ is replaced by the single-site response $\Sstar(\omega)$. Second, the power spectrum of the filtered drive, $|\mathcal{T}(\omega)|^2\,\sigma^2_\xi(\omega)$, is replaced by $\CstarDelta(\omega)$ of Eq.~\eqref{eq:Cdelta_def}. We denote this ansatz by $\bar{\bm{C}}^\phi(\omega)$,
\begin{tcolorbox}[prxmathbox,title={Linear-equivalent ansatz (Shen and Hu, 2025)}]
\begin{align}
    \bar{\bm{C}}^\phi(\omega) &= \CstarDelta(\omega)\;\bm{M}(\omega)\,\bm{M}(\omega)^\dagger,
    \label{eq:ansatz} \\
    \text{where} \quad \bm{M}(\omega) &= \bigl(\bm{I}_N - \Sstar(\omega)\,\bm{J}\bigr)^{-1},
    \label{eq:resolvent}\\
    \CstarDelta(\omega) &= (1 - g^2|\Sstar(\omega)|^2)\, \Cstar(\omega).
    \label{eq:Cdelta_def}
\end{align}
\end{tcolorbox}
\noindent We call $\bm{M}(\omega)$ the \emph{resolvent} and call the linear network with these two replacements the \emph{linear-equivalent network}. In Sec.~\ref{sec:method1}, we identify its external drive, filtered by $\Sstar(\omega)$, with a \emph{residual} $\Delta_i(t)$, the part of the activity of unit $i$ not captured by the linear response to its local field. The power spectrum of the residual is $\CstarDelta(\omega)$, the \emph{residual power spectrum}. An analogous linear equivalence, the \emph{response-matrix ansatz}, holds for $\bm{S}^\phi(\omega)$ (derivation in Appendix~\ref{app:response}). As $N \to \infty$, the eigenvalues of $\Sstar(\omega)\,\bm{J}$ fill the disk $|\lambda| \leq g\,|\Sstar(\omega)|$, with $g\,|\Sstar(\omega)| < 1$~\footnote{The inequality $g\,|\Sstar(\omega)| < 1$ follows, in the single-site DMFT, from Cauchy--Schwarz, $|\sstavg{\phi^\free_0(\omega)\,\etac_0(\omega)^*}|^2 \leq \sstavg{|\phi^\free_0(\omega)|^2}\,\sstavg{|\etac_0(\omega)|^2} = g^2\,\Cstar(\omega)^2$, paired with the Furutsu--Novikov theorem (derivation in Appendix~\ref{app:price}), $\sstavg{\phi^\free_0(\omega)\,\etac_0(\omega)^*} = g^2\,\Sstar(\omega)\,\Cstar(\omega)$. Equality would require $\phi^\free_0(\omega)$ to depend on $\etac_0(\omega)$ in a linear and noiseless manner, which we exclude (\emph{nondegeneracy condition}), so the inequality is strict. Moreover, because $\Sstar(\tau)$ is causal and integrable (Sec.~\ref{sec:dmft}), $\Sstar(\omega)$ is analytic and bounded in the lower half of the complex $\omega$-plane, and the maximum-modulus principle extends the bound $g\,|\Sstar(\omega)| < 1$ there, so $\bm{M}(\omega)$ has no poles in this half-plane.}, so $\bm{M}(\omega)$ is well-defined and the linear-equivalent network is stable.

In the SCS model, the single-site response is $\Sstar(\omega) = \beta_\star/(1 + i\omega)$, with the \emph{gain} $\beta_\star = \sstavg{f'(x^\free_0(t))}$, where $x^\free_0(t)$ is the preactivation of the cavity unit driven by $\etac_0(t) + \xi_0(t)$ (SCS formulas in Appendix~\ref{app:sompolinsky_specialization}). At $\omega = 0$, the bound $g\,|\Sstar(\omega)| < 1$ becomes $g_{\mathrm{eff}} < 1$, where $g_{\mathrm{eff}} = g\,\beta_\star$ is the \emph{linear-equivalent coupling strength}. We formulate the theory in terms of activities, rather than the preactivations that are more common for this model, because the linear-equivalent network, Eq.~\eqref{eq:linear_dynamics}, is a closed equation for the activities. For Gaussian external drive, the ansatz for the preactivation covariance matrix follows from the ansatz for the activities and the response-matrix ansatz (Appendix~\ref{app:sompolinsky_specialization}).

\begin{figure*}
    \centering
    \includegraphics[width=0.92\textwidth]{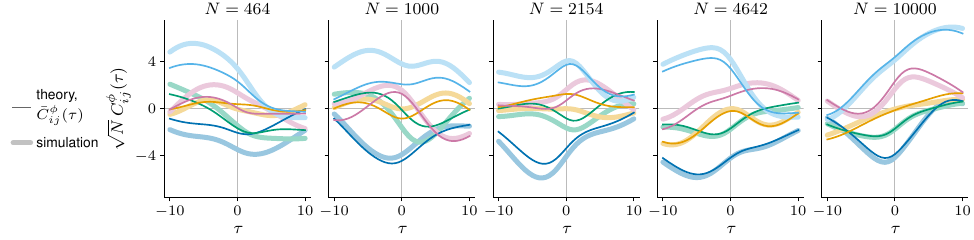}
    \caption{%
    Cross covariances of the SCS model in simulation and theory. All panels show $g = 2.5$ and sampling ratio $\alpha = 800$ (Sec.~\ref{sec:numerics}), with one realization of $\bm{J}$ per $N$. Each panel shows five off-diagonal pairs $(i,j)$, chosen at, and colored by, the $0.1, 0.3, \ldots, 0.9$ quantiles of the simulated $C^\phi_{ij}(\tau = 0)$ over pairs of the first $200$ sites.}
    \label{fig:cij_tau}
\end{figure*}

The linear-network covariance formula, Eq.~\eqref{eq:linear_cov}, is exact for any $\bm{J}$ for which the linear network is stable. The ansatz, Eq.~\eqref{eq:ansatz}, is, by contrast, approximate, holding for typical quenched $\bm{J}$ at large $N$. The diagonal elements are of order one and are approximated by $\bar{\bm{C}}^\phi(\omega)$ to $\order{N^{-1/2}}$ accuracy, while the off-diagonal elements are of order $N^{-1/2}$ and are approximated to $\order{N^{-1}}$ accuracy in the root-mean-square (RMS) sense. The typical off-diagonal error is thus $\order{N^{-1/2}}$ relative to the typical off-diagonal element. These scalings also apply to the time-domain matrices, provided that the frequency-domain bounds hold uniformly in $\omega$ and the errors decay integrably at large $|\omega|$ (numerical tests at $\tau = 0$ in Sec.~\ref{sec:numerics} and at nonzero lags in Appendix~\ref{app:sim_lags}). In simulations, the ansatz tracks individual off-diagonal elements $C^\phi_{ij}(\tau)$ with increasing fidelity as $N$ increases (Fig.~\ref{fig:cij_tau}).

The diagonal precision holds because the linear-equivalent network is constructed to reproduce the single-site order parameters of the nonlinear network. In particular, making the two replacements of the ansatz, $\mathcal{T}(\omega) \to \Sstar(\omega)$ and $|\mathcal{T}(\omega)|^2\sigma^2_\xi(\omega) \to \CstarDelta(\omega)$, in Eq.~\eqref{eq:linear_op_cov} and inserting the definition of $\CstarDelta(\omega)$, Eq.~\eqref{eq:Cdelta_def}, yields the order parameter $\Cstar(\omega)$ of the nonlinear network. Thus, both $C^\phi_{ii}(\omega)$ and $\bar{C}^\phi_{ii}(\omega)$ concentrate around $\Cstar(\omega)$ with $\order{N^{-1/2}}$ fluctuations, giving $C^\phi_{ii}(\omega) - \bar{C}^\phi_{ii}(\omega) = \order{N^{-1/2}}$. An analogous argument applies to the response-matrix ansatz (Appendix~\ref{app:response}).

The nontrivial content of the ansatz concerns the off-diagonal elements. We will establish the RMS scaling
\begin{equation}
    \sqrt{\frac{1}{N(N-1)}\sum_{\substack{i,j=1 \\ i \neq j}}^N |C^\phi_{ij}(\omega) - \bar{C}^\phi_{ij}(\omega)|^2} = \order{N^{-1}}.
    \label{eq:ansatz_precision}
\end{equation}
Together, the diagonal and off-diagonal bounds imply that the error of the ansatz, $\bm{\mathcal{E}}(\omega) = \bm{C}^\phi(\omega) - \bar{\bm{C}}^\phi(\omega)$, has Frobenius norm $\lVert \bm{\mathcal{E}}(\omega)\rVert_F = \order{1}$. Finally, since both $\bm{C}^\phi(\omega)$ and $\bar{\bm{C}}^\phi(\omega)$ are Hermitian, their eigenvalues are real, and the Hoffman--Wielandt inequality bounds the mean squared distance between the optimally matched sets of eigenvalues by $N^{-1} \lVert \bm{\mathcal{E}}(\omega)\rVert_F^2 = \order{N^{-1}}$. The two matrices therefore have the same limiting eigenvalue density.

To demonstrate the off-diagonal precision, we now review the two-site extension of the cavity method.

\section{Review of the two-site cavity method}
\label{sec:cavity}

At a single site, the local field is Gaussian at leading order, with non-Gaussian corrections of $\order{N^{-1/2}}$ that do not affect the single-site DMFT (Sec.~\ref{sec:dmft}). At a pair of sites, one might hope for analogous leading-order \emph{joint} Gaussianity, but some higher-order cross-cumulants are $\order{N^{-1/2}}$, the same scale as the cross covariances themselves, and thus cannot be neglected. This non-Gaussianity arises because the recurrent dynamics correlate the couplings $J_{ij}$ with the activities $\phi_j(t)$ in the sum defining each local field, and these dynamics are nonlinear (in a linear network driven by Gaussian noise, the same correlations arise, but the activities remain jointly Gaussian). These non-Gaussian effects are visible in simulations, where a fourth-order reservoir cross-cumulant is $\order{N^{-1/2}}$ and Price's theorem fails at leading order for reservoir variables (numerical tests in Appendix~\ref{app:cavity_numerics}).

The two-site cavity method separates the joint statistics of activities at a pair of sites into Gaussian and non-Gaussian parts. Cavity fields capture the Gaussian part, while an interaction kernel mediates the non-Gaussian part. Both objects are random projections of high-dimensional unperturbed reservoir trajectories, with the projecting couplings independent of the trajectories being projected. Under the time average at typical quenched $\bm{J}$, this independence suppresses higher-order cumulants, as in the central limit theorem. The result is two leading-order properties, namely, joint Gaussianity of the cavity fields (Property~1) and decoupling of the kernel from cavity-field functionals (Property~2), which are stated below and derived in Appendix~\ref{app:cavity_properties} under the commuting-limits assumption of Appendix~\ref{app:commuting_limits}. Both properties, which \citet{clark2023dimension} assumed, are what one obtains by applying the central limit theorem without distinguishing the time average from the disorder average.

Extending the single-site analysis to a second cavity unit labeled $0'$, we define the cavity field at site $\mu$ as
\begin{equation}
    \etac_\mu(t) = \sum_{i=1}^N J_{\mu i}\,\tilde{\phi}_i(t),
    \label{eq:cavity_field_def}
\end{equation}
where the tildes again denote reservoir trajectories with the cavity units absent, and $\phi^\free_\mu(t)$ and $R^\free_\mu(t,t')$ are the cavity-field-dependent quantities of Eq.~\eqref{eq:cavity_quantities} at site $\mu$. Throughout, $\mu,\nu,\rho\in\{0,0'\}$.

Under the time average at typical quenched $\bm{J}$, the cavity fields $\etac_0(t)$ and $\etac_{0'}(t)$ are jointly Gaussian at leading order (Property~1). In particular, cumulants of a single field beyond second order are $\order{N^{-1/2}}$, and cumulants involving both cavity fields beyond second order are $\order{N^{-1}}$, below the $\order{N^{-1/2}}$ scale of the cross covariance. These fields have mean zero (zero-mean assumption), autocovariance $g^2\,\Cstar(\tau)$, and cross covariance
\begin{align}
    C^{\etac}_{00'}(\tau) &= \tavg{\etac_0(t+\tau)\,\etac_{0'}(t)}_t \nonumber \\
    &= \order{N^{-1/2}}.
    \label{eq:cavity_field_crosscov}
\end{align}
Consequently, Price's theorem and the Furutsu--Novikov theorem (derivation in Appendix~\ref{app:price}) apply to smooth functionals of the cavity fields at distinct sites with $\order{N^{-1}}$ error.

Introducing the cavity units perturbs the reservoir. The perturbed activity of the unit at reservoir site $i$ is
\begin{align}
    \hspace{-1em} \phi_i(t) &= \tilde{\phi}_i(t) + \delta\phi_i(t) + \order{N^{-1}},
    \label{eq:reservoir_perturbation} \\
    \hspace{-1em} \delta\phi_i(t) &= \sum_{j=1}^N \int_0^\infty \dd s\, \tilde{S}^\phi_{ij}(t, t\!-\!s) \sum_{\mu \in \{0,0'\}} J_{j\mu}\, \phi^\free_\mu(t\!-\!s).
    \label{eq:reservoir_perturbation_expr}
\end{align}
Here, $\delta\phi_i(t) = \order{N^{-1/2}}$ is the first-order perturbation, with $\tilde{S}^\phi_{ij}(t,t')$ the unperturbed reservoir response function~\footnote{These expansions require the burn-in time $T_\mathrm{burn}$, after which statistics are taken, to be $\order{1}$ (commuting-limits assumption, Appendix~\ref{app:commuting_limits}), since in the chaotic regime $\tilde{S}^\phi_{ij}(t,t-s)$ grows exponentially in $s$ along a single trajectory. The $\order{N^{-1}}$ terms come from using $\phi^\free_\mu$ in place of $\phi_\mu$, which differ by $\order{N^{-1/2}}$ (Eq.~\eqref{eq:phi_cavity_expansion}), and from the second-order response of the reservoir to the cavity units. The latter is $\order{N^{-1}}$ by the counting of Appendix~\ref{app:reservoir_scaling}, applied to second-order response functions, and remains so in Eq.~\eqref{eq:full_local_field}, where it is summed with the couplings $J_{\mu i}$, because these are independent of $J_{i\mu}$.}. In addition to its direct input from the cavity units, each cavity unit receives input from the reservoir activities, which are themselves perturbed in response to the cavity units. Consequently, the full local field at cavity unit $\mu$ is
\begin{align}
    \eta_\mu(t) &= \etac_\mu(t) + \delta\eta_\mu(t) + \order{N^{-1}},
    \label{eq:full_local_field} \\
    \delta\eta_\mu(t) &= \frac{1}{\sqrt{N}} \sum_{\nu \in \{0,0'\}} \int_0^\infty \dd s\, F_{\mu\nu}(t,t\!-\!s)\, \phi^\free_\nu(t\!-\!s).
    \label{eq:delta_eta_def}
\end{align}
Here, $\delta \eta_\mu(t) = \order{N^{-1/2}}$ is the perturbation to the local field at cavity unit $\mu$, mediated by the $\order{1}$ interaction kernel, which we call the \emph{$F$-kernel},
\begin{align}
    F_{\mu\nu}(t, t') &= F^R_{\mu\nu}(t, t') + \sqrt{N}\, J_{\mu\nu}\, \delta(t - t'),
    \label{eq:F_decomp} \\
    F^R_{\mu\nu}(t,t') &= \sqrt{N} \sum_{i,j=1}^N J_{\mu i}\, J_{j\nu}\, \tilde{S}^\phi_{ij}(t, t').
    \label{eq:reverb_kernel_def}
\end{align}
The \emph{reverberation kernel} $F^R_{\mu\nu}(t,t')$ captures propagation from cavity unit $\nu$ through the reservoir to cavity unit $\mu$, and $\sqrt{N}\, J_{\mu\nu}\, \delta(t - t')$ captures the direct coupling, with the delta function contributing its full weight at $s = 0$ in Eq.~\eqref{eq:delta_eta_def}.

Under the time average at typical quenched $\bm{J}$, the $F$-kernel is uncorrelated at leading order with any smooth $\order{1}$ functional of the cavity fields (Property~2). That is, for such a functional $G(t)$,
\begin{multline}
    \tavg{F_{\mu\nu}(t,t-s)\,G(t)}_t \\
    = \tavg{F_{\mu\nu}(t,t-s)}_t\,\tavg{G(t)}_t
    + \order{N^{-1/2}}.
    \label{eq:property2_factorization}
\end{multline}
In time-averaged products with cavity-field quantities, the kernel can thus be replaced by its time average, $F_{\mu\nu}(s) = \tavgshort{F_{\mu\nu}(t,t-s)}_t$, with Fourier transform $F_{\mu\nu}(\omega)$.

Finally, the expansion for the activities of the cavity units is
\begin{align}
    \phi_\mu(t) &= \phi^\free_\mu(t) + \delta \phi_\mu(t)  + \order{N^{-1}}, \nonumber \\
    \delta \phi_\mu(t) &=  \int_0^\infty \dd s\, R^\free_\mu(t,t-s)\,\delta\eta_\mu(t-s).
    \label{eq:phi_cavity_expansion}
\end{align}
The activities of the full network have exactly zero temporal mean (zero-mean assumption). Since $\phi^\free_\mu(t)$ differs from $\phi_\mu(t)$ by $\order{N^{-1/2}}$, Eq.~\eqref{eq:phi_cavity_expansion} gives $\tavgshort{\phi^\free_\mu(t)}_t = \order{N^{-1/2}}$~\footnote{The cavity field $\etac_\mu(t)$ is linear in the reservoir activities and has exactly zero mean, but $\phi^\free_\mu(t)$ is a nonlinear functional of it. Under the sign-flip symmetry assumption, for odd $\mathcal{T}[h](t)$ and $\mathcal{P}_\xi(\xi) = \mathcal{P}_\xi(-\xi)$, $\tavgshort{\phi^\free_\mu(t)}_t$ vanishes exactly. The zero-mean assumption alone gives only this bound.}.

\section{Derivation of linear equivalence}
\label{sec:method1}

Following \citet{shen2025covariance}, we decompose each activity into a linear response to the local field and a residual, and show that the cross covariances of the residuals are small. Here and in Sec.~\ref{sec:method2}, we make the zero-mean assumption of Sec.~\ref{sec:model}. To accomplish this decomposition, we define the residual
\begin{equation}
    \Delta_i(t) = \phi_i(t) - [\Sstar \ast \eta_i](t),
    \label{eq:residual_def}
\end{equation}
where $\ast$ denotes convolution~\footnote{The residual $\Delta_i(t)$ includes any external-drive contribution.}. In Fourier space, Eq.~\eqref{eq:residual_def} reads
\begin{equation}
    \phi_i(\omega) = \Sstar(\omega) \sum_{j=1}^N J_{ij}\, \phi_j(\omega) + \Delta_i(\omega),
    \label{eq:decomp_fourier}
\end{equation}
with the closed-form solution $\bm{\phi}(\omega) = \bm{M}(\omega)\, \bm{\Delta}(\omega)$, where $\bm{M}(\omega)$ is the resolvent of Eq.~\eqref{eq:resolvent}. Consequently, the covariance matrix takes the form
\begin{equation}
    \bm{C}^\phi(\omega) = \bm{M}(\omega)\, \bm{C}^\Delta(\omega)\, \bm{M}(\omega)^\dagger,
    \label{eq:cov_from_residual}
\end{equation}
where $\bm{C}^\Delta(\tau)$ is the covariance matrix of the residuals, with elements $C^\Delta_{ij}(\tau) = \tavg{\Delta_i(t+\tau)\, \Delta_j(t)}_t$.

The ansatz $\bar{\bm{C}}^\phi(\omega)$ holds with the off-diagonal precision of Eq.~\eqref{eq:ansatz_precision} given two conditions on $\bm{C}^\Delta(\omega)$. First, the diagonal elements $C^\Delta_{ii}(\omega)$ must concentrate around $\CstarDelta(\omega)$ with $\order{N^{-1/2}}$ fluctuations. Second, the off-diagonal elements $C^\Delta_{ij}(\omega)$ must be of $\order{N^{-1}}$ for $i \neq j$, below the $\order{N^{-1/2}}$ scale of the activity cross covariances. We now demonstrate both conditions in turn.

\textbf{Diagonal elements of \texorpdfstring{$\bm{C}^\Delta(\omega)$}{C-Delta(omega)}.}
The elements $C^\Delta_{ii}(\omega)$ are computed within the single-site DMFT (Sec.~\ref{sec:dmft}). Specializing Eq.~\eqref{eq:residual_def} to the cavity units, with the cavity field in place of the local field, we define the cavity residual $\Delta^\free_\mu(t) = \phi^\free_\mu(t) - [\Sstar \ast \etac_\mu](t)$. Under the zero-mean assumption, a diagonal element concentrates, to $\order{N^{-1/2}}$, around
\begin{multline}
    \sstavg{|\Delta^\free_0(\omega)|^2} = \Cstar(\omega)  + |\Sstar(\omega)|^2\, g^2 \Cstar(\omega)     \\
    - \Sstar(\omega)^*\, \sstavg{\phi^\free_0(\omega)\,\etac_0(\omega)^*} - \text{c.c.}
    \label{eq:diagonal_four_terms}
\end{multline}
The Furutsu--Novikov theorem gives
\begin{equation}
    \sstavg{\phi^\free_0(\omega)\,\etac_0(\omega)^*} = g^2\,\Sstar(\omega)\,\Cstar(\omega).
\end{equation}
Consequently, Eq.~\eqref{eq:diagonal_four_terms} becomes
\begin{equation}
    \sstavg{|\Delta^\free_0(\omega)|^2} = (1 - g^2 |\Sstar(\omega)|^2)\,\Cstar(\omega),
\end{equation}
which is the residual power spectrum $\CstarDelta(\omega)$ of Eq.~\eqref{eq:Cdelta_def}.

\textbf{Off-diagonal elements of \texorpdfstring{$\bm{C}^\Delta(\omega)$}{C-Delta(omega)}.}
It remains to show that $C^\Delta_{ij}(\omega) = \order{N^{-1}}$ for $i \neq j$. To do so, we take $i$ and $j$ to be the cavity units $0$ and $0'$. The cavity expansions of Eqs.~\eqref{eq:full_local_field} and~\eqref{eq:phi_cavity_expansion} produce a corresponding expansion for the residuals at the cavity units,
\begin{align}
    \Delta_\mu(t) &= \Delta^\free_\mu(t) + \delta \Delta_\mu(t) + \order{N^{-1}}, \nonumber \\
    \delta \Delta_\mu(t) &= \int_0^\infty \dd s
    \left(R^\free_\mu(t, t-s) - \Sstar(s) \right) \delta \eta_\mu(t-s),
    \label{eq:Delta_cavity}
\end{align}
where $\delta \Delta_\mu(t) = \order{N^{-1/2}}$ is the first-order perturbation. From this expansion, the covariance of the residuals at the cavity units is
\begin{multline}
    C^\Delta_{00'}(\tau)
    =
    \tavg{\Delta^\free_0(t+\tau)\,
    \Delta^\free_{0'}(t)}_t
    +
    \tavg{\Delta^\free_0(t+\tau)\,
    \delta\Delta_{0'}(t)}_t \\ + \tavg{\delta \Delta_0(t+\tau)\,
    \Delta^\free_{0'}(t)}_t
    +
    \order{N^{-1}},
    \label{eq:Cdelta_offdiag_expand}
\end{multline}
where the product $\tavgshort{\delta\Delta_0(t+\tau)\,\delta\Delta_{0'}(t)}_t$ of the two first-order perturbations is absorbed into $\order{N^{-1}}$.

We first evaluate the cavity-field term in Eq.~\eqref{eq:Cdelta_offdiag_expand},
\begin{multline}
    \tavg{\Delta^\free_0(\omega)\,
    \Delta^\free_{0'}(\omega)^*}_t
    =
    \tavg{\phi^\free_0(\omega)\,
    \phi^\free_{0'}(\omega)^*}_t \\
    + |\Sstar(\omega)|^2 C^{\etac}_{00'}(\omega)
    - \Sstar(\omega)^*
    \tavg{\phi^\free_0(\omega)\,
    \etac_{0'}(\omega)^*}_t \\
    - \Sstar(\omega)
    \tavg{\etac_0(\omega)\,
    \phi^\free_{0'}(\omega)^*}_t.
    \label{eq:free_residual_four_terms}
\end{multline}
The marginal-product term of Price's theorem, $\tavgshort{\phi^\free_0(t)}_t\,\tavgshort{\phi^\free_{0'}(t)}_t$, is $\order{N^{-1}}$, since each factor is $\order{N^{-1/2}}$ (zero-mean assumption). By Property~1, Price's theorem and the Furutsu--Novikov theorem then give
\begin{align}
    \tavg{\phi^\free_0(\omega)\,
    \phi^\free_{0'}(\omega)^*}_t
    &=
    |\Sstar(\omega)|^2 C^{\etac}_{00'}(\omega)
    + \order{N^{-1}}, \nonumber \\
    \tavg{\phi^\free_0(\omega)\,
    \etac_{0'}(\omega)^*}_t
    &=
    \Sstar(\omega) C^{\etac}_{00'}(\omega)
    + \order{N^{-1}}, \nonumber \\
    \tavg{\etac_0(\omega)\,
    \phi^\free_{0'}(\omega)^*}_t
    &=
    \Sstar(\omega)^* C^{\etac}_{00'}(\omega)
    + \order{N^{-1}}.
    \label{eq:free_phi_phi_price}
\end{align}
Substituting into Eq.~\eqref{eq:free_residual_four_terms} produces a cancellation, $\tavgshort{\Delta^\free_0(\omega)\, \Delta^\free_{0'}(\omega)^*}_t = \order{N^{-1}}$.

We next evaluate the cross terms in Eq.~\eqref{eq:Cdelta_offdiag_expand}. First,
\begin{widetext}
\begin{multline}
    \tavg{\Delta^\free_0(t+\tau)\,
    \delta\Delta_{0'}(t)}_t \\
    =
    \frac{1}{\sqrt N}
    \sum_{\nu\in\{0,0'\}}
    \int_0^\infty \dd s \int_0^\infty \dd s'\,
    \tavg{
    \Delta^\free_0(t+\tau)
    \left(R^\free_{0'}(t,t-s)-\Sstar(s)\right)
    F_{0'\nu}(t-s,t-s-s')\,
    \phi^\free_\nu(t-s-s')
    }_t.
    \label{eq:one_deltaDelta_right}
\end{multline}
\end{widetext}
Property~2 replaces $F_{0'\nu}(t-s,t-s-s')$ by its time average, and Property~1 factorizes the remaining average into a product of a site-$0$ marginal and a site-$0'$ marginal. The errors of these two steps are $\order{N^{-1/2}}$ and combine with the $N^{-1/2}$ prefactor in Eq.~\eqref{eq:one_deltaDelta_right} to give $\order{N^{-1}}$. The two values of $\nu$ then contribute as follows.

For $\nu = 0$, $\phi^\free_0(t-s-s')$ belongs to the site-$0$ marginal alongside $\Delta^\free_0(t+\tau)$, leaving the site-$0'$ marginal $\tavgshort{R^\free_{0'}(t,t-s) - \Sstar(s)}_t$. Since $\Sstar(s)$ is the mean of the local response (Eq.~\eqref{eq:dmft_S_sc}), around which its time average at each site concentrates (zero-mean assumption), this marginal is $\order{N^{-1/2}}$.

For $\nu = 0'$, $\phi^\free_{0'}(t-s-s')$ belongs to the site-$0'$ marginal alongside $R^\free_{0'}(t,t-s) - \Sstar(s)$, leaving the site-$0$ marginal $\tavgshort{\Delta^\free_0(t+\tau)}_t$. Since $\tavgshort{\Delta_0(t)}_t = 0$ (zero-mean assumption), Eq.~\eqref{eq:Delta_cavity} shows that this marginal is $\order{N^{-1/2}}$.

In both cases, the $N^{-1/2}$ prefactor makes the contribution $\order{N^{-1}}$. The remaining term in Eq.~\eqref{eq:Cdelta_offdiag_expand}, $\tavgshort{\delta\Delta_0(t+\tau)\,\Delta^\free_{0'}(t)}_t$, follows by exchanging $0$ and $0'$. This completes the demonstration that the off-diagonal elements are suppressed.

This suppression does not depend on the specific form of the $F$-kernel. If the kernel were zero, the local fields would be jointly Gaussian at leading order, and the suppression would follow from Property~1 alone. A nonzero kernel adds the terms just evaluated, and each of these terms contains, at one of the two sites, only a residual or only a fluctuation of the local response around its mean. Both of these quantities have temporal means of $\order{N^{-1/2}}$, which combine with the $N^{-1/2}$ prefactor to make the added terms $\order{N^{-1}}$. Physically, Property~2 says that reverberations through the reservoir are uncorrelated at leading order with any smooth functional of the cavity fields.

\textbf{Error propagation.}
We now propagate these results to the error of the ansatz. Writing
\begin{equation}
    \bm{C}^\Delta(\omega) = \CstarDelta(\omega)\,\bm{I}_N + \bm{\mathcal{E}}^\Delta(\omega),  \label{eq:controlled_err_expr}
\end{equation}
where $\bm{\mathcal{E}}^\Delta(\omega)$ has diagonal elements of $\order{N^{-1/2}}$ and off-diagonal elements of $\order{N^{-1}}$, Eq.~\eqref{eq:cov_from_residual} gives
\begin{equation}
    \bm{C}^\phi(\omega) = \bar{\bm{C}}^\phi(\omega) + \bm{M}(\omega)\,\bm{\mathcal{E}}^\Delta(\omega)\,\bm{M}(\omega)^\dagger.
    \label{eq:method1_error_sandwich}
\end{equation}
The diagonal precision follows from the concentration of $C^\phi_{ii}(\omega)$ and $\bar{C}^\phi_{ii}(\omega)$ around $\Cstar(\omega)$ (Sec.~\ref{sec:ansatz}). For the off-diagonal precision, we use $\lVert \bm{\mathcal{E}}^\Delta(\omega)\rVert_F = \order{1}$ (since sites are exchangeable, the squared elements sum to $N\,\order{N^{-1}} + N(N-1)\,\order{N^{-2}}$) and the operator-norm bound $\lVert \bm{M}(\omega)\rVert_{\mathrm{op}} = \order{1}$~\footnote{The operator norm $\lVert \bm{M}(\omega)\rVert_{\mathrm{op}}$ is the inverse of the smallest singular value of $\bm{I}_N - \Sstar(\omega)\bm{J}$. For i.i.d.\ $\bm{J}$ and $g\,|\Sstar(\omega)| < 1$ (Sec.~\ref{sec:ansatz}), this singular value stays bounded away from zero at large $N$ (Proposition~3.1 of \citet{bordenave2016outlier}).}, which give
\begin{align}
    \lVert \bm{M}(\omega)\bm{\mathcal{E}}^\Delta(\omega)
    \bm{M}(\omega)^\dagger\rVert_F
    &\leq \lVert \bm{M}(\omega)\rVert_{\mathrm{op}}^2
    \lVert \bm{\mathcal{E}}^\Delta(\omega)\rVert_F \nonumber \\
    &= \order{1}.
\end{align}
Since the $N(N-1)$ off-diagonal elements have a sum of squares of at most $\order{1}$, the off-diagonal RMS error is $\order{N^{-1}}$, which is Eq.~\eqref{eq:ansatz_precision}.

\section{Self-consistent matrix equation}
\label{sec:method2}

In the derivation of Sec.~\ref{sec:method1}, non-Gaussianity appears as a nuisance that does not spoil the result. However, the non-Gaussian corrections are generated by the nonlinearity, which also generates the residual and therefore the effective drive of the linear-equivalent network. The alternative derivation we now give emphasizes the \emph{constructive} role of non-Gaussianity. Its goal is a self-consistent equation for $\bm{C}^\phi(\omega)$ at fixed $\bm{J}$, in which $F$-kernel-mediated contributions supply the cross terms that, in a linear network, would be produced by an i.i.d.\ external drive. We first review the corresponding equation for a linear network, since comparing the equations for the two networks identifies the solution for the nonlinear network.

Taking the outer product of the linear dynamics, Eq.~\eqref{eq:linear_dynamics}, with its conjugate transpose and averaging over time gives the self-consistent equation for the covariance matrix of a \emph{linear} network,
\begin{multline}
    \bm{C}^\phi(\omega)
    = |\mathcal{T}(\omega)|^2\bm{J}\bm{C}^\phi(\omega)\bm{J}^T \\
    + |\mathcal{T}(\omega)|^2\sigma^2_\xi(\omega)
    \bigl(\bm{J}\bm{S}^\phi(\omega)
    + \bm{S}^\phi(\omega)^\dagger\bm{J}^T + \bm{I}_N\bigr),
    \label{eq:linear_self_consistent}
\end{multline}
where the terms involving $\bm{S}^\phi(\omega)$ follow from $\tavgshort{\bm{\phi}(\omega)\,\bm{\xi}(\omega)^\dagger}_t = \sigma^2_\xi(\omega)\,\bm{S}^\phi(\omega)$, with $\bm{S}^\phi(\omega) = \mathcal{T}(\omega)\,(\bm{I}_N - \mathcal{T}(\omega)\,\bm{J})^{-1}$ the response matrix of the linear network. The cross terms $\bm{J}\,\bm{S}^\phi(\omega) + \bm{S}^\phi(\omega)^\dagger\,\bm{J}^T$ reflect the presence of the external drive.

Turning to the nonlinear network, we now attempt to derive an analogous self-consistent equation under the \emph{false} assumption of joint Gaussianity of the local fields at distinct sites and show that the resulting equation is not the correct one. For this naive derivation only, we set $\xi_i(t) = 0$, so that any effective drive in the linear-equivalent network must arise from the internal dynamics. Under joint Gaussianity and the zero-mean assumption, Price's theorem gives, for $i \neq j$,
\begin{equation}
    C^\phi_{ij}(\omega) \overset{?}{=} |\Sstar(\omega)|^2\,\tavg{\eta_i(\omega)\,\eta_j(\omega)^*}_t + \order{N^{-1}}.
\end{equation}
In matrix form,
\begin{equation}
    \bm{C}^\phi(\omega)
    \overset{?}{=} |\Sstar(\omega)|^2\bm{J}\bm{C}^\phi(\omega)\bm{J}^T
    + \CstarDelta(\omega)\bm{I}_N
    + \bm{\mathcal{E}}^C(\omega),
    \label{eq:naive_self_consistent}
\end{equation}
where the identity term enforces that $C^\phi_{ii}(\omega)$ concentrates around $\Cstar(\omega)$, and $\bm{\mathcal{E}}^C(\omega)$ has diagonal elements of $\order{N^{-1/2}}$ and off-diagonal elements of $\order{N^{-1}}$. We now see the issue: Eq.~\eqref{eq:naive_self_consistent} is missing the cross terms of Eq.~\eqref{eq:linear_self_consistent}, which also appear in the correct equation for the nonlinear network derived below, Eq.~\eqref{eq:nonlinear_self_consistent}. Therefore, Eq.~\eqref{eq:naive_self_consistent} cannot, for generic $\bm{J}$, be the covariance equation of a linear network with the same couplings, driven by independent noise.

As we now sketch, the two-site cavity method recovers the missing cross terms from the non-Gaussian joint statistics (details in Appendix~\ref{app:method2_details}). We multiply the cavity expansions of $\phi_0(t+\tau)$ and $\phi_{0'}(t)$, Eq.~\eqref{eq:phi_cavity_expansion}, and average over time. Applying Properties~1 and~2 and using $\tavgshort{\phi^\free_\mu(t)}_t = \order{N^{-1/2}}$ (zero-mean assumption) yields, in Fourier space,
\begin{widetext}
\begin{equation}
    C^\phi_{00'}(\omega)
    = |\Sstar(\omega)|^2 \sum_{i,j=1}^N J_{0i}\, J_{0'j}\,
    \tilde{C}^\phi_{ij}(\omega)
    + \frac{\Cstar(\omega)}{\sqrt{N}}\,\Sstar(\omega)\,F_{00'}(\omega)
    + \frac{\Cstar(\omega)}{\sqrt{N}}
    \bigl(\Sstar(\omega)\, F_{0'0}(\omega)\bigr)^*
    + \order{N^{-1}}.
    \label{eq:single_site_cav_eqn}
\end{equation}
\end{widetext}
This equation is Eq.~(8) of \citet{clark2023dimension}, written in the conventions of this paper~\footnote{There, $C^\phi_{ij}(\tau) = \tavgshort{\phi_i(t)\,\phi_j(t+\tau)}_t$, which exchanges the roles of $0$ and $0'$, and their Fourier-transform convention gives an overall factor of $2\pi$.}. The cross terms are now present. Indeed, using $N^{-1/2}\Sstar(\omega)\,F_{00'}(\omega) = [\bm{J}\,\bm{S}^\phi(\omega)]_{00'} + \order{N^{-1}}$ (Eq.~\eqref{eq:F_result_S}), they take the same form as in Eq.~\eqref{eq:linear_self_consistent}. However, their prefactor, the counterpart of $|\mathcal{T}(\omega)|^2\sigma^2_\xi(\omega)$, is $\Cstar(\omega)$, whereas the correct equation has $\CstarDelta(\omega) = (1 - g^2|\Sstar(\omega)|^2)\,\Cstar(\omega)$. If $\sum_{i,j=1}^N J_{0i}\, J_{0'j}\, \tilde{C}^\phi_{ij}(\omega)$ were simply replaced by $[\bm{J}\,\bm{C}^\phi(\omega)\,\bm{J}^T]_{00'}$, Eq.~\eqref{eq:single_site_cav_eqn} would thus overestimate the cross terms.

The resolution is that this replacement is not exact, and correcting it changes the prefactor. To see this, note that the first term on the right-hand side of Eq.~\eqref{eq:single_site_cav_eqn} contains the cavity-field cross covariance $C^{\etac}_{00'}(\omega) = \sum_{i,j=1}^N J_{0i}\, J_{0'j}\, \tilde{C}^\phi_{ij}(\omega)$. This quantity differs from $[\bm{J}\,\bm{C}^\phi(\omega)\,\bm{J}^T]_{00'}$, its counterpart in Eq.~\eqref{eq:linear_self_consistent}, in two ways. First, the unperturbed reservoir covariance $\tilde{C}^\phi_{ij}(\omega)$ is computed with the cavity units removed. Replacing it by the full $C^\phi_{ij}(\omega)$ generates a negative correction proportional to the reverberation kernel $F^R_{\mu\nu}(\omega)$. Second, the sums run only over the reservoir. Extending them to include the cavity indices generates a negative correction proportional to the direct coupling $\sqrt{N}\,J_{\mu\nu}$. By Eq.~\eqref{eq:F_decomp}, the two corrections form the full $F$-kernel, and, in Eq.~\eqref{eq:single_site_cav_eqn}, they equal $-g^2|\Sstar(\omega)|^2$ times the cross terms already present (Eq.~\eqref{eq:cavity_field_reexpressed}). Adding them converts the prefactor from $\Cstar(\omega)$ to $\CstarDelta(\omega)$. The result is the self-consistent equation for the covariance matrix of the nonlinear network,
\begin{multline}
    \bm{C}^\phi(\omega)
    = |\Sstar(\omega)|^2\bm{J}\bm{C}^\phi(\omega)\bm{J}^T \\
    + \CstarDelta(\omega)
    \bigl(\bm{J}\bm{S}^\phi(\omega)
    + \bm{S}^\phi(\omega)^\dagger\bm{J}^T + \bm{I}_N\bigr)
    + \bm{\mathcal{E}}^C(\omega),
    \label{eq:nonlinear_self_consistent}
\end{multline}
with an error matrix $\bm{\mathcal{E}}^C(\omega)$ of the same element-wise scaling as in Eq.~\eqref{eq:naive_self_consistent}. Equation~\eqref{eq:nonlinear_self_consistent} is Eq.~\eqref{eq:linear_self_consistent} under the two replacements of Sec.~\ref{sec:ansatz}, $\mathcal{T}(\omega) \to \Sstar(\omega)$ and $|\mathcal{T}(\omega)|^2\sigma^2_\xi(\omega) \to \CstarDelta(\omega)$. With $\bm{S}^\phi(\omega)$ given by its own self-consistent equation (Eq.~\eqref{eq:response_matrix_eq}), the unique solution of Eq.~\eqref{eq:nonlinear_self_consistent} agrees with the ansatz $\bar{\bm{C}}^\phi(\omega)$ to the precision of Eq.~\eqref{eq:ansatz_precision} (Appendices~\ref{app:method2_solution} and~\ref{app:method2_error}).

\section{Higher-order moments}
\label{sec:moment_decomp}

The ansatz is a second-moment result, but higher-order cross-site moments can also be evaluated. Because units are only weakly correlated, as quantified by the scaling of cumulants with the number of distinct sites, the moment-cumulant formula reduces even-order moments at leading order to products of single-site and two-site moments. This reduction uses the zero-mean assumption when all sites are distinct and the stronger sign-flip symmetry assumption of Sec.~\ref{sec:model} when sites repeat. These decompositions were first obtained by \citet{wakhloo2026solution} with diagrammatic techniques. We suppress time arguments in cumulants and moments, understood as time averages with relative times fixed.

These decompositions rest on two ingredients. The first ingredient is this cumulant scaling, derived in Appendix~\ref{app:reservoir_scaling} under the commuting-limits assumption of Appendix~\ref{app:commuting_limits}. Applied to activities, with $\kappa_t(\cdot, \ldots, \cdot)$ denoting a joint cumulant under the time average, it states that, for typical quenched $\bm{J}$,
\begin{equation}
    \kappa_t\bigl(\phi_{i_1},\ldots,\phi_{i_p}\bigr) = \order{N^{-(r-1)/2}},
    \label{eq:cumulant_scaling_main}
\end{equation}
where $r$ is the number of distinct sites in $\{i_1,\ldots,i_p\}$. The $p = r = 2$ case recovers the familiar $\order{N^{-1/2}}$ scaling of cross covariances, and each additional distinct site costs a further factor of $\order{N^{-1/2}}$, as one expects from couplings of size $\order{N^{-1/2}}$. The second ingredient is the standard moment-cumulant formula,
\begin{equation}
    \tavg{\phi_{i_1}\cdots\phi_{i_p}}_t
    =
    \sum_{\pi}
    \prod_{B \in \pi}
    \kappa_t\bigl(\phi_{i_b},\, b \in B\bigr),
    \label{eq:moment_cumulant_formula}
\end{equation}
where the sum runs over all partitions $\pi$ of $\{1,\ldots,p\}$ into $|\pi|$ blocks $B$.

We first treat the simplest case, even-order moments in which all sites are distinct. Equation~\eqref{eq:cumulant_scaling_main} then gives
\begin{equation}
    \prod_{B \in \pi}
    \kappa_t\bigl(\phi_{i_b},\, b \in B\bigr)
    =
    \order{N^{-(p - |\pi|)/2}}.
    \label{eq:partition_product_scaling}
\end{equation}
Leading-order contributions maximize $|\pi|$. Singleton blocks are means and vanish (zero-mean assumption), so every nonzero block contains at least two activities. The leading nonzero partitions are therefore the pairings ($|\pi| = p/2$), which give a Wick decomposition,
\begin{align}
    &\tavg{\prod_{a=1}^{p}\phi_{i_a}}_t \nonumber \\
    &= \sum_{\mathcal{P}\in\mathcal{P}_2(p)}
    \prod_{\{a,b\}\in\mathcal{P}} C^\phi_{i_a i_b}
    + \order{N^{-(p+2)/4}} \nonumber \\
    &= \sum_{\mathcal{P}\in\mathcal{P}_2(p)}
    \prod_{\{a,b\}\in\mathcal{P}} \bar{C}^\phi_{i_a i_b}
    + \order{N^{-(p+2)/4}},
    \label{eq:all_distinct_wick}
\end{align}
where $\mathcal{P}_2(p)$ denotes the set of pairings of $\{1,\ldots,p\}$. The second equality uses the off-diagonal precision of the ansatz, Eq.~\eqref{eq:ansatz_precision}, in its time-domain form (Sec.~\ref{sec:ansatz}), so this equality holds in the RMS sense over index tuples. For example, with $i, j, k, l$ all distinct, Eq.~\eqref{eq:all_distinct_wick} reads
\begin{align}
    &\tavg{\phi_i\phi_j\phi_k\phi_l}_t \nonumber \\
    &= C^\phi_{ij}\,C^\phi_{kl}
    + C^\phi_{ik}\,C^\phi_{jl}
    + C^\phi_{il}\,C^\phi_{jk}
    + \order{N^{-3/2}} \nonumber \\
    &= \bar{C}^\phi_{ij}\,\bar{C}^\phi_{kl}
    + \bar{C}^\phi_{ik}\,\bar{C}^\phi_{jl}
    + \bar{C}^\phi_{il}\,\bar{C}^\phi_{jk}
    + \order{N^{-3/2}}.
    \label{eq:four_point_wick}
\end{align}

We emphasize that the first equality in Eq.~\eqref{eq:all_distinct_wick}, the Wick decomposition, does not depend on the details of the network dynamics. It is a direct consequence of the moment-cumulant formula applied to weakly correlated, zero-mean variables, with the weak correlation quantified by Eq.~\eqref{eq:cumulant_scaling_main}.

For repeated-index even moments, the same two ingredients, together with the vanishing of every odd-order cumulant (sign-flip symmetry assumption), yield a Wick-like decomposition. The bookkeeping is more involved, but the argument is unchanged, namely, Eq.~\eqref{eq:cumulant_scaling_main} identifies the leading-order partitions in the moment-cumulant formula. Sites carrying an even power factor out as single-site moments, and sites carrying an odd power are paired, with a sum over all pairings. Each pair contributes a two-site moment of odd powers, such as $\tavg{\phi_i^3 \phi_j^5}_t$, which is of $\order{N^{-1/2}}$. These two-site moments are, in turn, given to the same $\order{N^{-1}}$ off-diagonal precision as the ansatz by an expression involving the resolvent $\bm{M}(\omega)$ and single-site DMFT quantities. The Wick-like decomposition and the expression for the two-site moments are derived in Appendix~\ref{app:pairwise_moments}, where the latter is also tested numerically.

\section{Numerics in the Sompolinsky--Crisanti--Sommers model}
\label{sec:numerics}

We verified the analytical results through simulations of the SCS model with the error-function nonlinearity $f(x) = \mathrm{erf}\!\left(\sqrt{\pi} x / 2 \right)$, i.i.d.\ Gaussian couplings satisfying Eq.~\eqref{eq:coupling_moments}, $J_{ii} = 0$, and no external drive, so that fluctuations arise purely from deterministic chaos (simulation details in Appendix~\ref{app:sim_details}, evaluation of the ansatz in Appendix~\ref{app:sim_ansatz}). This model satisfies both the zero-mean and the sign-flip symmetry assumptions. We take $g = 2.5$, in the chaotic regime. Network sizes are logarithmically spaced, from $N = 100$ to $N = 10000$, with total simulation time $T_{\mathrm{tot}} = \alpha N$ for the \emph{sampling ratios} $\alpha = 200$, $800$, and $3200$.

Throughout the paper, we summarize how a quantity scales with $N$ by a \emph{fitted exponent}, computed as the least-squares slope of the logarithm of its median over realizations of $\bm{J}$ against $\log N$. The standard error of a fitted exponent is the standard deviation of this slope when the realizations at each $N$ are resampled with replacement (a bootstrap). In the SCS model, the fit uses $N \geq 464$. Smaller sizes, which are not yet in the scaling regime, are shown but excluded from the fit (fit protocol in Appendix~\ref{app:sim_floor}).

Each covariance estimate differs from its infinite-time value by a sampling error of $\order{T_{\mathrm{tot}}^{-1/2}}$, whose size we measure by the \emph{sampling floor}, the off-diagonal RMS of half the difference between the estimates of a covariance matrix from two disjoint halves of the initial conditions. Because the estimates are unbiased, the squared off-diagonal RMS of an estimated matrix equals, on average, that of its infinite-time value plus the squared sampling floor. With $T_{\mathrm{tot}} = \alpha N$, the sampling floor scales as $N^{-1/2}$ at fixed $\alpha$, like the off-diagonal elements, whereas the error of the ansatz is $\order{N^{-1}}$. We choose $\alpha$ so that the sampling floor stays well below the error of the ansatz up to the largest $N$ (though this is not possible for the faster-decaying residual cross covariances).

Where the sampling floor is well below a quantity, as for the error of the ansatz, we show it for reference only. Where the sampling floor is comparable to a quantity, as for the residual cross covariances, we subtract it in quadrature, that is, we subtract the squared sampling floor from the squared off-diagonal RMS. In the undriven SCS model, we check both treatments by varying $\alpha$, where a larger $\alpha$ lowers the sampling floor. Without subtraction, fitted exponents that do not vary systematically with $\alpha$ show that the quantity is not limited by sampling. With subtraction, fitted exponents that do not vary systematically with $\alpha$ show that the subtraction removes the sampling contribution (details in Appendix~\ref{app:sim_floor}).

Figure~\ref{fig:scaling} shows how the error of the ansatz and the residual cross covariance scale with $N$ in the SCS model. The fitted exponents are collected in Table~\ref{tab:exponents}.

\begin{figure*}
    \centering
    \includegraphics[width=\textwidth]{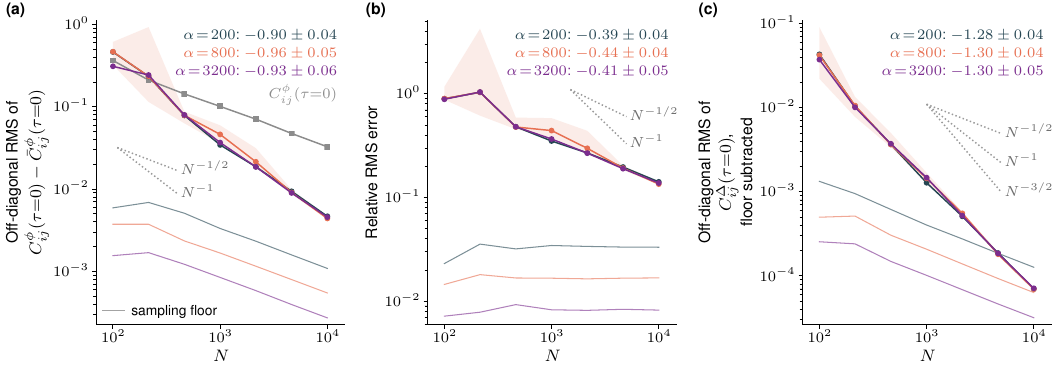}
    \caption{%
    Scaling of the error of the ansatz and of the residual cross covariances with network size in the SCS model. All panels show $g = 2.5$ at time lag $\tau = 0$, for three sampling ratios $\alpha$. Lines show medians over ten realizations of $\bm{J}$ per $N$ (five for $\alpha = 3200$), and the shaded region shows the interquartile range at $\alpha = 800$. Keys give the fitted exponents. Dotted gray lines show reference power laws.
    \textbf{(a)}~Off-diagonal RMS of the error of the ansatz, $C^\phi_{ij}(\tau = 0) - \bar{C}^\phi_{ij}(\tau = 0)$, together with the off-diagonal RMS of $C^\phi_{ij}(\tau = 0)$ at $\alpha = 800$.
    \textbf{(b)}~Error of (a) relative to the off-diagonal RMS of $C^\phi_{ij}(\tau = 0)$.
    \textbf{(c)}~Off-diagonal RMS of the residual cross covariance $C^\Delta_{ij}(\tau = 0)$, with $\Delta_i(t) = \phi_i(t) - \beta_\star x_i(t)$, after subtraction of the sampling floor.}
    \label{fig:scaling}
\end{figure*}

We first test the off-diagonal precision of the ansatz, Eq.~\eqref{eq:ansatz_precision}. Figure~\ref{fig:scaling}(a) shows the off-diagonal RMS of the error of the ansatz as a function of $N$, with fitted exponents between $-0.90$ and $-0.96$. Figure~\ref{fig:scaling}(b) shows the same error relative to the off-diagonal RMS of $C^\phi_{ij}(\tau = 0)$, with fitted exponents between $-0.39$ and $-0.44$. Neither fitted exponent depends systematically on $\alpha$, so these errors are not limited by sampling. These fitted exponents are slightly smaller in magnitude than the predicted exponents, $-1$ and $-1/2$, which we attribute to the proximity of the linear-equivalent network to the edge of stability, with linear-equivalent coupling strength $g_{\mathrm{eff}} = 0.96$ close to its critical value of $1$ (Sec.~\ref{sec:ansatz}). Driving the network with external noise lowers $g_{\mathrm{eff}}$ and brings the fitted exponents close to the predicted values, reaching $-0.99$ and $-0.49$ at $g_{\mathrm{eff}} = 0.79$ after subtraction of the sampling floor (numerical tests in Appendix~\ref{app:sim_noise}). Results at nonzero time lags and for the diagonal elements are given in Appendix~\ref{app:sim_lags}.

We next test the bound $C^\Delta_{ij}(\omega) = \order{N^{-1}}$ for $i \neq j$ (Sec.~\ref{sec:method1}). Figure~\ref{fig:scaling}(c) shows the off-diagonal RMS of $C^\Delta_{ij}(\tau = 0)$ as a function of $N$, with the residual in its zero-drive form, $\Delta_i(t) = \phi_i(t) - \beta_\star x_i(t)$ (Eq.~\eqref{eq:sompolinsky_residual_zero_drive}). After subtraction of the sampling floor, the off-diagonal RMS of $C^\Delta_{ij}(\tau = 0)$ decays faster than the $\order{N^{-1}}$ bound over the simulated range of $N$, with fitted exponents between $-1.28$ and $-1.30$ that agree across $\alpha$.

This faster decay is not explained by sign-flip symmetry alone, since the residual cross covariances of the noise-driven networks (Appendix~\ref{app:sim_noise}), which are also sign-flip symmetric, decay as $N^{-1}$. The models of Sec.~\ref{sec:other_models}, which lack the symmetry, also show $N^{-1}$ decay. The faster decay does not improve the error of the ansatz, since the diagonal elements of $\bm{C}^\Delta(\omega)$ still fluctuate by $\order{N^{-1/2}}$ around $\CstarDelta(\omega)$ (Sec.~\ref{sec:method1}). Correspondingly, the error of the ansatz in Fig.~\ref{fig:scaling}(a) does not decay faster than $N^{-1}$. The origin of the faster decay is open.

\section{Nonzero mean activities and other models}
\label{sec:other_models}

Because the theory is formulated for a general single-unit functional $\mathcal{T}[h](t)$, linear equivalence extends to a large number of models beyond the SCS model. Unlike the SCS model, however, these models generically violate the zero-mean assumption. Examples include recurrent networks with nonnegative firing rates, which are more biologically realistic, and ecological models with nonnegative species abundances. To treat such models, we now lift the zero-mean assumption for the linear equivalence of the covariance matrix and test the resulting theory in several models.

Since mean activities are now allowed to be nonzero, we also generalize the dynamics of Sec.~\ref{sec:model} by adding site-specific biases $b_i$,
\begin{align}
    \phi_i(t) &= \mathcal{T}[h_i](t), \nonumber \\
    h_i(t) &= \eta_i(t) + b_i + \xi_i(t), \nonumber \\
    \eta_i(t) &= \sum_{j=1}^N J_{ij}\,\phi_j(t),
    \label{eq:biased_dynamics}
\end{align}
where the biases are drawn i.i.d.\ across units from a distribution $\mathcal{P}_b(b)$. The external drives are drawn from $\mathcal{P}_\xi(\xi)$ as in Sec.~\ref{sec:model}, which we now take to have zero mean, since a nonzero mean can be absorbed into the bias. The functional $\mathcal{T}[h](t)$ need not be odd. The assumptions of Sec.~\ref{sec:model} on the couplings, the initial conditions, and the stationary, ergodic state are unchanged.

We denote the temporal mean of a signal by a bar and its centered part by a breve,
\begin{align}
    \bar{\phi}_i &= \tavg{\phi_i(t)}_t, \nonumber \\
    \breve{\phi}_i(t) &= \phi_i(t) - \bar{\phi}_i,
    \label{eq:bar_breve_def}
\end{align}
and likewise for other signals. In contrast to the zero-mean case, the temporal mean local field and the temporal mean input,
\begin{align}
    \bar{\eta}_i &= \sum_{j=1}^N J_{ij}\,\bar{\phi}_j, \label{eq:mean_local_field} \\
    \bar{h}_i &= \bar{\eta}_i + b_i, \label{eq:mean_input}
\end{align}
are $\order{1}$ and differ across sites, which makes the temporal mean activities site-specific. The centered parts obey $\breve{\eta}_i(t) = \sum_{j=1}^N J_{ij}\,\breve{\phi}_j(t)$ and $\breve{h}_i(t) = \breve{\eta}_i(t) + \xi_i(t)$. The covariance matrix decomposes into a \emph{disconnected part} and a \emph{connected part}, also called the connected covariance matrix,
\begin{align}
    C^\phi_{ij}(\tau) &= \bar{\phi}_i \bar{\phi}_j + C^{\breve{\phi}}_{ij}(\tau), \nonumber \\
    C^{\breve{\phi}}_{ij}(\tau) &= \tavg{\breve{\phi}_i(t+\tau)\,\breve{\phi}_j(t)}_t.
    \label{eq:centered_covariance}
\end{align}
The disconnected part $\bar{\phi}_i \bar{\phi}_j$ is $\order{1}$ for both diagonal and off-diagonal elements, while the connected part $C^{\breve{\phi}}_{ij}(\tau)$ is $\order{1}$ for diagonal elements and $\order{N^{-1/2}}$ for off-diagonal elements.

The theory for both parts rests on a single-site DMFT with site-specific temporal mean inputs (derivation in Appendix~\ref{app:nonzero_mean}). For a site with temporal mean input $\bar{h}$, this theory provides its temporal mean activity $\mathcal{M}_\star(\bar{h})$, its response $S_\star(\omega \,|\, \bar{h})$, and its residual power spectrum $C^{\breve{\Delta}}_\star(\omega \,|\, \bar{h})$ (Eqs.~\eqref{eq:conditional_mean}, \eqref{eq:conditional_S}, and~\eqref{eq:centered_residual_spectrum}). We collect the responses $S_\star(\omega \,|\, \bar{h}_i)$ of the individual sites, the \emph{site-specific responses}, into the diagonal matrix $\bm{D}^S(\omega)$, and the residual power spectra $C^{\breve{\Delta}}_\star(\omega \,|\, \bar{h}_i)$ into the diagonal matrix $\bm{D}^{\breve{\Delta}}(\omega)$. The resolvent of Eq.~\eqref{eq:resolvent} then generalizes to $\breve{\bm{M}}(\omega) = \bigl(\bm{I}_N - \bm{D}^S(\omega)\,\bm{J}\bigr)^{-1}$ (Eq.~\eqref{eq:centered_resolvent}).

The disconnected part follows from an equation similar to the Thouless--Anderson--Palmer (TAP) equations for spin glasses~\cite{thouless1977solution}, which we call the \emph{TAP-like equation} (derivation in Appendix~\ref{app:nzm_single_site}). Unlike the original TAP equations for spin glasses, it has no Onsager reaction term, since $J_{ij}$ and $J_{ji}$ are independent, as in asymmetric kinetic Ising models~\cite{mezard2011exact}. The connected part follows from a linear equivalence, in parallel with Sec.~\ref{sec:ansatz},
\begin{tcolorbox}[prxmathbox,title={Linear-equivalent ansatz, nonzero means}]
\begin{align}
    \bar{\phi}_i &= \mathcal{M}_\star\Bigl(b_i + \sum_{j=1}^N J_{ij}\,\bar{\phi}_j\Bigr),
    \label{eq:tap_equation} \\
    \bar{\bm{C}}^{\breve{\phi}}(\omega)
    &=
    \breve{\bm{M}}(\omega)\,
    \bm{D}^{\breve{\Delta}}(\omega)\,
    \breve{\bm{M}}(\omega)^\dagger.
    \label{eq:centered_ansatz}
\end{align}
\end{tcolorbox}
\noindent As in Eq.~\eqref{eq:ansatz}, the bar on $\bar{\bm{C}}^{\breve{\phi}}(\omega)$ denotes the ansatz rather than a temporal mean. Assuming that $\lVert \breve{\bm{M}}(\omega)\rVert_{\mathrm{op}} = \order{1}$, the counterpart of the bound on $\bm{M}(\omega)$ in Sec.~\ref{sec:method1}, the TAP-like equation, Eq.~\eqref{eq:tap_equation}, has a solution that differs from the temporal mean activities of the network by $\order{N^{-1/2}}$ per element in the RMS sense, since these activities satisfy it up to an error of the same order. Under the same assumption, the ansatz of Eq.~\eqref{eq:centered_ansatz} holds with RMS error $\order{N^{-1/2}}$ on the diagonal and $\order{N^{-1}}$ off the diagonal, as in Eq.~\eqref{eq:ansatz_precision}, whether its site-specific functions are evaluated at the temporal mean inputs of the network or at those of this solution (Appendix~\ref{app:nzm_error}). The disconnected part, by contrast, inherits the $\order{N^{-1/2}}$ error of this solution, the same order as the off-diagonal elements of the connected covariance matrix. The off-diagonal elements of the full covariance matrix, Eq.~\eqref{eq:centered_covariance}, are therefore predicted to $\order{N^{-1}}$ only when the disconnected part is formed from the measured temporal mean activities rather than from the solution of Eq.~\eqref{eq:tap_equation}.

The centered activities $\breve{\phi}_i(t)$ have zero mean and obey the cumulant scaling of Eq.~\eqref{eq:cumulant_scaling_main} (Appendix~\ref{app:reservoir_scaling}). Thus, the distinct-site result of Sec.~\ref{sec:moment_decomp}, Eq.~\eqref{eq:all_distinct_wick}, holds for them, with the connected covariances and their ansatz, Eq.~\eqref{eq:centered_ansatz}, in place of the covariances and their ansatz. The repeated-index results of Sec.~\ref{sec:moment_decomp}, which use the sign-flip symmetry assumption, do not extend to nonzero mean activities.

We consider the following three models.
\begin{itemize}
    \item \textbf{Recurrent network with sigmoid units.} The units obey $\partial_t x_i(t) = -x_i(t) + \eta_i(t) + b_i$ with nonnegative activities $\phi_i(t) = \tfrac{1}{2}[1 + \tanh x_i(t)]$. We simulate this network without biases and with $b_i \sim \mathcal{N}(0, 5^2)$.
    \item \textbf{Recurrent network with Hebbian plasticity.} In the model of \citet{clark2024theory}, Hebbian modifications are added to the quenched couplings, giving total couplings $W_{ij}(t) = J_{ij} + A_{ij}(t)$, with $p\,\partial_t A_{ij}(t) = -A_{ij}(t) + \frac{k}{N}\,\phi_i(t)\,\phi_j(t)$ and $A_{ij}(0) = 0$. This plasticity can be absorbed exactly into the single-unit dynamics, so that the quenched couplings $\bm{J}$ are the only couplings between units. This gives $\phi(t) = f(x(t))$ with
    \begin{align*}
        \partial_t x(t) 
        &= -x(t) + h(t) \\
        &\quad + k \int_0^t \frac{\dd t'}{p}\, e^{-(t-t')/p}\, Q^\phi(t,t')\, \phi(t'),
    \end{align*}
    where $Q^\phi(t,t') = N^{-1} \sum_{i=1}^N \phi_i(t)\,\phi_i(t')$ is the empirical two-time correlation function. At large $N$, $Q^\phi(t,t')$ concentrates around $\Cstar(t - t')$ of Eq.~\eqref{eq:uncentered_dmft}, and we treat the single-unit functional $\mathcal{T}[h](t)$ as fixed by replacing $Q^\phi(t,t')$ by $\Cstar(t - t')$~\footnote{The $\order{N^{-1/2}}$ fluctuations of $Q^\phi(t,t')$ about $\Cstar(t - t')$ change the off-diagonal elements of the connected covariance matrix by $\order{N^{-1}}$ and the temporal mean activities by $\order{N^{-1/2}}$, the orders of the errors of the ansatz, Eq.~\eqref{eq:centered_ansatz}, and of the TAP-like equation, Eq.~\eqref{eq:tap_equation}, respectively.}. This functional thus depends on the order parameters of the network, which the single-site theory supplies self-consistently. We take $f(x) = \tfrac{1}{2}[1 + \tanh x]$, as for the sigmoid units, with $p = 2$, $k = \pm 2$, and no biases.
    \item \textbf{Saturating Lotka--Volterra model.} This model is a variant of the generalized Lotka--Volterra equations, which are used frequently in ecology to model multi-species populations~\cite{bunin2017ecological, roy2019numerical}. These equations have single-unit dynamics $\partial_t n(t) = n(t)\,[1 - n(t) - h(t)]$, where $n(t) \geq 0$ is a species abundance, and the abundances themselves mediate the interactions, $\phi(t) = n(t)$. The saturating variant is $\partial_t n_i(t) = n_i(t)\,[1 - n_i(t) - \eta_i(t)] + \epsilon$, with immigration rate $\epsilon = 0.01$, and a bounded function of the abundance mediates the interactions, $\phi_i(t) = n_i(t)/[1 + n_i(t)]$. These bounded interactions remove the unbounded-growth instability of the generalized model~\cite{sidhom2020ecological, suweis2024generalized, zenari2026generalized}. At the coupling strength we simulate, the largest Lyapunov exponent is positive in every realization, indicating chaos.
\end{itemize}

\begin{figure*}
    \centering
    \includegraphics[width=0.87\textwidth]{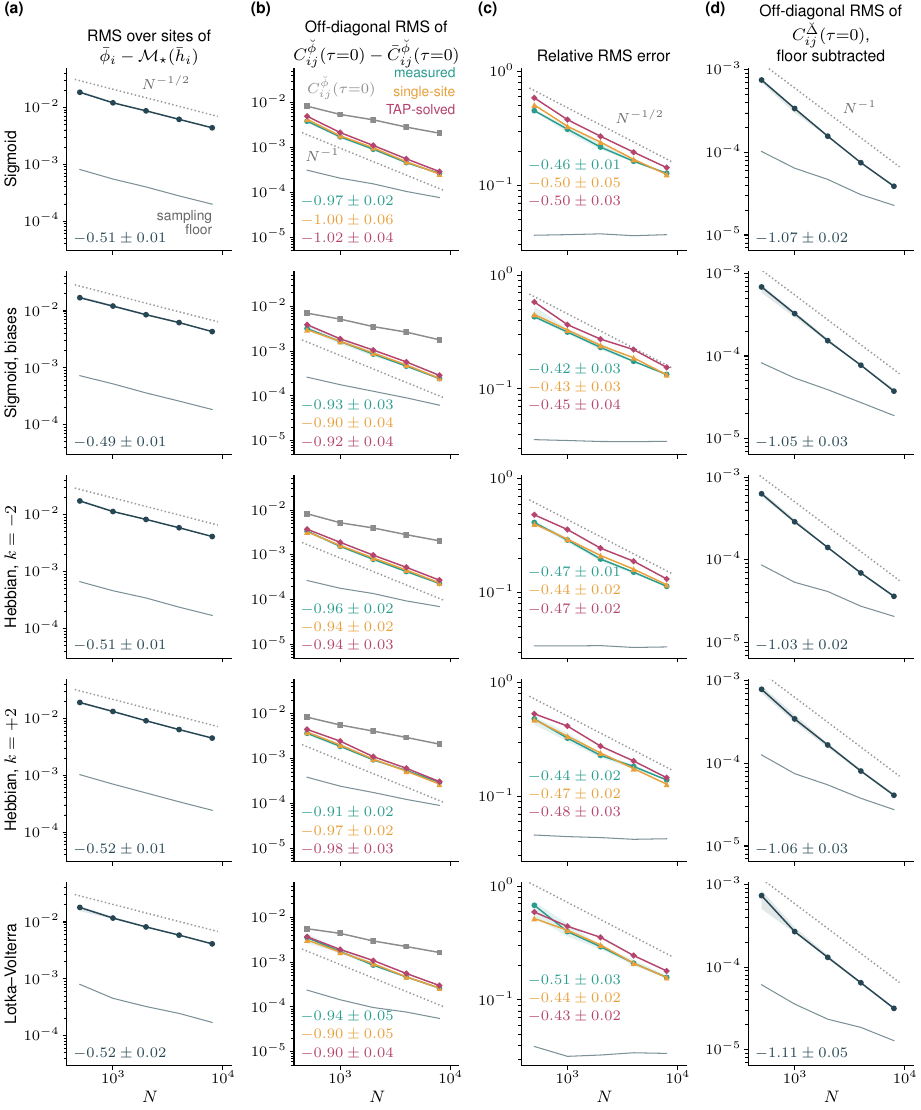}
    \caption{%
    Scaling of the error of the TAP-like equation, the error of the ansatz, and the residual cross covariances with network size in the models with nonzero mean activities (Sec.~\ref{sec:other_models}). All panels show $g = 20$ at time lag $\tau = 0$ and sampling ratio $\alpha = 200$. In (b) and (c), the ansatz is evaluated in three ways. The first way uses the site-specific responses and residual power spectra measured in the network (measured), the second computes them from the single-site theory at the measured temporal mean inputs (single-site), and the third computes them from the single-site theory at the temporal mean inputs of the solved TAP-like equation (TAP-solved). Lines show medians over eight realizations of $\bm{J}$ per $N$, and shaded regions show interquartile ranges, in (b) and (c) only for the first way. Keys give the fitted exponents. Dotted gray lines show reference power laws.
    \textbf{(a)}~RMS over sites of $\bar{\phi}_i - \mathcal{M}_\star(\bar{h}_i)$, the error of the TAP-like equation, Eq.~\eqref{eq:tap_equation}, at the measured temporal mean activities.
    \textbf{(b)}~Off-diagonal RMS of the error of the ansatz, $C^{\breve{\phi}}_{ij}(\tau = 0) - \bar{C}^{\breve{\phi}}_{ij}(\tau = 0)$, for the three ways, together with the off-diagonal RMS of $C^{\breve{\phi}}_{ij}(\tau = 0)$.
    \textbf{(c)}~Error of (b) relative to the off-diagonal RMS of $C^{\breve{\phi}}_{ij}(\tau = 0)$.
    \textbf{(d)}~Off-diagonal RMS of the residual cross covariance $C^{\breve{\Delta}}_{ij}(\tau = 0)$, formed with the site-specific responses measured in the network, after subtraction of the sampling floor.}
    \label{fig:nzm_scaling}
\end{figure*}

We verified the TAP-like equation and the ansatz, Eqs.~\eqref{eq:tap_equation} and~\eqref{eq:centered_ansatz}, through simulations of the three models above, with i.i.d.\ Gaussian couplings satisfying Eq.~\eqref{eq:coupling_moments}, $g = 20$, and no external drive (simulation details in Appendix~\ref{app:sim_nzm}). Network sizes are logarithmically spaced, from $N = 500$ to $N = 8000$, with total simulation time $T_{\mathrm{tot}} = \alpha N$ for the sampling ratio $\alpha = 200$.

Figure~\ref{fig:nzm_scaling} shows how the error of the TAP-like equation, the error of the ansatz, and the residual cross covariance scale with $N$. The fitted exponents are collected in Table~\ref{tab:exponents} and follow the protocol of Sec.~\ref{sec:numerics}, with all sizes included in the fit.

We first test the TAP-like equation, Eq.~\eqref{eq:tap_equation}. Figure~\ref{fig:nzm_scaling}(a) shows the RMS over sites of its error at the measured temporal mean activities, with fitted exponents between $-0.49$ and $-0.52$, close to the predicted $-1/2$.

We next test the off-diagonal precision of the ansatz for the connected covariance matrix, Eq.~\eqref{eq:centered_ansatz}, which we evaluate in three ways, labeled \emph{measured}, \emph{single-site}, and \emph{TAP-solved} in Fig.~\ref{fig:nzm_scaling} and Table~\ref{tab:exponents}. The first uses site-specific responses and residual power spectra measured in the network (measurement method in Appendix~\ref{app:sim_nzm}), which tests the structure of Eq.~\eqref{eq:centered_ansatz} alone. The second uses the site-specific functions above, evaluated at the measured $\bar{h}_i$ and with order parameters measured in the network (the mean connected autocovariance across sites and, for the network with Hebbian plasticity, also the mean of $\bar{\phi}_i^2$ across sites) in place of the solutions of Eqs.~\eqref{eq:closure_disconnected} and~\eqref{eq:closure_connected}, which also tests the single-site theory. The third uses no site-specific measurements, solving the TAP-like equation for the temporal mean activities, given $\bm{J}$ and the biases, and evaluating the site-specific functions at the resulting temporal mean inputs, with the same measured order parameters as the second. These inputs differ from the measured ones by an RMS over sites that decays as $N^{-1/2}$, as predicted, with fitted exponents between $-0.46$ and $-0.53$ (not shown). Figure~\ref{fig:nzm_scaling}(b) shows the off-diagonal RMS of the error of the ansatz, with fitted exponents between $-0.90$ and $-1.02$ for all three ways. Figure~\ref{fig:nzm_scaling}(c) shows the same error relative to the off-diagonal RMS of $C^{\breve{\phi}}_{ij}(\tau = 0)$, with fitted exponents between $-0.42$ and $-0.51$. The second and third ways agree closely in exponent, with a somewhat larger prefactor for the third. The error of the first way decays slightly faster than those of the other two in some models and slightly slower in others (Table~\ref{tab:exponents}), so using the single-site theory in place of the measured responses and residual power spectra does not systematically change the scaling.

Figure~\ref{fig:nzm_scaling}(d) shows the off-diagonal RMS of $C^{\breve{\Delta}}_{ij}(\tau = 0)$, the cross covariance of the centered residuals of Eq.~\eqref{eq:centered_residual}, formed with the site-specific responses measured in the network in place of $S_\star(\omega \,|\, \bar{h}_i)$. After subtraction of the sampling floor, this off-diagonal RMS decays near the $\order{N^{-1}}$ bound, with fitted exponents between $-1.03$ and $-1.11$. This decay is similar to that in the noise-driven SCS networks (Appendix~\ref{app:sim_noise}) and contrasts with the faster decay in the undriven SCS model (Sec.~\ref{sec:numerics}).

\section{Discussion}
\label{sec:discussion}

We have shown via the two-site cavity method that, at large $N$ and for typical quenched $\bm{J}$, the covariance matrix of a nonlinear recurrent network approximates that of a linear network with the same couplings, driven by independent noise.

\textbf{Relation to Shen and Hu.}
\citet{shen2025covariance} conjectured the linear-equivalent ansatz, motivated by the observation that the frequency-dependent participation-ratio dimension takes the same form in a nonlinear network as in a linear network with appropriate effective parameters~\footnote{Specifically, $D^\phi(\omega) = N^{-1}(\Tr\,\bm{C}^\phi(\omega))^2/\Tr[\bm{C}^\phi(\omega)^2]$, which approaches $\Cstar(\omega)^2/[\psi^\phi(\omega,-\omega) + \Cstar(\omega)^2]$ at large $N$, with $\psi^\phi(\omega_1,\omega_2)$ the Fourier transform, in both time arguments, of $\psi^\phi(\tau_1,\tau_2)$ of Eq.~\eqref{eq:psi_def}.}. They gave numerical evidence by introducing the residual, denoted $\zeta_i(t)$ in their paper, and showing that its cross covariances were small. In the second version of their preprint, they also gave a heuristic two-site cavity argument that these cross covariances are $\order{N^{-1}}$ (Appendix~L.1 of \citet{shen2025covariance}). That argument replaces the local response of each cavity unit in the $\order{N^{-1/2}}$ terms by its mean, $\Sstar(s)$, so that these terms cancel in the residual, and evaluates the remaining cavity-field term with Price's theorem. The present derivation adds three elements. First, it retains the fluctuation $R^\free_\mu(t,t-s) - \Sstar(s)$ of the local response, which is $\order{1}$ at each time, and shows that its contribution, Eq.~\eqref{eq:one_deltaDelta_right}, to the residual cross covariance $C^\Delta_{00'}(\tau)$ is nonetheless $\order{N^{-1}}$. Second, it derives in Appendix~\ref{app:cavity_properties} the two properties these steps require, namely, joint Gaussianity of the cavity fields (Property~1) and $F$-kernel decoupling (Property~2). Third, it propagates the residual cross covariances to the $\order{N^{-1}}$ error of the ansatz, Eq.~\eqref{eq:ansatz_precision}. Moreover, the self-consistent equation of Sec.~\ref{sec:method2} gives a second derivation that does not use the residual.

We also state the theory for the general single-unit dynamics of Eqs.~\eqref{eq:dynamics} and~\eqref{eq:local_field}, so it applies beyond the SCS model. The linear-equivalent network depends on the single-unit dynamics only through the single-site response $\Sstar(\omega)$ and the residual power spectrum $\CstarDelta(\omega)$ under the zero-mean assumption, and through site-specific functions of the temporal mean inputs otherwise (Sec.~\ref{sec:other_models}).

\textbf{Relation to Wakhloo.}
In concurrent work, \citet{wakhloo2026solution} derives the linear-equivalent ansatz, the response-matrix ansatz, and the higher-moment decompositions of Sec.~\ref{sec:moment_decomp} using diagrammatic techniques, for the SCS model with odd $f(x)$ and Gaussian external drive, including the undriven network as the limit of vanishing drive. The approach uses a path integral to expand a given joint moment as a power series in $\bm{J}$ around the DMFT background. A counting argument shows that the Wick decomposition holds at leading order, and the two-point factors are then explicitly resummed. Embedding the diagrams as ribbon graphs shows that only planar diagrams are of leading order, specifically ``chains'' and ``rainbows.'' Rainbows are shown by induction to cancel against diagrams in which bare rainbows are replaced by ``bubbles,'' leaving only chains. These chains are products of $\bm{J}$ and $\bm{J}^T$ with indices constrained to not repeat, and summing them via a ``no-repeats lemma'' for products of an i.i.d.\ random matrix yields the two-point result, completing the resummation.

Our approach is organized around cavity quantities, random projections of reservoir trajectories whose independence from the projecting couplings suppresses higher cumulants (Sec.~\ref{sec:cavity}). We derive this suppression from a diagrammatic calculation of cumulant scalings (Appendix~\ref{app:cavity_properties}). Given this suppression, demonstrating linear equivalence, analogous to the resummation step in \citet{wakhloo2026solution}, reduces to showing that the residual cross covariances satisfy $C^\Delta_{ij}(\tau) = \order{N^{-1}}$ for $i \neq j$ (Sec.~\ref{sec:method1}). Additionally, the Wick decomposition of Sec.~\ref{sec:moment_decomp} follows from the activity cumulant scaling of Eq.~\eqref{eq:cumulant_scaling_main} together with the zero-mean assumption, and thus holds for any zero-mean variables whose cumulants obey this scaling. \citet{mato2026statistics} derived a large-$N$ factorization of disorder-averaged correlation statistics at zero frequency, whereas the decompositions here hold at fixed $\bm{J}$ and at all frequencies.

\textbf{Relation to linear-response approximations.}
For networks of stochastic spiking or binary neurons, the covariance matrix for fixed couplings has been approximated by that of a linear network~\cite{ginzburg1994theory, trousdale2012impact, grytskyy2013unified, dahmen2019second}. In this approach, the output of each neuron is expanded to first order in the fluctuations of its local field about its mean. The zeroth-order term is the output that the neuron would produce if its local field were held fixed at this mean. In the absence of common external input, this term fluctuates purely due to the stochasticity of the neuron itself, so it is independent across neurons. The first-order term is the linear response of the neuron to the fluctuations of its local field, with the response function of the neuron when its local field is held fixed at this mean~\cite{lindner2005theory}. A refinement also averages this response function over the fluctuations of the local field~\cite{grytskyy2013unified}. The expansion is justified when the fluctuations of the local field are small compared to those generated by the stochasticity of the neuron~\cite{trousdale2012impact}. For couplings of order $N^{-1}$, this holds at large $N$~\cite{ginzburg1994theory}. For couplings of order $N^{-1/2}$, as here, the fluctuations of the local field are $\order{1}$, and the approximation is generally not controlled and must be tested in simulations.

Our starting point for deriving linear equivalence, the decomposition of each activity in Eq.~\eqref{eq:decomp_fourier}, has the same form as this expansion, namely, a linear response to the local field plus a residual. However, it is not a perturbative expansion: the nonlinear effect of the local field is $\order{1}$ and is contained in the residual. Consequently, unlike the zeroth-order terms of the expansion above, the residuals are not independent by construction, and their decorrelation at leading order must be derived (Sec.~\ref{sec:method1}). The $\order{N^{-1}}$ cross covariances of the residuals, together with the $\order{N^{-1/2}}$ fluctuations of their power spectra across sites, bound the error of the ansatz, which is thus $\order{N^{-1}}$ off the diagonal, below the $\order{N^{-1/2}}$ scale of the cross covariances of the activities (Eq.~\eqref{eq:ansatz_precision}), so the ansatz is controlled at large $N$. Since nothing is expanded about a mean local field, linear equivalence also applies to networks without external drive, such as the SCS model in its chaotic phase, whose fluctuations arise purely from deterministic chaos and which therefore lack the stochasticity on which the expansion above relies.

\textbf{Comparing the linear and nonlinear views.}
Near the chaotic transition, the nonlinear and linear-equivalent views are dual descriptions of the same critical point. In the SCS model without external drive, the network is quiescent for $g < 1$ and chaotic for $g > 1$, with linear-equivalent coupling strength $g_{\mathrm{eff}} = g\,\beta_\star$. As $g \to 1^+$ (the onset of chaos), $g_{\mathrm{eff}} \to 1^-$ (the edge of stability). The amplitude of the chaotic fluctuations vanishes in this limit, as do $\CstarDelta(\omega)$ and the noise-induced fluctuations of the linear-equivalent network. In both the nonlinear and the linear-equivalent views, the single-site~\cite{sompolinsky1988chaos} and cross-site~\cite{clark2023dimension} correlation timescales diverge as $g \to 1^+$, while the participation-ratio dimension, as a fraction of $N$, vanishes~\cite{clark2023dimension}.

In noise-driven linear networks near the edge of stability, the eigenvalue spectrum of the covariance matrix decays as a power law~\cite{hu2022spectrum}, with exponent set by the density of coupling-matrix eigenvalues just inside the stability boundary~\cite{tiberi2023hidden}. By linear equivalence, the same power-law spectrum arises in the autonomous nonlinear network as $g \to 1^+$~\cite{shen2025covariance}, with fluctuations generated by deterministic chaos rather than external noise. Such edge-of-stability statistics generically require fine-tuning in a linear network; \citet{chen2024searching} showed that activity-dependent plasticity can reach the edge automatically, but with only a single long timescale rather than a full power-law spectrum. By contrast, the nonlinear network approaches edge-of-stability statistics without fine-tuning~\cite{shen2025covariance}, since $g_{\mathrm{eff}}$ stays close to $1$ throughout the chaotic regime (e.g., $g_{\mathrm{eff}} \to 1/\sqrt{\pi-2} \approx 0.94$ as $g \to \infty$ for $f(x) = \tanh(x)$)~\cite{clark2023dimension}.

Despite sharing second-order statistics, the nonlinear network may have functional advantages over its linear-equivalent counterpart. For instance, it can generate fluctuations autonomously, without an external drive. Neural circuits in sensory areas may be largely input-driven, but circuits further from the sensory periphery, such as motor and cognitive areas, presumably generate intrinsic dynamics. Additionally, the nonlinear network has positive Lyapunov exponents, in fact an extensive number of them~\cite{engelken2023lyapunov, clark2026lyapunov}, and can therefore perform pattern separation by amplifying small differences between trajectories exponentially, which a stable linear system cannot. The same positive exponents bring sensitivity to initial conditions that could compromise reliable computation, but this sensitivity is mitigated in the weakly chaotic regime $g \to 1^+$.

The edge-of-stability statistics of the nonlinear network bear on the interpretation of experimental data. \citet{hu2022spectrum} modeled population activity in larval zebrafish as noise-driven linear dynamics with random couplings, and \citet{pachitariu2026critical} interpreted population activity in mouse cortex as such dynamics near the edge of stability. In both cases, the signature of these dynamics is the eigenvalue spectrum of the covariance matrix. As \citet{shen2025covariance} proposed, linear equivalence offers an alternative, in which the same signature arises from a chaotic nonlinear network generating its own fluctuations, with $g_{\mathrm{eff}}$ close to $1$ without fine-tuning. \citet{pachitariu2026critical} specifically considered symmetric couplings; seeking a linear equivalence in the symmetric case is a natural next step, discussed below.

\textbf{Related equivalence results.}
Our result is analogous to a growing body of linear-equivalence results in machine learning. An early example concerned Gram matrices with $\order{N^{-1/2}}$ off-diagonal elements, acted on element-wise by a nonlinearity~\cite{elkaroui2010spectrum}. The resulting matrix can be approximated in operator norm by the linear term in its Taylor expansion together with a multiple of the identity and low-rank terms. Another regime is when the nonlinearity acts on $\order{1}$ elements. In this case, several results have shown equivalence to an appropriate linear model with noise. Examples include the eigenvalue spectra of random-feature model Gram matrices~\cite{pennington2017nonlinear, louart2018random} and the training and test errors of random-feature models~\cite{goldt2022gaussian, hu2022universality}.

Like these works, we derive an equivalent linear-plus-noise system, starting from a nonlinearity acting on $\order{1}$ elements. There are two differences. First, whereas most prior results characterize low-dimensional observables such as eigenvalue spectra or scalar order parameters, we establish equivalence for individual covariance-matrix elements. Second, these results, often called Gaussian equivalence in this literature, have concerned feedforward architectures, in which the inputs to each layer are, by construction, independent of the quenched weights acting on them. In the recurrent setting, the couplings act on activity that they themselves generate, an aspect that we handled using the cavity method.

\textbf{Extensions to structured connectivity.}
The present derivations treat i.i.d.\ couplings, with the linear equivalence holding for typical quenched $\bm{J}$. Non-i.i.d.\ couplings, such as those with low-rank structure or correlations, would invalidate these derivations. The situation is analogous to the TAP equations for spin glasses~\cite{thouless1977solution}, which give self-consistent equations for magnetizations on a single realization of the couplings but rely on that realization being a draw from the i.i.d.\ ensemble (symmetric, in the spin-glass case). Both the linear equivalence and the TAP-like equation, Eq.~\eqref{eq:tap_equation}, rely on the i.i.d.\ couplings in this way.

A natural question is whether similar ansätze hold for other coupling structures. Candidate extensions include products of random matrices~\cite{clark2025connectivity}, connectivity with multiple cell types or regions~\cite{aljadeff2015transition, clark2025structure}, and (partially) symmetric couplings, with exact symmetry corresponding to the spin-glass case~\cite{marti2018correlations}. For excitatory-inhibitory networks obeying Dale's law and for partially symmetric couplings, \citet{shen2025covariance} found numerically that an ansatz of the same form holds, with a frequency-dependent effective gain in the partially symmetric case, which suggests that the present derivation can be extended to these cases.

\section*{Code availability}

{\raggedright Code is available at \href{https://github.com/davidclark1/Linear-Equivalence}{\nolinkurl{github.com/davidclark1/Linear-Equivalence}}.\par}

\section*{Acknowledgments}

It is a pleasure to thank Ashok Litwin-Kumar, L.~F. Abbott, Jacob Zavatone-Veth, Sebastian Mizera, Yu Hu, Xuanyu Shen, Yue M. Lu, Alexander van Meegen, Blake Bordelon, and Cengiz Pehlevan for helpful discussions. The author is especially grateful to Albert Wakhloo, whose diagrammatic treatment preceded the calculations presented here, for exchanges on this topic.

During the initial stages of this work, the author was supported by a Kempner Institute Research Fellowship. This work has been made possible in part by a gift from the Chan Zuckerberg Initiative Foundation to establish the Kempner Institute at Harvard University. The author is now at the Flatiron Institute, a division of the Simons Foundation.

\clearpage

\onecolumngrid
\appendix
\addtocontents{toc}{\protect\apptocstart}
\appendixcontents
\clearpage

\section{AI methodology}
\label{app:ai}

The author produced both derivations of linear equivalence (Secs.~\ref{sec:method1} and~\ref{sec:method2}) without AI assistance in early 2026. As the work continued over the year, AI models (Claude and GPT, in both chat and coding-agent form) became more capable, and the author used them increasingly. Early on, these models assisted with the derivation of the reservoir and cavity cumulant scalings (Appendix~\ref{app:cavity_properties}) by suggesting components of the diagrammatic arguments used there. Later, during revision, AI models wrote almost all of the code, including the plotting code, and helped draft and revise the text. They also drafted, from the author's notes, an initial mathematical treatment of nonzero mean activities (Sec.~\ref{sec:other_models} and Appendix~\ref{app:nonzero_mean}). In addition, they suggested some references and standardized the formatting of the bibliography. The author verified every reference and added each one by hand. The present version of the paper is thus a collaboration between the author and AI models.

To ensure correctness, in addition to the author checking every result, the paper and its code were audited extensively by Claude Opus 5.5 and GPT-6.1-Sol, each run as many independent agents. These audits checked every statement for correctness, scope, and consistency with the rest of the text. In particular, the agents re-derived each equation and claim independently, checked calculations symbolically, tested scaling claims in small simulations, and recomputed reported quantities from simulation data. Separate agents then tried to refute each finding, which filtered out many of them. The author reviewed the remaining findings and the model-proposed changes, and decided which changes to make and how to implement them. This process was repeated over several rounds, each with fresh agents that had no access to earlier audit reports, until no further substantive errors arose.

\section{Cumulant expansions}
\label{app:price}

Throughout the paper, we Taylor-expand expectations in cumulants. Here, we derive the expansion we use and its two Gaussian special cases, Price's theorem and the Furutsu--Novikov theorem.

\subsection{Cumulant-derivative formula}

The building block is the derivative of an expectation with respect to a cumulant. For a random vector $\bm{Z} = (Z_1, \ldots, Z_n)$ with cumulants $\kappa(Z_{i_1}, \ldots, Z_{i_p})$, the logarithm of the characteristic function is
\begin{equation}
    \ln \tavg{e^{i \bm{k} \cdot \bm{Z}}}_{\bm{Z}} = \sum_{p \geq 1} \frac{i^p}{p!} \sum_{i_1, \ldots, i_p = 1}^n \kappa(Z_{i_1}, \ldots, Z_{i_p})\, k_{i_1} \cdots k_{i_p}.
    \label{eq:log_characteristic}
\end{equation}
The inner sum runs over ordered index tuples, and each cumulant is symmetric in its arguments. A cumulant in which index $a$ occurs $m_a$ times therefore appears $p!/\prod_a m_a!$ times, the multinomial coefficient. For a smooth function $H(\bm{Z})$ with Fourier transform $H(\bm{k}) = \int \dd^n z\, e^{-i\bm{k}\cdot\bm{z}}\,H(\bm{z})$, writing $H(\bm{Z})$ as an inverse Fourier transform and averaging over $\bm{Z}$ gives $\tavgshort{H(\bm{Z})}_{\bm{Z}}$ as an integral against the characteristic function,
\begin{equation}
    \tavg{H(\bm{Z})}_{\bm{Z}} = \int \frac{\dd^n k}{(2\pi)^n}\, H(\bm{k})\, \tavg{e^{i \bm{k} \cdot \bm{Z}}}_{\bm{Z}}.
    \label{eq:fourier_expectation}
\end{equation}
We differentiate Eq.~\eqref{eq:fourier_expectation} with respect to the value of a cumulant $\kappa(Z_{i_1}, \ldots, Z_{i_p})$ in which index $a$ occurs $m_a$ times. This brings down $i^p k_{i_1} \cdots k_{i_p}/\prod_a m_a!$, which acts on $H(\bm{Z})$ as $\partial_{i_1} \cdots \partial_{i_p}/\prod_a m_a!$, so
\begin{equation}
    \frac{\partial \tavg{H(\bm{Z})}_{\bm{Z}}}{\partial\,\kappa(Z_{i_1},\ldots,Z_{i_p})} = \frac{1}{\prod_a m_a!}\,\tavg{\partial_{i_1}\cdots \partial_{i_p}\,H(\bm{Z})}_{\bm{Z}}.
    \label{eq:cumulant_derivative}
\end{equation}
For distinct indices, the prefactor is one. A cumulant in which one index occurs $p$ times and another $q$ times carries the factor $1/(p!\,q!)$, which appears in Eq.~\eqref{eq:cross_cumulant_expansion} below.

\subsection{Cross-cumulant expansion}

We now apply Eq.~\eqref{eq:cumulant_derivative} to a product of functions of two variables. For random variables $X, Y$ and smooth functions $u(X), v(Y)$, we Taylor-expand $\tavgshort{u(X)\,v(Y)}_{X, Y}$ in the cross-cumulants of $X$ and $Y$ around the product of the marginal distributions, at which all cross-cumulants vanish and the expectation equals $\tavgshort{u(X)}_X\,\tavg{v(Y)}_Y$. There, $X$ and $Y$ are independent, so, by Eq.~\eqref{eq:cumulant_derivative}, the derivative of the expectation with respect to a cross-cumulant in which $X$ occurs $p$ times and $Y$ occurs $q$ times is $\tavgshort{u^{(p)}(X)}_X\,\tavgshort{v^{(q)}(Y)}_Y/(p!\,q!)$. Consequently,
\begin{multline}
    \tavg{u(X)\,v(Y)}_{X, Y} \\
    = \tavg{u(X)}_X\,\tavg{v(Y)}_Y
    +\!\sum_{p, q\geq 1}\!\frac{1}{p!\,q!}\,
    \tavg{u^{(p)}(X)}_X\,\tavg{v^{(q)}(Y)}_Y \,
    \kappa\bigl(\underbrace{X,\ldots,X}_{p\text{ times}},\,
    \underbrace{Y,\ldots,Y}_{q\text{ times}}\bigr)
    + \order{\substack{\text{products of $\geq 2$}\\\text{cross-cumulants}}}.
    \label{eq:cross_cumulant_expansion}
\end{multline}
The same argument gives the analogous expansion for random vectors $X$ and $Y$, with derivatives with respect to their components and the prefactor $1/\prod_a m_a!$ of Eq.~\eqref{eq:cumulant_derivative} in place of $1/(p!\,q!)$, and for functionals of random processes, with functional derivatives integrated over their time arguments. For linear $v(Y)$, only the terms with $q = 1$ remain, and the remainder vanishes, since the higher derivatives of $v(Y)$ are zero.

\subsection{Price's theorem and the Furutsu--Novikov theorem}

For jointly Gaussian $(X, Y)$, cumulants of order $\geq 3$ vanish, and Eq.~\eqref{eq:cross_cumulant_expansion} reduces, at first order in $\kappa(X, Y)$, to its $(p, q) = (1, 1)$ term. This gives Price's theorem,
\begin{equation}
    \tavg{u(X)\,v(Y)}_{X, Y}
    = \tavg{u(X)}_X\,\tavg{v(Y)}_Y
    + \kappa(X, Y)\,\tavg{u'(X)}_X\,\tavg{v'(Y)}_Y
    + \order{\kappa(X, Y)^2}.
    \label{eq:price_scalar}
\end{equation}
For $v(Y) = Y$, all terms beyond first order in $\kappa(X, Y)$ vanish exactly, which yields the Furutsu--Novikov theorem,
\begin{equation}
    \tavg{u(X)\,Y}_{X, Y}
    = \tavg{u(X)}_X\,\tavg{Y}_Y
    + \kappa(X, Y)\,\tavg{u'(X)}_X.
    \label{eq:fn_scalar}
\end{equation}

We apply Eqs.~\eqref{eq:price_scalar} and~\eqref{eq:fn_scalar} to stationary, zero-mean Gaussian fields at leading order. We therefore state their functional forms in Fourier space. Consider such Gaussian fields $\eta_1(t)$, $\eta_2(t)$ and functionals $H[\eta_1](t)$ and $G[\eta_2](t)$, with at least one functional having zero mean, so that the marginal product vanishes. Price's theorem then reads
\begin{equation}
    \tavg{H[\eta_1](\omega)\,G[\eta_2](\omega)^*}_{\eta_1, \eta_2}
    = \tavg{\frac{\delta H[\eta_1]}{\delta \eta_1}}_{\eta_1}\!(\omega)\,
       \tavg{\frac{\delta G[\eta_2]}{\delta \eta_2}}_{\eta_2}\!(\omega)^*\,
       C^\eta_{12}(\omega)
       + \order{(C^\eta_{12})^2},
    \label{eq:price_fourier}
\end{equation}
and the Furutsu--Novikov theorem reads
\begin{equation}
    \tavg{H[\eta_1](\omega)\,\eta_2(\omega)^*}_{\eta_1, \eta_2}
    = \tavg{\frac{\delta H[\eta_1]}{\delta \eta_1}}_{\eta_1}\!(\omega)\, C^\eta_{12}(\omega),
    \label{eq:FN_fourier}
\end{equation}
where $C^\eta_{12}(\omega)$ is the cross-spectrum of $\eta_1(t)$ and $\eta_2(t)$, and $\tavgshort{\delta H[\eta_1]/\delta \eta_1}_{\eta_1}(\omega)$ is the Fourier transform in $s$ of $\tavgshort{\delta H[\eta_1](t)/\delta \eta_1(t-s)}_{\eta_1}$.

\section{Cumulant scalings and cavity properties}
\label{app:cavity_properties}

Here, we derive the cumulant scalings underlying the two-site cavity method and, from them, Properties~1 and~2 of Sec.~\ref{sec:cavity}. We then test these results in simulations. \citet{clark2023dimension} assumed both properties, which follow from the central limit theorem when the time average is, naively, not distinguished from the disorder average (Sec.~\ref{sec:cavity}). Deriving them is of particular interest here, since the ansatz is a statement about a single realization of $\bm{J}$.

\subsection{Outline and setting}
\label{app:cavity_scope}

The derivation proceeds in two stages. The first stage (Appendix~\ref{app:reservoir_scaling}) treats cumulants of reservoir variables (activities and responses). Each additional distinct site suppresses the cumulant by $N^{-1/2}$. The second stage (Appendix~\ref{app:cavity_scaling}) treats cumulants of cavity variables (cavity fields and $F$-kernels). Each additional factor suppresses the cumulant by $N^{-1/2}$, even when it repeats a site already present. There is one further factor of $N^{-1/2}$ when some cavity row or column index occurs in only one factor.
The second-stage scaling renders the cavity fields jointly Gaussian and decouples the $F$-kernel from cavity-field functionals, both at leading order. These are Properties~1 and~2 (Appendices~\ref{app:property_1} and~\ref{app:property_2}).

Weaker cavity scalings can be obtained from the reservoir cumulant scalings alone, since the cavity suppression is in essence a central limit theorem for quenched random projections of weakly correlated variables, with the weak correlation quantified by the reservoir cumulant scalings. However, because this approach does not exploit the random signs of reservoir cumulants along certain index combinations, the resulting scalings are, in general, too weak to establish Property~1. The scalings we derive exploit these signs by expanding the reservoir cumulants in the couplings.

Both stages use the same method, consisting of the following steps.
\begin{enumerate}
    \item Iterate the functional Taylor expansion of the dynamics, which expresses each variable as a sum over rooted directed multigraphs whose edges carry $\bm{J}$ elements. Because we compute a cumulant, only connected diagrams survive, which is the principle underlying the linked-cluster theorem.
    \item Group the diagrams into topologies, each carrying an inner sum over labelings of its internal nodes by network sites.
    \item Compute the size of a single topology by squaring it and averaging over the disorder. Because the couplings have zero mean, site labels must coincide, which limits the number of free indices and thus sets the $N$-scaling.
    \item Assume that the sum over topologies inherits the scaling of leading-order topologies. We show that this amounts to commuting the large-$N$ limit with the limits that take the burn-in time and the Taylor order to infinity (Appendix~\ref{app:commuting_limits}).
\end{enumerate}

The setting is a typical quenched realization of $\bm{J}$ in the dynamics of Eqs.~\eqref{eq:dynamics} and~\eqref{eq:local_field}. We write $\kappa_t(\cdot, \ldots, \cdot)$ for a joint cumulant under the stationary-state time average $\tavgshort{\cdots}_t$, where each slot carries its time arguments, held fixed as $N \to \infty$ and suppressed for brevity. A scaling $\order{\cdot}$ of a cumulant refers to its root-mean-square size across realizations of $\bm{J}$. By Chebyshev's inequality, the same scaling holds in probability, that is, for typical quenched $\bm{J}$; furthermore, in this sense, a product of finitely many cumulants scales as the product of their scalings, as used in Sec.~\ref{sec:moment_decomp} and Appendices~\ref{app:property_1} and~\ref{app:pairwise_moments}. We assume that the same product-scaling rule holds for the RMS size across realizations of $\bm{J}$, which the Frobenius-norm bounds on the error matrices use (Sec.~\ref{sec:method1} and Appendices~\ref{app:response}, \ref{app:method2_details}, \ref{app:pairwise_moments}, and~\ref{app:nzm_error}). The derivations of the scalings and of Properties~1 and~2 use neither the zero-mean assumption nor the sign-flip symmetry assumption.

We also make the role of initial conditions explicit, which is important in the deterministic-chaos setting. We indicate the initial conditions in the dynamics functional,
\begin{equation}
    \phi_i(t) = \mathcal{T}[\eta_i + \xi_i \,;\, \bm{z}_i](t),
\end{equation}
where $\bm{z}_i$ collects the initial conditions for unit $i$ at $t = 0$ (e.g., $d$ numbers for $d$-th-order single-unit dynamics). As with the external drives, we draw the initial conditions i.i.d.\ across units, $\bm{z}_i \overset{\mathrm{i.i.d.}}{\sim} \mathcal{P}_{\bm{z}}$. The stationary state is reached after a burn-in time $T_\mathrm{burn}$. Stationary-state averages are then computed by setting $t = T_\mathrm{burn}$ and averaging over $\bm{z}_i$ and $\xi_i(t)$.

\subsection{Reservoir-variable cumulant scaling}
\label{app:reservoir_scaling}

We now carry out the first stage, whose result is the following scaling.

\begin{tcolorbox}[prxmathbox,title={Reservoir-variable cumulant scaling}]
For typical quenched $\bm{J}$, the joint cumulants of activity and response variables satisfy
\begin{equation}
    \kappa_t\bigl(S^\phi_{i_1j_1},\ldots,S^\phi_{i_mj_m},
    \phi_{k_1},\ldots,\phi_{k_n}\bigr)
    = \order{N^{-(r-1)/2}},
    \label{eq:cumulant_scaling_app}
\end{equation}
where $r$ is the number of distinct sites in $\{i_1, j_1, \ldots, i_m, j_m, k_1, \ldots, k_n\}$.
\end{tcolorbox}
The case $m = 0$ is Eq.~\eqref{eq:cumulant_scaling_main} of Sec.~\ref{sec:moment_decomp}, and $m \geq 1$ extends it to cumulants involving responses. Intuitively, distinct sites communicate only through couplings of size $\order{N^{-1/2}}$, so each additional distinct site costs one factor of $N^{-1/2}$.

\textbf{Multigraph expansion.} Let $Y = \kappa_t(S^\phi_{i_1 j_1}, \ldots, S^\phi_{i_m j_m}, \phi_{k_1}, \ldots, \phi_{k_n})$ denote the cumulant in Eq.~\eqref{eq:cumulant_scaling_app}. To express $Y$ in terms of the couplings, we expand $\mathcal{T}[\eta_i + \xi_i \,;\, \bm{z}_i](t)$ in powers of $\eta_i(t)$, which gives
\begin{equation}
    \phi_i(t)
    = \sum_{k=0}^\infty \frac{1}{k!}
    \int_0^t \!\dd s_1 \cdots \int_0^t \dd s_k\,
    \Pi_i^{(k)}(t;s_1,\ldots,s_k)
    \sum_{l_1,\ldots,l_k=1}^N
    J_{il_1}\cdots J_{il_k}\,
    \phi_{l_1}(s_1)\cdots\phi_{l_k}(s_k),
    \label{eq:depth1_app}
\end{equation}
where
\begin{equation}
    \Pi_i^{(k)}(t; s_1, \ldots, s_k) = \left. \frac{\delta^k \mathcal{T}[h \, ; \, \bm{z}_i](t)}{\delta h(s_1) \cdots \delta h(s_k)}\right|_{h = \xi_i}
\end{equation}
is the $k$-th functional derivative of the single-unit dynamics around the external drive $\xi_i(t)$ with initial condition $\bm{z}_i$. Iterating the same expansion on each $\phi_{l_a}(s_a)$ expresses $\phi_i(t)$ as a sum of terms, each a product of $\bm{J}$ elements and of one factor $\Pi^{(k)}_\bullet/k!$ per occurrence of a site in the iterated expansion, with the time arguments integrated according to the causal structure of the expansion. The $\bm{J}$ elements of a term form a rooted directed multigraph $\mathcal{G}$, with one node per distinct site, the root labeled by site $i$, the other nodes labeled by sites in $\{1, \ldots, N\}$, and edges carrying the $\bm{J}$ elements. Grouping the terms by $\mathcal{G}$ gives
\begin{align}
    \phi_i(t) &= \sum_{\mathcal{G}} J_{\mathcal{G}}\,\Pi_{\mathcal{G}}(t), \nonumber \\
    J_{\mathcal{G}} &= \prod_{\text{edges of }\mathcal{G}} J_{\bullet\bullet},
    \label{eq:multigraph_expansion}
\end{align}
where $\Pi_{\mathcal{G}}(t)$ is the sum, over the terms with multigraph $\mathcal{G}$, of their time-integrated products of factors $\Pi^{(k)}_\bullet/k!$. Differentiating Eq.~\eqref{eq:multigraph_expansion} with respect to $\xi_j(t')$ gives an analogous expansion for $S^\phi_{ij}(t, t')$, in which one node of each multigraph is pinned to site $j$. Thus, $j$ acts as an additional fixed external site in the labeling below.

Multilinearity of the cumulant then gives
\begin{equation}
    Y = \sum_{\mathcal{G}_1, \ldots, \mathcal{G}_{m+n}} J_{\mathcal{G}_1} \cdots J_{\mathcal{G}_{m+n}} \, \kappa_t\bigl(\Pi_{\mathcal{G}_1}, \ldots, \Pi_{\mathcal{G}_{m+n}}\bigr).
    \label{eq:Y_multigraph_sum}
\end{equation}
Each $\Pi_{\mathcal{G}_a}(t)$ depends on external drives and initial conditions only at the sites labeling its nodes. Because these are sampled i.i.d.\ across units, the cumulant in Eq.~\eqref{eq:Y_multigraph_sum} vanishes unless $\mathcal{G}_1 \cup \cdots \cup \mathcal{G}_{m+n}$ is a single connected multigraph.

\textbf{Topologies and labelings.} We reorganize Eq.~\eqref{eq:Y_multigraph_sum} as a sum over topologies $\theta$, each carrying an inner sum over labelings $\Lambda(\theta)$. A \emph{topology} specifies the abstract directed connectivity of $\mathcal{G}_1, \ldots, \mathcal{G}_{m+n}$, including the nodes they share, the edge multiplicities, and the assignment of $r$ external nodes to the distinct sites among $\{i_1, j_1, \ldots, k_n\}$. A \emph{labeling} assigns sites in $\{1, \ldots, N\}$ to the internal nodes, distinct from each other and from the $r$ external sites. Then,
\begin{align}
    Y &= \sum_\theta J_\theta \kappa_\theta, \nonumber \\
    J_\theta &= \sum_{\Lambda(\theta)} J_{\Lambda(\theta)},
\end{align}
where $J_{\Lambda(\theta)}$ is the product of $\bm{J}$ elements along edges, and $\kappa_{\theta} = \kappa_t(\Pi_{\mathcal{G}_1}, \ldots, \Pi_{\mathcal{G}_{m+n}})$. Because of the i.i.d.\ sampling, $\kappa_{\theta}$ depends only on the topology and not on the labeling. It is $\bm{J}$- and $N$-independent and $\order{1}$. All of the $N$-dependence therefore resides in $J_\theta$. With the quenched biases of Eq.~\eqref{eq:biased_dynamics}, $\kappa_\theta$ also depends on the biases at the labeled sites, but, since these are independent of $\bm{J}$ and $\kappa_\theta$ remains $\order{1}$ for every labeling, the counting of free indices below is unchanged.

To find the size of $J_\theta$, we square it and average over the disorder,
\begin{equation}
    \tavg{J_\theta^2}_{\bm{J}} = \sum_{\Lambda(\theta), \Lambda'(\theta)} \tavg{J_{\Lambda(\theta)} \, J_{\Lambda'(\theta)}}_{\bm{J}}.
\end{equation}
Because the couplings have zero mean, the dominant contributions to this sum have a reduced number of free indices, which we call $\mathcal{F}$. Let $E$ denote the number of $\bm{J}$ elements per copy. Each term of the sum contains $2E$ elements and is thus $\order{N^{-E}}$ by the moment scaling $\tavgshort{J_{ij}^m}_{\bm{J}} = \order{N^{-m/2}}$ (Sec.~\ref{sec:model}), while the dominant terms number $\order{N^{\mathcal{F}}}$. Consequently, $J_\theta = \order{N^{-(E - \mathcal{F})/2}}$, and it remains to bound $\mathcal{F}$.

\begin{figure}
    \centering
    \includegraphics[width=3in]{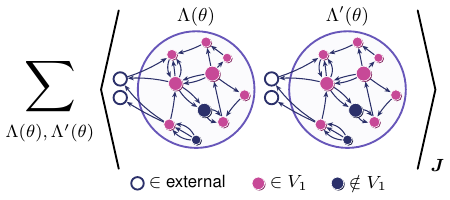}
    \caption{Disorder average $\tavg{J_\theta^2}_{\bm{J}}$ for a topology $\theta$, as a sum over pairs of labelings $\Lambda(\theta)$ and $\Lambda'(\theta)$ of two copies of $\theta$. A node $v \in V_1$ is incident to at least one multiplicity-1 bunch. Such nodes reduce the number of free indices $\mathcal{F}$, because the zero-mean property of the couplings implies that the site assigned to $v$ in one copy must also be assigned to a node in the other copy.}
    \label{fig:labeling}
\end{figure}

\textbf{Counting free indices.} To bound $\mathcal{F}$, we classify the internal nodes of $\theta$ by the edges incident to them. A \emph{bunch} is the set of parallel edges between an ordered pair of nodes. Let $V_\mathrm{int}$ denote the set of internal nodes, and let $V_1 \subseteq V_\mathrm{int}$ denote the subset of internal nodes incident to at least one multiplicity-1 bunch (Fig.~\ref{fig:labeling}).

We first bound $\mathcal{F}$ in terms of $|V_1|$. Labeling distinctness within each copy forbids two distinct bunches of that copy from sharing a $\bm{J}$ element. Consequently, each $\bm{J}$ element of a multiplicity-1 bunch, which has zero mean, must coincide with a $\bm{J}$ element in the opposite copy. This coincidence locks together the source nodes across copies, and likewise the target nodes. Thus, each $V_1$ node in one copy must share its site with a node in the opposite copy. Each $V_1$ node therefore contributes at most one free index per pair of copies, and each non-$V_1$ node at most two, so that $\mathcal{F} \leq |V_1| + 2(|V_\mathrm{int}| - |V_1|)$, or
\begin{equation}
    \mathcal{F} \leq 2|V_\mathrm{int}| - |V_1|.
    \label{eq:labeling_bound}
\end{equation}
When the odd moments of $J_{ij}$ are nonzero, this bound is attained by the ``mirror'' pairing, which assigns the same sites to the $V_1$ nodes of both copies and leaves the remaining internal nodes free (the pairing need not be unique).

We next lower-bound $|V_1|$ via a spanning-tree argument. Pick an undirected spanning tree of $\theta$, and root it at one of the $r$ fixed external nodes. For $v \in V_\mathrm{int} \setminus V_1$, every bunch incident to $v$ has multiplicity $\geq 2$. Thus, the bunch containing the tree edge from $v$ to its parent also contains a non-tree edge, which, together with the tree edge, forms a length-two cycle from $v$ to its parent and back. Distinct $v \in V_\mathrm{int} \setminus V_1$ have parent edges in distinct bunches, so the resulting cycles are independent. Let $L$ denote the number of independent cycles of $\theta$. We have shown that $L \geq |V_\mathrm{int}| - |V_1|$, that is, $|V_1| \geq |V_\mathrm{int}| - L$. Substituting Euler's formula $L = E - V + 1$, with $V = r + |V_\mathrm{int}|$, gives
\begin{equation}
    |V_1| \geq 2|V_\mathrm{int}| - E + r - 1.
    \label{eq:V1_lower_bound}
\end{equation}

Finally, we combine the two bounds. The slack in Eq.~\eqref{eq:V1_lower_bound} is the number of independent cycles beyond those the argument above requires,
\begin{equation}
    \ell_\theta = L - |V_\mathrm{int}| + |V_1| \geq 0.
    \label{eq:ell_theta}
\end{equation}
Inserting Eq.~\eqref{eq:ell_theta} and Euler's formula into Eq.~\eqref{eq:labeling_bound} yields $\mathcal{F} \leq E - (r - 1) - \ell_\theta$, so
\begin{equation}
    \tavg{J_\theta^2}_{\bm{J}} = \order{N^{-(r-1) - \ell_\theta}}.
    \label{eq:J_theta_scaling}
\end{equation}
Leading-order topologies, with $\ell_\theta = 0$, contribute at $\order{N^{-(r-1)/2}}$. Summing over topologies then gives $Y = \order{N^{-(r-1)/2}}$, which is Eq.~\eqref{eq:cumulant_scaling_app}.

\subsection{Commuting-limits assumption}
\label{app:commuting_limits}

The last step above, the sum over topologies, assumes that this sum introduces no additional powers of $N$. Here, we show that this amounts to a certain \emph{commuting-limits assumption}. Topologies proliferate along two axes. \emph{Depth} is the number of recursions in the iterated expansion, equal to the number of effective recurrent steps (exactly so in discrete time). \emph{Breadth} is the Taylor order retained at each node, which sets the branching factor there (in the SCS model, the highest derivative order of $f(x)$ retained). We regulate depth by a cutoff $D \propto T_\mathrm{burn}$ and breadth by a cutoff $K$. Writing $Y_{D, K}$ for the cumulant truncated at depth $D$ and breadth $K$, we have
\begin{align}
    \lim_{N \to \infty} N^{r-1} \tavg{Y_{D, K}^2}_{\bm{J}} &= \sum_{\theta, \theta'\,\leq\, D,K} C_{\theta\theta'}\, \kappa_\theta \kappa_{\theta'}, \nonumber \\
    C_{\theta\theta'} &= \lim_{N \to \infty} N^{r-1} \tavg{J_\theta J_{\theta'}}_{\bm{J}},
\end{align}
where, by Cauchy--Schwarz and Eq.~\eqref{eq:J_theta_scaling}, $C_{\theta\theta'}$ is $\order{1}$ for pairs of leading-order topologies and zero otherwise (strictly, these limits exist when $J_{ij} = g X_{ij}/\sqrt{N}$ with $X_{ij}$ drawn i.i.d.\ from an $N$-independent distribution, as in our simulations). The right-hand side is a finite sum of $\order{1}$ terms, so $\tavg{Y_{D,K}^2}_{\bm{J}} = \order{N^{-(r-1)}}$ holds rigorously at each $D$ and $K$.

The physical order of limits is $\lim_{N \to \infty} \lim_{D \to \infty} \lim_{K \to \infty}$. We assume that, at the level of the formal sum over topologies, we can instead take $\lim_{D \to \infty} \lim_{K \to \infty} \lim_{N \to \infty}$. This assumption has two parts. The first part, commutation of $\lim_{N \to \infty}$ and $\lim_{D \to \infty}$, requires the cumulants to reach their stationary values, at leading order in $N$, within a burn-in time $T_\mathrm{burn}$ that does not grow with $N$. In the SCS model, this is consistent with the Lyapunov spectrum approaching a fixed shape at large $N$~\cite{engelken2023lyapunov, clark2026lyapunov}. The second part, commutation of $\lim_{N \to \infty}$ and $\lim_{K \to \infty}$, is the remaining assumption. It is not needed in the special case of polynomial $f(x)$, for which $K$ remains finite. Given this assumption,
\begin{equation}
    \lim_{N \to \infty} N^{r-1} \tavg{Y^2}_{\bm{J}} = \sum_{\theta, \theta'} C_{\theta \theta'} \kappa_\theta \kappa_{\theta'},
\end{equation}
where the topology sum is formal in general, but $\bm{J}$- and $N$-independent. We conclude that $\tavg{Y^2}_{\bm{J}} = \order{N^{-(r-1)}}$. We make the same assumption for the Taylor orders $p$ and $q$ of the cross-cumulant expansions behind Properties~1 and~2 (Appendices~\ref{app:property_1} and~\ref{app:property_2}). Likewise, this part of the assumption is not needed for polynomial functionals.

\subsection{Cavity-variable cumulant scaling}
\label{app:cavity_scaling}

We now carry out the second stage, whose result is the following scaling.

\begin{tcolorbox}[prxmathbox,title={Cavity-variable cumulant scaling}]
For typical quenched $\bm{J}$, the joint cumulants of the $F$-kernel and the cavity fields satisfy
\begin{equation}
    \kappa_t\bigl(F_{\mu_1\nu_1},\ldots,F_{\mu_m\nu_m},
    \etac_{\rho_1},\ldots,\etac_{\rho_n}\bigr)
    = \order{N^{-(m+n+U-2)/2}},
    \label{eq:cavity_master_bound}
\end{equation}
where $U$ is defined as follows. The cavity field $\etac_\rho$ involves row $\rho$ of $\bm{J}$, namely, the couplings $J_{\rho k}$ (Eq.~\eqref{eq:cavity_field_def}), and the $F$-kernel $F_{\mu\nu}$ involves row $\mu$ and column $\nu$, namely, the couplings $J_{\mu i}$ and $J_{j\nu}$ (Eq.~\eqref{eq:reverb_kernel_def}). We set $U = 1$ if some cavity row or column index appears in exactly one of the $m + n$ slots, and $U = 0$ otherwise.
\end{tcolorbox}

Apart from the factor $N^{-U/2}$, each additional $F$-kernel or cavity-field factor suppresses the cumulant by $N^{-1/2}$, regardless of whether the indices of that factor duplicate those already present. By contrast, in the reservoir-variable cumulant scaling of Eq.~\eqref{eq:cumulant_scaling_app}, only distinct external sites contribute to the suppression. The reason is that each cavity variable is a random projection of high-dimensional unperturbed reservoir trajectories that are \emph{independent of the projecting couplings}, and such projections suppress higher cumulants, as in the central limit theorem. For reservoir variables, the projecting couplings are correlated with the activity, and removing this dependence is the purpose of the cavity construction. Both Properties~1 and~2 follow from Eq.~\eqref{eq:cavity_master_bound}, as we show in Appendices~\ref{app:property_1} and~\ref{app:property_2}.

\textbf{Expansion in reservoir cumulants.} Let $Z = \kappa_t(F_{\mu_1 \nu_1}, \ldots, F_{\mu_m \nu_m}, \etac_{\rho_1}, \ldots, \etac_{\rho_n})$ denote the cumulant in Eq.~\eqref{eq:cavity_master_bound}. We substitute Eqs.~\eqref{eq:cavity_field_def}, \eqref{eq:F_decomp}, and~\eqref{eq:reverb_kernel_def} into $Z$ to express it in terms of the unperturbed reservoir variables. The direct-coupling piece of the $F$-kernel is time-independent at fixed $\bm{J}$, so it drops out of cumulants for $m + n \geq 2$. For $(m, n) = (1, 0)$, it contributes $\sqrt{N}\,J_{\mu\nu}\,\delta(t-t') = \order{1}$, which satisfies Eq.~\eqref{eq:cavity_master_bound}, and we omit it below. This yields
\begin{multline}
    Z = N^{m/2}\sum_{i_1,\ldots,i_m=1}^N\,\sum_{j_1,\ldots,j_m=1}^N\,\sum_{k_1,\ldots,k_n=1}^N
    J_{\mu_1i_1}\!\cdots\!J_{\mu_mi_m}
    J_{j_1\nu_1}\!\cdots\!J_{j_m\nu_m} 
    \, J_{\rho_1k_1}\!\cdots\!J_{\rho_nk_n} \\
    \times \kappa_t\!\bigl(\tilde S^\phi_{i_1j_1},\ldots,
    \tilde S^\phi_{i_mj_m}, 
    \tilde\phi_{k_1},\ldots,\tilde\phi_{k_n}\bigr).
    \label{eq:cavity_kappa_expansion}
\end{multline}
The reservoir cumulant in Eq.~\eqref{eq:cavity_kappa_expansion} has the same form as $Y$. We therefore extend the multigraph analysis of $Y$ to handle the outer cavity sums, with two changes. First, the cavity couplings $J_{0 i}$, $J_{0' i}$, $J_{i 0}$, and $J_{i 0'}$ form four independent sets, which we treat as four separate external sites in the multigraph for $Z$. Second, the cumulant-slot sites $i_1, \ldots, i_m$, $j_1, \ldots, j_m$, and $k_1, \ldots, k_n$, which were external nodes for $Y$, become internal nodes for $Z$ after the cavity sums. Let the \emph{reservoir subnetwork} be the multigraph obtained by removing the cavity external sites and their incident edges. As in the first stage, in the multigraph expansion of the reservoir cumulant, Eq.~\eqref{eq:Y_multigraph_sum}, each cumulant $\kappa_t(\Pi_{\mathcal{G}_1}, \ldots, \Pi_{\mathcal{G}_{m+n}})$ vanishes unless the reservoir subnetwork is connected.

\textbf{Counting free indices.} The bound of Eq.~\eqref{eq:labeling_bound} on $\mathcal{F}$ carries over, so we again need a lower bound on $|V_1|$. To obtain it, we use a spanning-tree argument tailored to the cavity setup. Take a spanning tree of the reservoir subnetwork, with $|V_\mathrm{int}| - 1$ tree edges, and let $a$ be the number of these edges occurring in bunches of multiplicity 1. Each tree edge occurs in its own bunch. A multiplicity-1 bunch contributes one edge, and each higher-multiplicity bunch contributes at least two, so the number of edges $E_\mathrm{res}$ of the reservoir subnetwork satisfies $E_\mathrm{res} \geq a + 2(|V_\mathrm{int}| - 1 - a)$, or
\begin{equation}
    E_\mathrm{res} \geq 2(|V_\mathrm{int}| - 1) - a.
    \label{eq:res_edge_bound}
\end{equation}
We bound $|V_1|$ in two cases.
\begin{itemize}
    \item For $a \geq 1$, the $a$ multiplicity-1 bunches define an $a$-edge subforest of the spanning tree with at least $a + 1$ nodes, all of which lie in $V_1$. Hence, $|V_1| \geq a + 1$. With Eq.~\eqref{eq:res_edge_bound}, this gives $|V_1| \geq 2|V_\mathrm{int}| - E_\mathrm{res} - 1$, and the right-hand side only decreases on adding $U - 1 \leq 0$.
    \item For $a = 0$, each cavity index occurring in exactly one slot carries a single coupling, which is a multiplicity-1 bunch whose reservoir endpoint lies in $V_1$. Hence, $|V_1| \geq U$. Moreover, Eq.~\eqref{eq:res_edge_bound} gives $2|V_\mathrm{int}| - E_\mathrm{res} - 2 \leq 0$.
\end{itemize}
Both cases therefore give the same bound,
\begin{equation}
    |V_1| \geq 2|V_\mathrm{int}| - E_\mathrm{res} - 2 + U.
    \label{eq:reservoir_tree_bound}
\end{equation}

Combining Eqs.~\eqref{eq:labeling_bound} and~\eqref{eq:reservoir_tree_bound} gives $\mathcal{F} \leq E_\mathrm{res} + 2 - U$. Each $F$-kernel adds two cavity couplings, and each cavity field adds one, so $E - E_\mathrm{res} = 2m + n$, and thus $\mathcal{F} - E \leq 2 - (2m + n) - U$. Each topology therefore contributes $\order{N^{(2 - (2m + n) - U)/2}}$, and the $N^{m/2}$ prefactor in Eq.~\eqref{eq:cavity_kappa_expansion} promotes this to $\order{N^{(2 - m - n - U)/2}}$. Summing over topologies, under the commuting-limits assumption (Appendix~\ref{app:commuting_limits}), gives Eq.~\eqref{eq:cavity_master_bound}.

\subsection{Joint Gaussianity of cavity fields (Property 1)}
\label{app:property_1}

We now specialize Eq.~\eqref{eq:cavity_master_bound} to the cavity fields alone ($m = 0$).

\begin{tcolorbox}[prxmathbox,title={Joint Gaussianity of cavity fields (Property 1)}]
For typical quenched $\bm{J}$, the joint cumulants of the cavity fields satisfy
\begin{equation}
    \kappa_t\!\bigl(\underbrace{\etac_0, \ldots, \etac_0}_{p \text{ times}}, \underbrace{\etac_{0'}, \ldots, \etac_{0'}}_{q \text{ times}}\bigr) = \order{N^{-(p + q + U - 2)/2}},
    \label{eq:cavity_field_cumulants}
\end{equation}
where $U = 1$ if $p = 1$ or $q = 1$, and $U = 0$ otherwise.
\end{tcolorbox}

The one-cavity-unit statement used in Sec.~\ref{sec:dmft} is the case $q = 0$. For $p, q \geq 1$, Eq.~\eqref{eq:cavity_field_cumulants} has two consequences. First, every cross-cumulant other than $\kappa_t(\etac_0, \etac_{0'})$ is $\order{N^{-1}}$. Second, every cross-cumulant is at most $\order{N^{-1/2}}$, so products of two or more cross-cumulants are $\order{N^{-1}}$.
Inserting these scalings into the cross-cumulant expansion of Eq.~\eqref{eq:cross_cumulant_expansion}, under the commuting-limits assumption (Appendix~\ref{app:commuting_limits}), gives, for smooth functionals $u[\etac_{0}](t)$ and $v[\etac_{0'}](t)$,
\begin{multline}
    \tavg{u[\etac_0](t)\,v[\etac_{0'}](t)}_t \\
    = \tavg{u[\etac_0](t)}_t\,
    \tavg{v[\etac_{0'}](t)}_t
    + \int_0^\infty \!\dd s \int_0^\infty \!\dd s'\,
    \tavg{\frac{\delta u[\etac_0](t)}{\delta \etac_0(t\!-\!s)}}_t
    \tavg{\frac{\delta v[\etac_{0'}](t)}{\delta \etac_{0'}(t\!-\!s')}}_t \kappa_t\bigl(\etac_0(t\!-\!s), \etac_{0'}(t\!-\!s')\bigr)
    + \order{N^{-1}}.
    \label{eq:property_1}
\end{multline}
This has the same form as Price's theorem for Gaussian variables (Eq.~\eqref{eq:price_scalar}). The same holds at unequal times, with $u[\etac_0](t+\tau)$ and $\etac_0(t+\tau-s)$ in place of $u[\etac_0](t)$ and $\etac_0(t-s)$, and the Fourier transform in $\tau$ of this relation has the form of Eq.~\eqref{eq:price_fourier}. Properties~1 and~2 also hold for functionals that depend on the external drive and initial conditions of their own cavity unit, since these are independent across units and of the reservoir.

In the first line of Eq.~\eqref{eq:free_phi_phi_price}, $u[\etac_0](t)$ and $v[\etac_{0'}](t)$ are $\phi^\free_0(t)$ and $\phi^\free_{0'}(t)$, respectively, and the marginal-product term is $\order{N^{-1}}$. In the other two lines, one of the two functionals is a cavity field itself, and the marginal-product term vanishes. In all three lines, $\kappa_t\bigl(\etac_0(t-s), \etac_{0'}(t-s')\bigr) = C^{\etac}_{00'}(s'-s)$, since the cavity fields have mean zero (zero-mean assumption).

\subsection{\texorpdfstring{$F$}{F}-kernel decoupling (Property 2)}
\label{app:property_2}

We now specialize Eq.~\eqref{eq:cavity_master_bound} to cumulants containing one $F$-kernel.

\begin{tcolorbox}[prxmathbox,title={$F$-kernel decoupling (Property 2)}]
For typical quenched $\bm{J}$ and a smooth functional $G[\etac_0, \etac_{0'}](t) = \order{1}$ of the cavity fields,
\begin{equation}
    \kappa_t\bigl(F_{\mu\nu},\, G[\etac_0, \etac_{0'}]\bigr) = \order{N^{-1/2}}.
    \label{eq:property_2}
\end{equation}
\end{tcolorbox}

The direct-coupling piece $\sqrt{N}\, J_{\mu\nu}\,\delta(t-t')$ of the $F$-kernel (Eq.~\eqref{eq:F_decomp}) is time-independent at fixed $\bm{J}$, so its cumulant with $G[\etac_0, \etac_{0'}](t)$ vanishes. The cumulant in Eq.~\eqref{eq:property_2} therefore reduces to $\kappa_t(F^R_{\mu\nu}, G[\etac_0, \etac_{0'}])$. Since this cumulant is linear in $F^R_{\mu\nu}$, the expansion of Eq.~\eqref{eq:cross_cumulant_expansion}, with the pair $(\etac_0, \etac_{0'})$ as $X$ and $F^R_{\mu\nu}$ as $Y$, is exact and gives
\begin{equation}
    \kappa_t\!\bigl(F^R_{\mu\nu},G[\etac_0,\etac_{0'}]\bigr)
    = \!\sum_{\substack{p,q\geq 0 \\ p+q\geq 1}}\!
    \frac{1}{p!\,q!}
    \tavg{\partial_{\etac_0}^p\partial_{\etac_{0'}}^q
    G[\etac_0,\etac_{0'}](t)}_t 
    \kappa_t\!\bigl(F^R_{\mu\nu},
    \underbrace{\etac_0,\ldots,\etac_0}_{p\text{ times}},
    \underbrace{\etac_{0'},\ldots,\etac_{0'}}_{q\text{ times}}\bigr).
    \label{eq:F_R_G_expansion}
\end{equation}
Here, $\partial_{\etac_0}^p\partial_{\etac_{0'}}^q$ denotes functional differentiation, with the time integrals of Eq.~\eqref{eq:property_1} suppressed. The cumulants on the right correspond to $m = 1$, $n = p + q$, and $U = 1$, since the cavity column $\nu$ appears in exactly one slot. Equation~\eqref{eq:cavity_master_bound} thus gives
\begin{equation}
    \kappa_t\!\bigl(F^R_{\mu \nu}, \underbrace{\etac_0, \ldots, \etac_0}_{p \text{ times}}, \underbrace{\etac_{0'}, \ldots, \etac_{0'}}_{q \text{ times}}\bigr) = \order{N^{-(p + q)/2}},
    \label{eq:F_eta_eta_bound}
\end{equation}
which is $\order{N^{-1/2}}$ at $p + q = 1$ and smaller otherwise. Summing in Eq.~\eqref{eq:F_R_G_expansion}, under the commuting-limits assumption (Appendix~\ref{app:commuting_limits}), gives Eq.~\eqref{eq:property_2}.

\subsection{Numerical verification}
\label{app:cavity_numerics}

We now test the reservoir-variable and cavity-variable cumulant scalings, Eqs.~\eqref{eq:cumulant_scaling_app} and~\eqref{eq:cavity_master_bound}, together with Properties~1 and~2, in large-$N$ simulations of the SCS model at $g = 20$, where non-Gaussian effects on network statistics are most pronounced~\cite{clark2023dimension}.

The simulated network plays the role of the reservoir. To construct \emph{cavity-analogue variables}, we draw two sets of random projections, $W^{\mathrm{in}}_{\mu i}$ and $W^{\mathrm{out}}_{i \mu}$. Here, $\mu \in \{1, \ldots, n_{\mathrm{cav}}\}$ indexes $n_{\mathrm{cav}} = 64$ analogue cavity units, in place of the two cavity units $0$ and $0'$, and $i \in \{1, \ldots, N\}$ indexes reservoir sites. The elements $W^{\mathrm{in}}_{\mu i}$ and $W^{\mathrm{out}}_{i \mu}$ are i.i.d.\ Gaussian, jointly independent of $\bm{J}$, with the same first and second moments as the elements of $\bm{J}$. They define the analogue cavity fields
\begin{equation}
    \hat\eta^\free_\mu(t) = \sum_{i=1}^N W^{\mathrm{in}}_{\mu i}\, \phi_i(t),
    \label{eq:hat_etac}
\end{equation}
the counterparts of the cavity fields of Eq.~\eqref{eq:cavity_field_def}. The analogue preactivations $\hat{x}^\free_\mu(t)$ and analogue activities $\hat\phi^\free_\mu(t)$ follow by passing $\hat\eta^\free_\mu(t)$ through the single-unit dynamics of Eq.~\eqref{eq:sompolinsky}, namely, $(1 + \partial_t)\,\hat{x}^\free_\mu(t) = \hat\eta^\free_\mu(t)$ and $\hat\phi^\free_\mu(t) = f(\hat{x}^\free_\mu(t))$. The analogue equal-time $F$-kernel, matching the causal equal-time limit of the reverberation kernel of Eq.~\eqref{eq:reverb_kernel_def}, is
\begin{equation}
    \hat F_{\mu \nu}(t) = \sqrt{N}\, \sum_{i=1}^N W^{\mathrm{in}}_{\mu i}\, W^{\mathrm{out}}_{i \nu}\, f'(x_i(t)).
    \label{eq:F_hat}
\end{equation}
The reservoir counterparts are the local fields $\eta_i(t) = \sum_{j=1}^N J_{ij}\,\phi_j(t)$, preactivations $x_i(t)$, and activities $\phi_i(t)$. The cavity-analogue and reservoir variables differ only in that $W^{\mathrm{in}}_{\mu i}$ is independent of $\phi_i(t)$, whereas $J_{ij}$ is correlated with $\phi_j(t)$.

\begin{figure*}
    \centering
    \includegraphics[width=\textwidth]{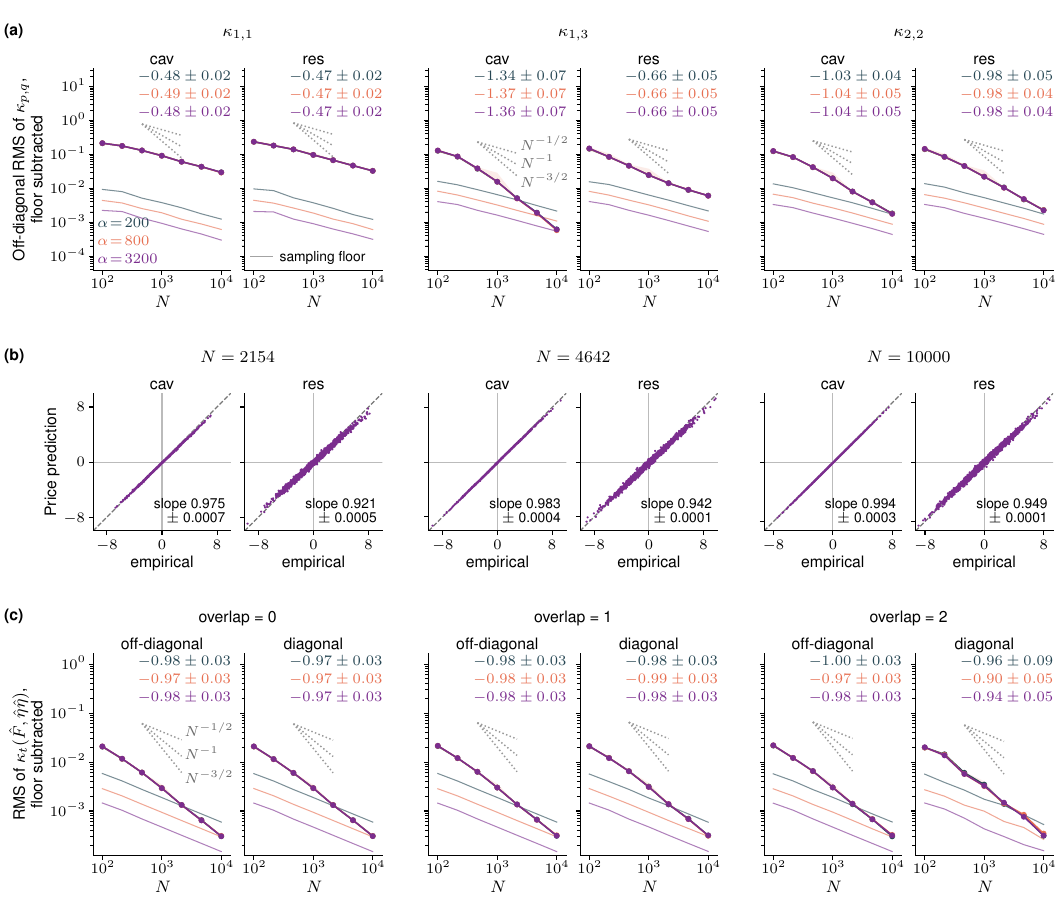}
    \caption{%
    Numerical tests of the cumulant scalings and of Properties~1 and~2 in the SCS model. All panels show $g = 20$ at time lag $\tau = 0$. In (a) and (c), for three sampling ratios $\alpha$, lines show medians over three realizations of $\bm{J}$ per $N$ after subtraction of the sampling floor, and the shaded region shows the interquartile range at $\alpha = 800$. Keys give the fitted exponents. Dotted gray lines show reference power laws.
    \textbf{(a)}~Off-diagonal RMS of the joint cumulant $\kappa_{p,q}$ of $p$ copies of $\hat\eta^\free_\mu(t)$ and $q$ copies of $\hat\eta^\free_\nu(t)$ at $\mu \neq \nu$, normalized by standard deviations of $\hat\eta^\free_\mu(t)$ and $\hat\eta^\free_\nu(t)$ raised to the $p$-th and $q$-th powers, respectively, for $(p, q) \in \{(1, 1),\, (1, 3),\, (2, 2)\}$. For each $(p, q)$, the left subpanel (cav) uses the analogue cavity field $\hat\eta^\free_\mu(t)$ of Eq.~\eqref{eq:hat_etac}, and the right (res) uses the reservoir local field $\eta_i(t)$.
    \textbf{(b)}~Price prediction $\tavgshort{f'(\hat{x}^\free_\mu)}_t\,\tavgshort{f'(\hat{x}^\free_\nu)}_t\,\kappa_t(\hat{x}^\free_\mu, \hat{x}^\free_\nu)$ against the empirical activity cross covariance $\kappa_t(\hat\phi^\free_\mu, \hat\phi^\free_\nu)$ (both scaled by $\sqrt{N}$) at the three largest $N$, for one realization of $\bm{J}$ at $\alpha = 3200$, with dashed identity lines and the slopes of least-squares fits through the origin, given with their sampling floors, half the difference of the slopes fitted on the two halves of the initial conditions. For each $N$, the left subpanel uses cavity-analogue variables $\hat{x}^\free_\mu(t),\, \hat\phi^\free_\mu(t)$, and the right uses reservoir variables $x_i(t),\, \phi_i(t)$.
    \textbf{(c)}~RMS of $\kappa_t(\hat F_{\mu \nu}(t),\, \hat\eta^\free_a(t)\, \hat\eta^\free_b(t))$, normalized by standard deviations of $\hat F_{\mu \nu}(t)$ and $\hat\eta^\free_a(t)\, \hat\eta^\free_b(t)$, organized by the overlap, namely, the number of shared cavity indices between $(\mu, \nu)$ and $(a, b)$. For each overlap, the left subpanel restricts to off-diagonal kernel components ($\mu \neq \nu$), and the right to diagonal components ($\mu = \nu$). The pair $(a, b)$ has $a \neq b$, except in the right subpanel at overlap $2$, where $a = b = \mu$.}
    \label{fig:appendix_numerics}
\end{figure*}

Statistics are estimated by time averaging at fixed $\bm{J}$. Network sizes are logarithmically spaced, from $N = 100$ to $N = 10000$, with total simulation time $T_{\mathrm{tot}} = \alpha N$ for the sampling ratios $\alpha = 200$, $800$, and $3200$ (simulation parameters in Appendix~\ref{app:sim_params}, fit protocol in Appendix~\ref{app:sim_floor}). Sign-flip symmetry holds in this simulation, so odd-order cumulants of the fields vanish, and $\hat F_{\mu\nu}(t)$ is even under the sign flip. These facts determine which statistics can be tested. Figure~\ref{fig:appendix_numerics} shows the tests of the cavity-field cumulant scaling, Price's theorem, and $F$-kernel decoupling, in panels (a), (b), and (c), respectively.

We first test the cavity-field cumulant scaling, Eq.~\eqref{eq:cavity_field_cumulants}, which predicts $N^{-1/2}$, $N^{-3/2}$, and $N^{-1}$ at $(p, q) = (1, 1)$, $(1, 3)$, and $(2, 2)$, respectively. Figure~\ref{fig:appendix_numerics}(a) shows the off-diagonal RMS of these cumulants of the analogue cavity fields as a function of $N$. After subtraction of the sampling floor, the fitted exponents at $\alpha = 3200$ are $-0.48 \pm 0.02$, $-1.36 \pm 0.07$, and $-1.04 \pm 0.05$ and agree across $\alpha$ within the standard errors. Every tested cumulant beyond $(1, 1)$ is therefore below the $\order{N^{-1/2}}$ cross covariance, as joint Gaussianity at leading order requires. The $(1, 3)$ exponent is somewhat smaller in magnitude than the predicted $-3/2$ but well beyond the $-1$ required by the $\order{N^{-1}}$ error of Eq.~\eqref{eq:property_1}.

Figure~\ref{fig:appendix_numerics}(a) also shows the same cumulants of the reservoir local fields, which the counting behind Eq.~\eqref{eq:cumulant_scaling_app}, with $r = 2$, bounds by $N^{-1/2}$ in all three cases. The fitted exponents at $\alpha = 3200$ are $-0.47 \pm 0.02$, $-0.66 \pm 0.05$, and $-0.98 \pm 0.04$, and the decay at $(1, 3)$ slows toward $N^{-1/2}$ at the largest $N$. The reservoir cumulants thus scale differently from cavity-field cumulants only at $(1, 3)$, where the reservoir cumulant is $\order{N^{-1/2}}$ rather than $\order{N^{-3/2}}$. This cumulant is of the same order as the cross covariance, which is why joint Gaussianity fails at leading order for reservoir variables.

We next test Property~1 through Price's theorem. By Eq.~\eqref{eq:property_1}, the empirical activity cross covariance $\kappa_t(\hat\phi^\free_\mu, \hat\phi^\free_\nu)$, $\mu \neq \nu$, of the cavity-analogue variables equals the Price prediction $\tavgshort{f'(\hat{x}^\free_\mu)}_t\,\tavgshort{f'(\hat{x}^\free_\nu)}_t\,\kappa_t(\hat{x}^\free_\mu, \hat{x}^\free_\nu)$ with $\order{N^{-1}}$ error. Figure~\ref{fig:appendix_numerics}(b) compares the two, both scaled by $\sqrt{N}$, at the three largest $N$, together with the corresponding quantities for the reservoir variables. The cavity-analogue points should therefore converge to the identity. These points lie on the identity, with regression slope tending to $1$ as $N$ grows ($0.994$ at $N = 10000$). The reservoir points scatter around the identity, with slope between $0.92$ and $0.95$ at all three sizes ($0.949$ at $N = 10000$), and neither the scatter nor the slope deficit appears to vanish as $N$ grows. Thus, as expected from the $\order{N^{-1/2}}$ reservoir cumulant at $(1, 3)$, which enters Eq.~\eqref{eq:cross_cumulant_expansion} at the order of the cross covariance, Price's theorem holds for the cavity-analogue variables but fails at leading order for the reservoir variables.

Finally, we test Property~2, Eq.~\eqref{eq:property_2}, which requires the cumulant of the $F$-kernel with a cavity-field functional to be $\order{N^{-1/2}}$. We use the functional $\hat{\eta}^\free_a(t)\, \hat{\eta}^\free_b(t)$, the lowest-order choice that is even under the sign flip, since the cumulant vanishes for odd functionals. For this functional, the $p + q = 1$ terms of Eq.~\eqref{eq:F_R_G_expansion} vanish, since they are proportional to the means of the cavity fields, and the only remaining term has $p + q = 2$, which Eq.~\eqref{eq:F_eta_eta_bound} makes $\order{N^{-1}}$, below the required $\order{N^{-1/2}}$. This prediction holds for all numbers of shared indices and for both off-diagonal ($\mu \neq \nu$) and diagonal ($\mu = \nu$) kernel components, since $U = 1$ in every case. Figure~\ref{fig:appendix_numerics}(c) shows the RMS of $\kappa_t(\hat{F}_{\mu\nu}(t),\, \hat{\eta}^\free_a(t)\, \hat{\eta}^\free_b(t))$ in each of these cases. After subtraction of the sampling floor, the fitted exponents at $\alpha = 3200$ are between $-0.94$ and $-0.98$ (standard errors $0.03$--$0.05$) for all of these cases and agree across $\alpha$. These exponents agree with the $\order{N^{-1}}$ of Eq.~\eqref{eq:F_eta_eta_bound} and satisfy the $\order{N^{-1/2}}$ bound that Property~2 requires.

\section{Response-matrix ansatz}
\label{app:response}

Here, we show that the response matrix $\bm{S}^\phi(\omega)$ is approximated by the response-matrix ansatz $\bar{\bm{S}}^\phi(\omega)$ with the same precision as the linear-equivalent ansatz $\bar{\bm{C}}^\phi(\omega)$, namely, $\order{N^{-1/2}}$ on the diagonal and $\order{N^{-1}}$ in the RMS sense off the diagonal (Sec.~\ref{sec:ansatz}).

For the linear network of Eq.~\eqref{eq:linear_dynamics}, the response matrix is $\bm{S}^\phi(\omega) = \mathcal{T}(\omega)\,(\bm{I}_N - \mathcal{T}(\omega)\,\bm{J})^{-1}$. The response-matrix ansatz of the nonlinear network takes this form with $\mathcal{T}(\omega)$ replaced by $\Sstar(\omega)$,
\begin{equation}
    \bar{\bm{S}}^\phi(\omega) = \Sstar(\omega)\,\bm{M}(\omega),
    \label{eq:response_ansatz}
\end{equation}
where $\bm{M}(\omega)$ is the resolvent of Eq.~\eqref{eq:resolvent}.

\subsection{Self-consistent equation for the response matrix}

We first derive a self-consistent equation for $\bm{S}^\phi(\omega)$. To accomplish this, we differentiate $\phi_i(t) = \mathcal{T}[\eta_i + \xi_i](t)$ with respect to $\xi_j(t-\tau)$, which gives
\begin{equation}
    \frac{\delta\phi_i(t)}{\delta\xi_j(t-\tau)}
    = \int_{0}^\infty \dd s\, R_i(t,t-s) \left(\frac{\delta\eta_i(t-s)}{\delta\xi_j(t-\tau)} + \delta_{ij}\,\delta(\tau - s)\right).
    \label{eq:response_chain_rule}
\end{equation}
Averaging over time gives
\begin{equation}
    S^\phi_{ij}(\tau)
    = \int_{0}^\infty \dd s\, \tavg{R_i(t, t\!-\!s)\, \frac{\delta\eta_i(t\!-\!s)}{\delta\xi_j(t-\tau)}}_t + \delta_{ij}\, \Sstar(\tau) + \delta_{ij}\,\order{N^{-1/2}},
    \label{eq:response_averaged}
\end{equation}
where the $\delta_{ij}\,\order{N^{-1/2}}$ error results from replacing $\tavgshort{R_i(t, t-\tau)}_t$ by $\Sstar(\tau)$ in the $i = j$ contribution. This replacement follows from the single-site DMFT (Eq.~\eqref{eq:dmft_S_sc}), with $\order{N^{-1/2}}$ site-to-site fluctuations (zero-mean assumption).

We now evaluate the remaining average in Eq.~\eqref{eq:response_averaged}. For $i \neq j$, the two factors decouple under the time average at leading order. To see this, choose $i$ and $j$ as the two cavity units of Sec.~\ref{sec:cavity}. Then, the local response $R_i(t,t\!-\!s)$ is, at leading order, a functional only of the cavity field and external drive at site $i$, whereas the response $\delta\eta_i(t\!-\!s)/\delta\xi_j(t\!-\!\tau)$ describes a perturbation that propagates from site $j$ to site $i$, mediated by the $F$-kernel $F_{ij}(t,t')$ acting on the local response at site $j$ (Eq.~\eqref{eq:delta_eta_def}).
Consequently, Properties~1 and~2 apply (Sec.~\ref{sec:cavity}, Appendix~\ref{app:cavity_properties}), which yields
\begin{equation}
    \tavg{R_i(t, t\!-\!s)\, \frac{\delta\eta_i(t\!-\!s)}{\delta\xi_j(t\!-\!\tau)}}_t 
    = \Sstar(s)\, \tavg{\frac{\delta\eta_i(t\!-\!s)}{\delta\xi_j(t-\tau)}}_t + \delta_{ij}\, \order{N^{-1/2}} + (1 - \delta_{ij})\,\order{N^{-1}}.
    \label{eq:response_decoupling}
\end{equation}
For $i \neq j$, the $\order{N^{-1}}$ error combines the decoupling error with the error of replacing $\tavgshort{R_i(t,t-s)}_t$ by $\Sstar(s)$ (Eq.~\eqref{eq:dmft_S_sc}). Each of these is $\order{N^{-1/2}}$ relative to the $\order{N^{-1/2}}$ response of $\eta_i(t)$ to $\xi_j(t)$. For $i = j$, the two factors involve the same site and need not decouple, so the error is set by that response, Eq.~\eqref{eq:delta_eta_def}, which is itself $\order{N^{-1/2}}$.

Using $\eta_i(t) = \sum_{k=1}^N J_{ik}\,\phi_k(t)$, the time-averaged response of $\eta_i(t)$ to $\xi_j(t)$ is
\begin{equation}
    \tavg{\frac{\delta\eta_i(t\!-\!s)}{\delta \xi_j(t-\tau)}}_t = \sum_{k=1}^N J_{ik}\, S^\phi_{kj}(\tau - s).
    \label{eq:eta_response_to_xi}
\end{equation}
Substituting Eqs.~\eqref{eq:response_decoupling} and~\eqref{eq:eta_response_to_xi} into Eq.~\eqref{eq:response_averaged} and transforming to Fourier space gives the self-consistent equation
\begin{equation}
    \bm{S}^\phi(\omega) = \Sstar(\omega)\, \bm{J}\, \bm{S}^\phi(\omega) + \Sstar(\omega)\, \bm{I}_N + \bm{\mathcal{E}}^S(\omega),
    \label{eq:response_matrix_eq}
\end{equation}
where $\bm{\mathcal{E}}^S(\omega)$ has diagonal elements of $\order{N^{-1/2}}$ and off-diagonal elements of $\order{N^{-1}}$.

\subsection{Solution and precision}

We now solve Eq.~\eqref{eq:response_matrix_eq} and bound the error of the response-matrix ansatz. Solving for $\bm{S}^\phi(\omega)$ gives
\begin{equation}
    \bm{S}^\phi(\omega) = \bar{\bm{S}}^\phi(\omega) + \bm{M}(\omega)\,\bm{\mathcal{E}}^S(\omega).
    \label{eq:response_solution}
\end{equation}
The two precisions then follow as for the ansatz $\bar{\bm{C}}^\phi(\omega)$ (Sec.~\ref{sec:method1}). For the diagonal elements, the linear-equivalent network has single-site response $\Sstar(\omega)$ by construction. Thus, both $S^\phi_{ii}(\omega)$ and $\bar{S}^\phi_{ii}(\omega)$ concentrate around $\Sstar(\omega)$ with $\order{N^{-1/2}}$ fluctuations. For the off-diagonal elements, since $\lVert \bm{M}(\omega) \rVert_\mathrm{op} = \order{1}$ and $\lVert \bm{\mathcal{E}}^S(\omega)\rVert_F = \order{1}$,
\begin{align}
    \lVert \bm{M}(\omega)\,\bm{\mathcal{E}}^S(\omega)\rVert_F &\leq \lVert \bm{M}(\omega)\rVert_{\mathrm{op}}\,\lVert \bm{\mathcal{E}}^S(\omega)\rVert_F \nonumber \\
    &= \order{1},
\end{align}
which gives an off-diagonal RMS error of $\order{N^{-1}}$.

\section{Derivation of the self-consistent matrix equation}
\label{app:method2_details}

Here, we give the steps omitted from the sketch in Sec.~\ref{sec:method2}. We derive the self-consistent equation for the covariance matrix $\bm{C}^\phi(\omega)$, show that its solution is the ansatz $\bar{\bm{C}}^\phi(\omega)$, and bound the error. In this appendix only, all matrices ($\bm{J}$, $\bm{C}^\phi(\omega)$, $\bm{S}^\phi(\omega)$, $\bm{F}(\omega)$, $\bm{M}(\omega)$) are $(N+2) \times (N+2)$, with indices running over $\{0, 0'\} \cup \{1, \ldots, N\}$. In the final results, we relabel $N+2$ as $N$.

\subsection{Cavity expression for the cross-spectrum}

We start from the cavity expression for the cross-spectrum $C^\phi_{00'}(\omega)$, Eq.~\eqref{eq:single_site_cav_eqn} of Sec.~\ref{sec:method2}, which is Eq.~(8) of \citet{clark2023dimension} in the conventions of this paper. Its first term, $|\Sstar(\omega)|^2\,C^{\etac}_{00'}(\omega)$, contains the cavity-field cross covariance $C^{\etac}_{00'}(\omega) = \sum_{i,j=1}^N J_{0i}\,J_{0'j}\,\tilde{C}^\phi_{ij}(\omega)$ and arises from the jointly Gaussian cavity fields via Price's theorem. Its cross terms contain the non-Gaussian contributions from the $F$-kernel. The first term and the cross terms are both $\order{N^{-1/2}}$.

Equation~\eqref{eq:single_site_cav_eqn} resembles a self-consistent equation for $\bm{C}^\phi(\omega)$, with $F_{00'}(\omega)$ and $F_{0'0}(\omega)$ to be determined later (Eq.~\eqref{eq:F_result_S}). To interpret it as such, we must express $C^{\etac}_{00'}(\omega)$ in terms of $[\bm{J}\,\bm{C}^\phi(\omega)\,\bm{J}^T]_{00'}$, which requires two corrections. First, the right-hand side contains the unperturbed reservoir covariance $\tilde{C}^\phi_{ij}(\omega)$ rather than the full covariance $C^\phi_{ij}(\omega)$. Second, the sums run only over the reservoir indices $\{1, \ldots, N\}$, excluding the cavity indices $\{0, 0'\}$.

\subsection{From the unperturbed to the full reservoir covariance}

To make the first correction, we use $\tilde{\phi}_i(t) = \phi_i(t) - \delta\phi_i(t) + \order{N^{-1}}$ (Eq.~\eqref{eq:reservoir_perturbation}), which gives
\begin{multline}
    \sum_{i,j=1}^N J_{0i}\,J_{0'j}\,\tilde{C}^\phi_{ij}(\tau) \\
    = \sum_{i,j=1}^N J_{0i}\,J_{0'j}\,C^\phi_{ij}(\tau) - \tavg{\Bigl(\sum_{i=1}^N J_{0i}\,\delta\phi_i(t+\tau)\Bigr)\etac_{0'}(t)}_t - \tavg{\etac_0(t+\tau)\Bigl(\sum_{j=1}^N J_{0'j}\,\delta\phi_j(t)\Bigr)}_t + \order{N^{-1}}.
    \label{eq:unperturbed_to_perturbed}
\end{multline}
We evaluate the first subtracted term by substituting $\delta\phi_i(t+\tau)$ from Eq.~\eqref{eq:reservoir_perturbation_expr} and recognizing $\sum_{i,j=1}^N J_{0i}\,J_{j\mu}\,\tilde{S}^\phi_{ij}(t,t') = N^{-1/2}\,F^R_{0\mu}(t,t')$, where $F^R_{0\mu}(t,t')$ is the reverberation kernel (Eq.~\eqref{eq:reverb_kernel_def}). This yields
\begin{equation}
    \tavg{\Bigl(\sum_{i=1}^N J_{0i}\,\delta\phi_i(t+\tau)\Bigr)\etac_{0'}(t)}_t = \frac{1}{\sqrt{N}}\,\sum_{\mu \in \{0, 0'\}} \int_0^\infty \dd s\;\tavg{F^R_{0\mu}(t+\tau, t+\tau-s)\,\phi^\free_\mu(t+\tau-s)\,\etac_{0'}(t)}_t + \order{N^{-1}}.
    \label{eq:step1_raw}
\end{equation}
This joint average decouples across cavity units and from $F^R_{0\mu}(t,t')$ (Properties~1 and~2, Appendix~\ref{app:cavity_properties}), with $\order{N^{-1/2}}$ error that combines with the $N^{-1/2}$ prefactor to give $\order{N^{-1}}$. The $\mu = 0$ contribution is then $\order{N^{-1}}$, since $\tavgshort{\etac_{0'}(t)}_t = 0$ (zero-mean assumption), and the $\mu = 0'$ contribution follows from the Furutsu--Novikov theorem (Eq.~\eqref{eq:FN_fourier}).
The result is
\begin{equation}
    \tavg{\Bigl(\sum_{i=1}^N J_{0i}\,\delta\phi_i(t+\tau)\Bigr)\etac_{0'}(t)}_t = \frac{1}{\sqrt{N}}\int_0^\infty \dd s\;F^R_{00'}(s)\,g^2\,[\Sstar \ast \Cstar](\tau - s) + \order{N^{-1}}.
    \label{eq:step1_factorized}
\end{equation}
In Fourier space, the $\order{N^{-1/2}}$ term of Eq.~\eqref{eq:step1_factorized} is
\begin{equation}
    \frac{g^2\,\Cstar(\omega)}{\sqrt{N}}\,\Sstar(\omega)\,F^R_{00'}(\omega).
    \label{eq:step1_reverb_term}
\end{equation}
The second subtracted term in Eq.~\eqref{eq:unperturbed_to_perturbed} gives $g^2\,\Cstar(\omega)\,N^{-1/2}\,\bigl(\Sstar(\omega)\,F^R_{0'0}(\omega)\bigr)^* + \order{N^{-1}}$ by exchanging $0$ and $0'$ and conjugating.

\subsection{From reservoir sums to sums over all indices}

To make the second correction, we extend the sum $\sum_{i,j=1}^N J_{0i}\,J_{0'j}\,C^\phi_{ij}(\tau)$ in Eq.~\eqref{eq:unperturbed_to_perturbed} to all indices and separate the terms involving cavity indices, which gives
\begin{equation}
    \sum_{i,j=1}^N J_{0i}\,J_{0'j}\,C^\phi_{ij}(\tau) = \bigl[\bm{J}\,\bm{C}^\phi(\tau)\,\bm{J}^T\bigr]_{00'} - \sum_{\mu \in \{0,0'\}} J_{0\mu}\,\sum_{j=1}^N J_{0'j}\,C^\phi_{\mu j}(\tau) - \sum_{\mu \in \{0,0'\}} J_{0'\mu}\,\sum_{i=1}^N J_{0i}\,C^\phi_{i\mu}(\tau) + \order{N^{-1}},
    \label{eq:extend_sums}
\end{equation}
where the term in which both $i$ and $j$ take values in $\{0, 0'\}$ is absorbed into $\order{N^{-1}}$. We evaluate the first subtracted term by substituting $C^\phi_{\mu j}(\tau) = \tavg{\phi_\mu(t+\tau)\,\phi_j(t)}_t$ and recognizing $\sum_{j=1}^N J_{0'j}\,\phi_j(t) = \eta_{0'}(t) + \order{N^{-1/2}}$. The $\order{N^{-1/2}}$ error, which comes from the terms $j \in \{0, 0'\}$ of $\eta_{0'}(t)$, combines with the prefactor $J_{0\mu} = \order{N^{-1/2}}$ to give $\order{N^{-1}}$, as does replacing $\phi_\mu(t+\tau)$ by $\phi^\free_\mu(t+\tau)$. This yields
\begin{equation}
    \sum_{\mu \in \{0,0'\}} J_{0\mu}\,\sum_{j=1}^N J_{0'j}\,C^\phi_{\mu j}(\tau) = \sum_{\mu \in \{0,0'\}} J_{0\mu}\,\tavg{\phi^\free_\mu(t+\tau)\,\eta_{0'}(t)}_t + \order{N^{-1}}.
\end{equation}
Replacing $\eta_{0'}(t)$ by $\etac_{0'}(t)$ and decoupling the joint average across cavity units (Property~1, Appendix~\ref{app:property_1}) each have $\order{N^{-1/2}}$ error, which combines with $J_{0\mu} = \order{N^{-1/2}}$ to give $\order{N^{-1}}$. The $\mu = 0$ contribution is then $\order{N^{-1}}$, since $\tavgshort{\etac_{0'}(t)}_t = 0$ (zero-mean assumption). The $\mu = 0'$ contribution is $J_{00'}\,g^2\,[\Sstar \ast \Cstar](\tau) + \order{N^{-1}}$, by the Furutsu--Novikov theorem (Eq.~\eqref{eq:FN_fourier}).
In Fourier space, the $\order{N^{-1/2}}$ term is $g^2\,\Cstar(\omega)\,\Sstar(\omega)\,J_{00'}$. The second subtracted term in Eq.~\eqref{eq:extend_sums} gives $g^2\,\Cstar(\omega)\,\Sstar(\omega)^*\,J_{0'0} + \order{N^{-1}}$ by exchanging $0$ and $0'$ and conjugating.

\subsection{Self-consistent equation with the \texorpdfstring{$F$}{F}-kernel}

We now combine the two corrections. The reverberation terms of the first correction and the direct-coupling terms of the second correction sum to the full $F$-kernel $F_{\mu\nu}(\omega) = F^R_{\mu\nu}(\omega) + \sqrt{N}\,J_{\mu\nu}$ (Eq.~\eqref{eq:F_decomp}), which gives
\begin{equation}
    C^{\etac}_{00'}(\omega) = \bigl[\bm{J}\,\bm{C}^\phi(\omega)\,\bm{J}^T\bigr]_{00'} - \frac{g^2\,\Cstar(\omega)}{\sqrt{N}}\,\Sstar(\omega)\,F_{00'}(\omega) - \frac{g^2\,\Cstar(\omega)}{\sqrt{N}}\,\bigl(\Sstar(\omega)\,F_{0'0}(\omega)\bigr)^* + \order{N^{-1}}.
    \label{eq:cavity_field_reexpressed}
\end{equation}
We substitute Eq.~\eqref{eq:cavity_field_reexpressed} into Eq.~\eqref{eq:single_site_cav_eqn} and collect the prefactors of the $F$-kernel terms. This converts $\Cstar(\omega)$ into $(1 - g^2|\Sstar(\omega)|^2)\,\Cstar(\omega) = \CstarDelta(\omega)$ (Eq.~\eqref{eq:Cdelta_def}), which gives
\begin{equation}
    C^\phi_{00'}(\omega) = |\Sstar(\omega)|^2\,\bigl[\bm{J}\,\bm{C}^\phi(\omega)\,\bm{J}^T\bigr]_{00'} + \frac{\CstarDelta(\omega)}{\sqrt{N}}\,\Bigl[\Sstar(\omega)\,F_{00'}(\omega) + \bigl(\Sstar(\omega)\,F_{0'0}(\omega)\bigr)^*\Bigr] + \order{N^{-1}}.
    \label{eq:Cphi_intermediate}
\end{equation}
The prefactor $\CstarDelta(\omega)$ of the $F$-kernel terms is the residual power spectrum, although this derivation does not use the residual of Eq.~\eqref{eq:residual_def}. Since the cavity pair can be any pair of units, Eq.~\eqref{eq:Cphi_intermediate} extends to the matrix equation
\begin{equation}
    \bm{C}^\phi(\omega) = |\Sstar(\omega)|^2\,\bm{J}\,\bm{C}^\phi(\omega)\,\bm{J}^T + \CstarDelta(\omega)\,\biggl(\frac{\Sstar(\omega)\,\bm{F}(\omega)}{\sqrt{N}} + \frac{\Sstar(\omega)^*\,\bm{F}(\omega)^\dagger}{\sqrt{N}} + \bm{I}_{N+2}\biggr) + \bm{\mathcal{E}}^C(\omega),
    \label{eq:matrix_eq_with_F}
\end{equation}
where $\bm{\mathcal{E}}^C(\omega)$ has diagonal elements of $\order{N^{-1/2}}$ and off-diagonal elements of $\order{N^{-1}}$. The identity term enforces that $C^\phi_{ii}(\omega)$ concentrates around $\Cstar(\omega)$, since $g^2\,|\Sstar(\omega)|^2\,\Cstar(\omega) + \CstarDelta(\omega) = \Cstar(\omega)$.

\subsection{\texorpdfstring{$F$}{F}-kernel in terms of the response matrix}

We now determine the matrix $\bm{F}(\omega)$. To accomplish this, we differentiate $\phi_\mu(t) = \mathcal{T}[\eta_\mu + \xi_\mu](t)$ with respect to $\xi_\nu(t - \tau)$, using the cavity expansion of $\eta_\mu(t)$ (Eqs.~\eqref{eq:full_local_field} and~\eqref{eq:delta_eta_def}). This gives
\begin{multline}
    \frac{\delta\phi_\mu(t)}{\delta\xi_\nu(t-\tau)} \\
    = \int_0^\infty \dd s\;R_\mu(t, t-s)\,\Biggl[\frac{1}{\sqrt{N}}\,\sum_{\rho \in \{0, 0'\}}\int_0^\infty \dd s'\;F_{\mu\rho}(t-s, t-s-s')\,\frac{\delta\phi^\free_\rho(t-s-s')}{\delta\xi_\nu(t-\tau)} + \delta_{\mu\nu}\,\delta(\tau - s)\Biggr] + \order{N^{-1}},
\end{multline}
where $\delta\phi^\free_\rho(t')/\delta\xi_\nu(t-\tau) = R^\free_\rho(t', t-\tau)\,\delta_{\nu\rho}$. We take the time average, use the decoupling of the two sites and of the $F$-kernel from cavity-field functionals at leading order (Properties~1 and~2, Appendix~\ref{app:cavity_properties}), and replace the time averages of the local responses $R_\mu(t, t-s)$ and $R^\free_\rho(t-s-s', t-\tau)$ by $\Sstar(s)$ and $\Sstar(\tau - s - s')$, respectively (zero-mean assumption). This yields
\begin{equation}
    S^\phi_{\mu\nu}(\tau) = \frac{1}{\sqrt{N}}\,\int_0^\infty \dd s\int_0^\infty \dd s'\;\Sstar(s)\,F_{\mu\nu}(s')\,\Sstar(\tau - s - s') + \delta_{\mu\nu}\,\Sstar(\tau) + \delta_{\mu\nu}\order{N^{-1/2}} + (1 - \delta_{\mu\nu})\order{N^{-1}},
\end{equation}
where the errors of these steps generate $\delta_{\mu\nu}\,\order{N^{-1/2}}$ on the diagonal and combine with the $N^{-1/2}$ prefactor of the $F$-kernel term to give $(1 - \delta_{\mu\nu})\,\order{N^{-1}}$ off the diagonal. Transforming to Fourier space yields
\begin{equation}
    S^\phi_{\mu\nu}(\omega) = \frac{\Sstar(\omega)^2}{\sqrt{N}}\,F_{\mu\nu}(\omega) + \delta_{\mu\nu}\,\Sstar(\omega) + \delta_{\mu\nu}\order{N^{-1/2}} + (1 - \delta_{\mu\nu})\order{N^{-1}}.
\end{equation}
Solving for $F_{\mu\nu}(\omega)$ at frequencies where $\Sstar(\omega) \neq 0$ and extending the result to the full $(N+2) \times (N+2)$ matrix, since the cavity pair can be any pair of units, gives
\begin{equation}
    \frac{\bm{F}(\omega)}{\sqrt{N}} = \frac{\bm{S}^\phi(\omega)}{\Sstar(\omega)^2} - \frac{\bm{I}_{N+2}}{\Sstar(\omega)} + \bm{\mathcal{E}}^F(\omega),
    \label{eq:F_matrix_full}
\end{equation}
where $\bm{\mathcal{E}}^F(\omega)$ has diagonal elements of $\order{N^{-1/2}}$ and off-diagonal elements of $\order{N^{-1}}$.

The self-consistent equation for the response matrix simplifies this result. In particular, Eq.~\eqref{eq:response_matrix_eq} gives $\bm{S}^\phi(\omega)/\Sstar(\omega)-\bm{I}_{N+2}=\bm{J}\,\bm{S}^\phi(\omega)+\bm{\mathcal{E}}^S(\omega)/\Sstar(\omega)$. Multiplying Eq.~\eqref{eq:F_matrix_full} by $\Sstar(\omega)$ and substituting this relation gives
\begin{equation}
    \frac{\Sstar(\omega)\,\bm{F}(\omega)}{\sqrt{N}} = \bm{J}\,\bm{S}^\phi(\omega) + \bm{\mathcal{E}}^F(\omega),
    \label{eq:F_result_S}
\end{equation}
where the error $\bm{\mathcal{E}}^S(\omega)/\Sstar(\omega)+\Sstar(\omega)\,\bm{\mathcal{E}}^F(\omega)$ has been absorbed into $\bm{\mathcal{E}}^F(\omega)$, at the same element-wise scaling.

\subsection{Solution}
\label{app:method2_solution}

We now show that the solution of the self-consistent equation for $\bm{C}^\phi(\omega)$ is the ansatz. Substituting Eq.~\eqref{eq:F_result_S} into Eq.~\eqref{eq:matrix_eq_with_F} gives
\begin{equation}
    \bm{C}^\phi(\omega) = |\Sstar(\omega)|^2\,\bm{J}\,\bm{C}^\phi(\omega)\,\bm{J}^T + \CstarDelta(\omega)\,\bigl(\bm{J}\,\bm{S}^\phi(\omega) + \bm{S}^\phi(\omega)^\dagger\,\bm{J}^T + \bm{I}_{N+2}\bigr) + \bm{\mathcal{E}}^C(\omega),
    \label{eq:self_consistent_with_S}
\end{equation}
which is Eq.~\eqref{eq:nonlinear_self_consistent} of Sec.~\ref{sec:method2}. Here, the substitution error has been absorbed into $\bm{\mathcal{E}}^C(\omega)$, which retains diagonal elements of $\order{N^{-1/2}}$ and off-diagonal elements of $\order{N^{-1}}$. The contributions mediated by the $F$-kernel thus enter in the form produced, in a linear network with transfer function $\Sstar(\omega)$, by an i.i.d.\ external drive for which the power spectrum of the filtered drive is $\CstarDelta(\omega)$. Consequently, nonlinearity generates an effective external drive in the linear-equivalent network, the same conclusion as the derivation of Sec.~\ref{sec:method1}.

The match with the linear-network equation, Eq.~\eqref{eq:linear_self_consistent}, is complete at leading order because $\bm{S}^\phi(\omega)$ is itself the linear-network response matrix under the replacement $\mathcal{T}(\omega) \to \Sstar(\omega)$, namely, $\bm{S}^\phi(\omega) = \Sstar(\omega)\,\bm{M}(\omega) + \bm{M}(\omega)\,\bm{\mathcal{E}}^S(\omega)$ (Eq.~\eqref{eq:response_solution}). The solution of Eq.~\eqref{eq:self_consistent_with_S} is therefore the ansatz $\bar{\bm{C}}^\phi(\omega) = \CstarDelta(\omega)\,\bm{M}(\omega)\,\bm{M}(\omega)^\dagger$, which is the linear-network covariance matrix, Eq.~\eqref{eq:linear_cov}, under the two replacements of Sec.~\ref{sec:ansatz}.

We confirm this solution directly, in three steps. First, we substitute $\bm{S}^\phi(\omega) = \Sstar(\omega)\,\bm{M}(\omega) + \bm{M}(\omega)\,\bm{\mathcal{E}}^S(\omega)$ into Eq.~\eqref{eq:self_consistent_with_S} and use the resolvent identity
\begin{equation}
    \Sstar(\omega)\,\bm{J}\,\bm{M}(\omega) = \bm{M}(\omega) - \bm{I}_{N+2}  \label{eq:simpler_identity}
\end{equation}
together with its conjugate transpose. These reduce the parenthesized factor in Eq.~\eqref{eq:self_consistent_with_S} to $\bm{M}(\omega) + \bm{M}(\omega)^\dagger - \bm{I}_{N+2}$, up to terms in $\bm{\mathcal{E}}^S(\omega)$ of Frobenius norm $\order{1}$ (since $\lVert \bm{J}\rVert_{\mathrm{op}} = \order{1}$ and $\lVert \bm{M}(\omega)\rVert_{\mathrm{op}} = \order{1}$). We absorb these terms into $\bm{\mathcal{E}}^C(\omega)$, whose Frobenius norm remains $\order{1}$. Second, multiplying Eq.~\eqref{eq:simpler_identity} by its conjugate transpose gives the identity
\begin{equation}
    \bm{M}(\omega) + \bm{M}(\omega)^\dagger - \bm{I}_{N+2} = \bm{M}(\omega)\,\bm{M}(\omega)^\dagger - |\Sstar(\omega)|^2\,\bm{J}\,\bm{M}(\omega)\,\bm{M}(\omega)^\dagger\,\bm{J}^T.
    \label{eq:M_identity}
\end{equation}
Third, inserting Eq.~\eqref{eq:M_identity} gives
\begin{equation}
    \bm{C}^\phi(\omega) = |\Sstar(\omega)|^2\,\bm{J}\,\bm{C}^\phi(\omega)\,\bm{J}^T + \CstarDelta(\omega)\,\bigl(\bm{M}(\omega)\,\bm{M}(\omega)^\dagger - |\Sstar(\omega)|^2\,\bm{J}\,\bm{M}(\omega)\,\bm{M}(\omega)^\dagger\,\bm{J}^T\bigr) + \bm{\mathcal{E}}^C(\omega).
    \label{eq:self_consistent_final}
\end{equation}
The ansatz $\bar{\bm{C}}^\phi(\omega)$ solves Eq.~\eqref{eq:self_consistent_final} when $\bm{\mathcal{E}}^C(\omega) = 0$.

\subsection{Error propagation}
\label{app:method2_error}

It remains to include $\bm{\mathcal{E}}^C(\omega)$ and to show that the solution is unique. Equation~\eqref{eq:self_consistent_final} is linear in $\bm{C}^\phi(\omega)$, and its solution, including the error, is
\begin{equation}
    \bm{C}^\phi(\omega) = \bar{\bm{C}}^\phi(\omega) + \sum_{n=0}^\infty \bigl(\Sstar(\omega)\,\bm{J}\bigr)^n\,\bm{\mathcal{E}}^C(\omega)\,\bigl[\bigl(\Sstar(\omega)\,\bm{J}\bigr)^n\bigr]^\dagger.
    \label{eq:method2_neumann_error}
\end{equation}
Since, as $N \to \infty$, the eigenvalues of $\Sstar(\omega)\,\bm{J}$ lie in the disk $|\lambda| \leq g\,|\Sstar(\omega)|$, with $g\,|\Sstar(\omega)| < 1$ (Sec.~\ref{sec:ansatz}), the powers $(\Sstar(\omega)\,\bm{J})^n$ decay geometrically as $n \to \infty$ (Gelfand's formula), despite possible transient growth of the powers of this non-normal matrix. The sum in Eq.~\eqref{eq:method2_neumann_error} therefore converges, and the solution of Eq.~\eqref{eq:self_consistent_final} is unique. The diagonal precision follows from the concentration of $C^\phi_{ii}(\omega)$ and $\bar{C}^\phi_{ii}(\omega)$ around $\Cstar(\omega)$ (Sec.~\ref{sec:ansatz}). For the off-diagonal precision, we bound the Frobenius norm of the sum in Eq.~\eqref{eq:method2_neumann_error}. For real $\theta$, the eigenvalues of $e^{i\theta}\,\Sstar(\omega)\,\bm{J}$ lie in the same disk, so the resolvent
\begin{align}
    \bm{M}_\theta(\omega) &= \bigl(\bm{I}_{N+2} - e^{i\theta}\,\Sstar(\omega)\,\bm{J}\bigr)^{-1} \nonumber \\
    &= \sum_{n=0}^\infty e^{in\theta}\,\bigl(\Sstar(\omega)\,\bm{J}\bigr)^n
\end{align}
exists for every $\theta$, and Parseval's theorem gives
\begin{equation}
    \sum_{n=0}^\infty \bigl(\Sstar(\omega)\,\bm{J}\bigr)^n\,\bm{\mathcal{E}}^C(\omega)\,\bigl[\bigl(\Sstar(\omega)\,\bm{J}\bigr)^n\bigr]^\dagger = \int_0^{2\pi} \frac{\dd\theta}{2\pi}\, \bm{M}_\theta(\omega)\,\bm{\mathcal{E}}^C(\omega)\,\bm{M}_\theta(\omega)^\dagger.
\end{equation}
Since $|e^{i\theta}\,\Sstar(\omega)| = |\Sstar(\omega)|$, the operator-norm bound of Sec.~\ref{sec:method1} gives $\lVert \bm{M}_\theta(\omega)\rVert_{\mathrm{op}} = \order{1}$, uniformly in $\theta$. The Frobenius norm of the sum is thus at most $\max_\theta \lVert \bm{M}_\theta(\omega)\rVert_{\mathrm{op}}^2\,\lVert \bm{\mathcal{E}}^C(\omega)\rVert_F = \order{1}$, which gives an off-diagonal RMS error of $\order{N^{-1}}$.

\section{Repeated-index moments}
\label{app:pairwise_moments}

Here, we derive the Wick-like decomposition of even-order moments with repeated indices and the expression for the two-site moments, both stated at the end of Sec.~\ref{sec:moment_decomp}, and test the expression numerically. Both results were first derived diagrammatically by \citet{wakhloo2026solution} for the SCS model with odd $f(x)$ and Gaussian external drive. Throughout, $\phi_i^{n}$ denotes a product $\phi_i(t_1)\cdots\phi_i(t_n)$ at fixed relative lags, with time arguments suppressed as in Sec.~\ref{sec:moment_decomp}.

\subsection{Wick-like decomposition}

Our goal is to reduce an even-order moment with repeated indices, at leading order, to single-site and two-site moments. The decomposition does not depend on details of the network dynamics, but follows from the cumulant scaling (Eq.~\eqref{eq:cumulant_scaling_main}), the moment-cumulant formula (Eq.~\eqref{eq:moment_cumulant_formula}), and the vanishing of every odd-order cumulant (sign-flip symmetry assumption).

Consider an even-order moment
$\tavg{\prod_{s=1}^r \phi_{i_s}^{n_s}}_t$, where
$i_1,\ldots,i_r$ are distinct sites. We order the sites such that
$n_1,\ldots,n_m$ are odd and $n_{m+1},\ldots,n_r$ are even.
Since the total order is even, $m$ is even.

We regard this moment as a moment of the $\sum_{s=1}^r n_s$ individual factors, with $\phi_{i_s}$ occupying $n_s$ slots, and apply the moment-cumulant formula, which sums over partitions $\pi$ of these slots. For a block $B\in\pi$, let $r_B$ be the number of distinct sites in $B$, and let $o_B$ be the number of those sites that occur an odd number of times in $B$. We say that these $o_B$ sites \emph{occur with odd multiplicity} in $B$. By Eq.~\eqref{eq:cumulant_scaling_main}, the contribution of partition $\pi$ is
$\order{N^{-\frac{1}{2}\sum_{B\in\pi}(r_B-1)}}$.

To find the least-suppressed partitions, we bound $\sum_{B\in\pi}(r_B-1)$ from below. Under the sign-flip symmetry assumption, every nonzero cumulant has even order, so only partitions into even-order blocks contribute. In such a block, $o_B$ is even. If $r_B=1$, then $o_B=0$, and if $r_B\geq2$, then $r_B-1\geq r_B/2\geq o_B/2$. Hence, $r_B-1\geq o_B/2$ for every even-order block, with equality only for $(r_B,o_B)=(1,0)$ or $(2,2)$. Moreover, each of $i_1,\ldots,i_m$ must occur with odd multiplicity in at least one block, since otherwise its total multiplicity would be even. Therefore,
\begin{equation}
    \sum_{B\in\pi}(r_B-1)
    \geq
    \frac{1}{2}\sum_{B\in\pi}o_B
    \geq
    \frac{m}{2}.
    \label{eq:repeated_partition_bound}
\end{equation}

A partition contributes at the least-suppressed scale $\order{N^{-m/4}}$ only if both inequalities in Eq.~\eqref{eq:repeated_partition_bound} are equalities. Every other partition is $\order{N^{-(m+2)/4}}$, since $\sum_{B\in\pi}(r_B-1)$ is an integer. Equality in the first inequality requires every block to be one of the two equality cases above, namely, a single-site block, $(r_B,o_B)=(1,0)$, or a two-site block in which both sites occur with odd multiplicity, $(r_B,o_B)=(2,2)$. Equality in the second inequality requires each of $i_1,\ldots,i_m$ to occur with odd multiplicity in exactly one block, and none of $i_{m+1},\ldots,i_r$ to occur with odd multiplicity in any block.
Together, these conditions fix the structure of the partition. Each of $i_1,\ldots,i_m$ lies in exactly one two-site block, with an odd number of its factors, and $i_{m+1},\ldots,i_r$ lie in no two-site block. That is, the two-site blocks pair the sites $i_1,\ldots,i_m$, and all remaining factors occur in even-order single-site blocks. A leading partition is thus specified by a pairing of the sites $i_1,\ldots,i_m$, together with independent choices of the blocks at each pair and at each remaining site. Summing over these choices gives
\begin{equation}
    \text{leading contribution}
    =
    \left[
    \prod_{s=m+1}^{r}
    \tavg{\phi_{i_s}^{n_s}}_t
    \right]
    \sum_{\mathcal{P}\in\mathcal{P}_2(m)}
    \prod_{\{s,s'\}\in\mathcal{P}}
    A_{ss'},
    \label{eq:repeated_partition_structure}
\end{equation}
where $\mathcal{P}_2(m)$ denotes the set of pairings of $\{1,\ldots,m\}$, with the empty pairing at $m=0$. At each site $i_s$ with $s > m$, a leading partition splits the $n_s$ factors of that site into even-order single-site blocks, in any way. Summing the products of the corresponding cumulants over these splits gives $\tavgshort{\phi_{i_s}^{n_s}}_t$ exactly, by the moment-cumulant formula, since the omitted partitions contain odd-order cumulants, which vanish. At each pair $\{i_s,i_{s'}\}$, a leading partition has exactly one two-site block on $\{i_s,i_{s'}\}$, with odd multiplicity at each site, together with even-order single-site blocks of the remaining factors at these two sites. The factor $A_{ss'}$ denotes the sum of the products of the corresponding cumulants over all such ways of forming these blocks.

Using the moment-cumulant formula once more, we now identify $A_{ss'}$ as the leading part of $\tavg{\phi_{i_s}^{n_s}\phi_{i_{s'}}^{n_{s'}}}_t$. To see this, we sort the terms in the expansion of this two-site moment by their number of two-site blocks.
\begin{itemize}
    \item Terms with no two-site block vanish, since $n_s$ and $n_{s'}$ are odd.
    \item The nonzero terms with exactly one two-site block are precisely those in $A_{ss'}$.
    \item Terms with two or more two-site blocks are $\order{N^{-1}}$, and hence suppressed.
\end{itemize}
Thus,
\begin{equation}
    \tavg{\prod_{s=1}^{r}\phi_{i_s}^{n_s}}_t
    =
    \left[
    \prod_{s=m+1}^{r}
    \tavg{\phi_{i_s}^{n_s}}_t
    \right]
    \sum_{\mathcal{P}\in\mathcal{P}_2(m)}
    \prod_{\{s,s'\}\in\mathcal{P}}
    \tavg{\phi_{i_s}^{n_s}\phi_{i_{s'}}^{n_{s'}}}_t
    +
    \order{N^{-(m+2)/4}}.
    \label{eq:repeated_index_wick}
\end{equation}
For example, with $i,j,k$ distinct, the cases $m=2$ and $m=0$ read
\begin{align}
    \tavg{\phi_i^3\phi_j\phi_k^2}_t
    &=
    \tavg{\phi_k^2}_t\tavg{\phi_i^3\phi_j}_t+\order{N^{-1}},
    \notag \\
    \tavg{\phi_i^2\phi_j^2}_t
    &=
    \tavg{\phi_i^2}_t\tavg{\phi_j^2}_t+\order{N^{-1/2}}.
    \notag
\end{align}

\subsection{Evaluating two-site moments}

The single-site moments in Eq.~\eqref{eq:repeated_index_wick} are determined by single-site DMFT (Sec.~\ref{sec:dmft}). It remains to evaluate, for $i\neq j$ and odd $p,q$,
\begin{equation}
    C^{(p,q)}_{ij}(\tau)
    =
    \tavg{\phi_i(t+\tau)^p\,\phi_j(t)^q}_t.
    \label{eq:power_moment_def}
\end{equation}
Our goal is an expression for these moments in terms of the resolvent $\bm{M}(\omega)$ and single-site DMFT quantities. To accomplish this, we extend the residual construction of Sec.~\ref{sec:method1} from activities to their odd powers. We first define the residual of a power, then show that the residuals are nearly uncorrelated across sites, as in Eq.~\eqref{eq:controlled_err_expr}, and finally propagate this result to the moments.

We begin with the residual. For any odd positive integer $a$, let $\bm{\phi}^{(a)}(t)$ denote the vector with components $\phi_i(t)^a$. By the chain rule, the local response of $\phi_i(t)^a$, namely, $\delta\phi_i(t)^a/\delta h_i(t')$, and its single-site average are
\begin{align}
    R_i^{(a)}(t,t')
    &=
    a\,\phi_i(t)^{a-1}R_i(t,t'),
 \nonumber \\
    S_\star^{(a)}(\tau)
    &=
    a \sstavg{\phi^\free_0(t+\tau)^{a-1}R^\free_0(t+\tau,t)}.
    \label{eq:power_response}
\end{align}
For a product of $a$ factors at fixed relative lags, $R_i^{(a)}(t,t')$ is instead the sum over factors of the derivative of that factor times the others, and the results below apply to either form. Generalizing Eq.~\eqref{eq:residual_def}, we define
\begin{equation}
    \Delta_i^{(a)}(t)
    =
    \phi_i(t)^a
    -
    [S_\star^{(a)}\ast\eta_i](t).
    \label{eq:power_residual}
\end{equation}
For $a=1$, $S_\star^{(1)}(\tau)=\Sstar(\tau)$ and
$\Delta_i^{(1)}(t)=\Delta_i(t)$. Using
$\bm{\phi}(\omega)=\bm{M}(\omega)\bm{\Delta}(\omega)$,
Eq.~\eqref{eq:power_residual} gives the exact relation
\begin{equation}
    \bm{\phi}^{(a)}(\omega)
    =
    S_\star^{(a)}(\omega)\,
    \bm{J}\bm{M}(\omega)\bm{\Delta}(\omega)
    +
    \bm{\Delta}^{(a)}(\omega).
    \label{eq:power_residual_dynamics}
\end{equation}

We next turn to the covariance of these residuals, with elements
$C^{\Delta^{(a)}\Delta^{(b)}}_{ij}(\tau)
=
\tavg{\Delta_i^{(a)}(t+\tau)\Delta_j^{(b)}(t)}_t$. As in Sec.~\ref{sec:method1}, we treat its diagonal and off-diagonal elements in turn.

The diagonal elements are computed within single-site DMFT. Let
\begin{equation}
    C_\star^{(a,b)}(\tau)
    =
    \sstavg{
    \phi_0^\free(t+\tau)^a\,
    \phi_0^\free(t)^b
    }
    \label{eq:single_site_power_moment}
\end{equation}
denote the single-site moment of powers. Specializing Eq.~\eqref{eq:power_residual} to the cavity unit $0$, with the cavity field in place of the local field, we define $\Delta^{\free\,(a)}_0(t) = \phi^\free_0(t)^a - [S_\star^{(a)}\ast\etac_0](t)$. The diagonal elements $C^{\Delta^{(a)}\Delta^{(b)}}_{ii}(\omega)$ concentrate, up to $\order{N^{-1/2}}$, around $C_\star^{\Delta^{(a)}\Delta^{(b)}}(\omega) = \sstavg{\Delta^{\free\,(a)}_0(\omega)\,\Delta^{\free\,(b)}_0(\omega)^*}$. Expanding the two residuals and applying the Furutsu--Novikov theorem (Eq.~\eqref{eq:FN_fourier}) gives
\begin{equation}
    C_\star^{\Delta^{(a)}\Delta^{(b)}}(\omega)
    =
    C_\star^{(a,b)}(\omega)
    -
    g^2
    S_\star^{(a)}(\omega)
    S_\star^{(b)}(\omega)^*
    \Cstar(\omega).
    \label{eq:power_residual_dmft}
\end{equation}

The off-diagonal elements $C^{\Delta^{(a)}\Delta^{(b)}}_{ij}(\omega)$, $i\neq j$, are $\order{N^{-1}}$. This follows from Properties~1 and~2 as in Sec.~\ref{sec:method1}, with two substitutions. First, the zero-mean assumption is replaced by $\tavgshort{\phi_i(t)^a}_t=0$, which holds for odd $a$ (sign-flip symmetry assumption). Consequently, the means of $\phi^\free_0(t)^a$ and $\Delta^{\free\,(a)}_0(t)$ are $\order{N^{-1/2}}$. Second, the averaged response is replaced by that of the power, $\tavgshort{R_i^{(a)}(t,t-\tau)}_t=S_\star^{(a)}(\tau)+\order{N^{-1/2}}$.
In summary, in analogy with Eq.~\eqref{eq:controlled_err_expr},
\begin{equation}
    \bm{C}^{\Delta^{(a)}\Delta^{(b)}}(\omega)
    =
    C_\star^{\Delta^{(a)}\Delta^{(b)}}(\omega)\,\bm{I}_N
    +
    \bm{\mathcal{E}}^{(a,b)}(\omega),
    \label{eq:power_residual_cov}
\end{equation}
where the diagonal elements of $\bm{\mathcal{E}}^{(a,b)}(\omega)$ are
$\order{N^{-1/2}}$ and its off-diagonal elements are
$\order{N^{-1}}$.

Finally, we propagate this result to the moments. Let $\bm{C}^{(p,q)}(\omega)$ be the matrix with elements given by the Fourier transforms of Eq.~\eqref{eq:power_moment_def}. Taking the cross-spectrum of Eq.~\eqref{eq:power_residual_dynamics} for $a=p$ and $a=q$, and substituting Eq.~\eqref{eq:power_residual_cov} for each of the four residual covariances that appear, gives $\bm{C}^{(p,q)}(\omega) = \bar{\bm{C}}^{(p,q)}(\omega) + \bm{\mathcal{E}}^{(p,q)}_C(\omega)$, where
\begin{multline}
    \bar{\bm{C}}^{(p,q)}(\omega)
    =
    C_\star^{\Delta^{(1)}\Delta^{(1)}}(\omega)\,
    S_\star^{(p)}(\omega)\,
    \bm{J}\bm{M}(\omega)\bm{M}(\omega)^\dagger\bm{J}^T\,
    S_\star^{(q)}(\omega)^*
    +
    C_\star^{\Delta^{(1)}\Delta^{(q)}}(\omega)\,
    S_\star^{(p)}(\omega)\bm{J}\bm{M}(\omega)
    \\*
    +
    C_\star^{\Delta^{(p)}\Delta^{(1)}}(\omega)\,
    \bm{M}(\omega)^\dagger\bm{J}^T
    S_\star^{(q)}(\omega)^*
    +
    C_\star^{\Delta^{(p)}\Delta^{(q)}}(\omega)\,\bm{I}_N,
    \label{eq:power_moment_ansatz}
\end{multline}
and $\bm{\mathcal{E}}^{(p,q)}_C(\omega)$ collects the four error matrices $\bm{\mathcal{E}}^{(a,b)}(\omega)$, each sandwiched between factors of operator norm $\order{1}$, since $\lVert \bm{J}\rVert_{\mathrm{op}} = \order{1}$ and $\lVert \bm{M}(\omega)\rVert_{\mathrm{op}} = \order{1}$. The error propagation of Sec.~\ref{sec:method1} therefore applies unchanged. Consequently, for $i\neq j$, $\bar{C}^{(p,q)}_{ij}(\omega)$ gives the two-site moment with the same $\order{N^{-1}}$ off-diagonal RMS error as the linear-equivalent ansatz. These off-diagonal elements recover the two-point formula of \citet{wakhloo2026solution} for odd powers.

As a consistency check, for $p=q=1$, Eq.~\eqref{eq:power_moment_ansatz} reduces to the ansatz $\bar{\bm{C}}^\phi(\omega)$ (Eq.~\eqref{eq:ansatz}). Indeed, $S_\star^{(1)}(\omega)=\Sstar(\omega)$ and
$C_\star^{\Delta^{(1)}\Delta^{(1)}}(\omega)=\CstarDelta(\omega)$, so that
\begin{align}
    \bar{\bm{C}}^{(1,1)}(\omega)
    &=
    \CstarDelta(\omega)
    \Big[
    |\Sstar(\omega)|^2
    \bm{J}\bm{M}(\omega)\bm{M}(\omega)^\dagger\bm{J}^T
    +
    \Sstar(\omega)\bm{J}\bm{M}(\omega)
    +
    \bm{M}(\omega)^\dagger\bm{J}^T\Sstar(\omega)^*
    +
    \bm{I}_N
    \Big]
    \notag \\*
    &=
    \CstarDelta(\omega)
    \Big[
    |\Sstar(\omega)|^2
    \bm{J}\bm{M}(\omega)\bm{M}(\omega)^\dagger\bm{J}^T
    +
    \bm{M}(\omega)
    +
    \bm{M}(\omega)^\dagger
    -
    \bm{I}_N
    \Big]
    \notag \\*
    &=
    \CstarDelta(\omega)\,
    \bm{M}(\omega)\bm{M}(\omega)^\dagger.
    \notag
\end{align}
The second equality uses Eq.~\eqref{eq:simpler_identity} and its conjugate transpose, while the third uses Eq.~\eqref{eq:M_identity} (both with $\bm{I}_N$ in place of $\bm{I}_{N+2}$).

We now combine Eqs.~\eqref{eq:repeated_index_wick} and~\eqref{eq:power_moment_ansatz}, and replace the single-site moments by their DMFT values. These replacements change the $\order{N^{-m/4}}$ leading term by a relative $\order{N^{-1/2}}$, since the replacement of each two-site moment has an $\order{N^{-1}}$ error on its $\order{N^{-1/2}}$ scale, and that of each single-site moment an $\order{N^{-1/2}}$ error on its $\order{1}$ scale. Thus,
\begin{equation}
    \tavg{\prod_{s=1}^{r}\phi_{i_s}^{n_s}}_t
    =
    \left[
    \prod_{s=m+1}^{r}
    \sstavg{\bigl(\phi_0^\free\bigr)^{n_s}}
    \right]
    \sum_{\mathcal{P}\in\mathcal{P}_2(m)}
    \prod_{\{s,s'\}\in\mathcal{P}}
    \bar{C}^{(n_s,n_{s'})}_{i_s i_{s'}}
    +
    \order{N^{-(m+2)/4}},
    \label{eq:repeated_index_final}
\end{equation}
where $\bar{C}^{(n_s,n_{s'})}_{i_s i_{s'}}$ is the time-domain form of the $(i_s,i_{s'})$ element of Eq.~\eqref{eq:power_moment_ansatz}, with time arguments suppressed. As in Eq.~\eqref{eq:all_distinct_wick}, the error holds in the RMS sense over index tuples, and Eq.~\eqref{eq:repeated_index_final} reduces to Eq.~\eqref{eq:all_distinct_wick} when all $n_s = 1$.

\subsection{Numerical verification}

\begin{figure}
    \centering
    \includegraphics[width=0.55\textwidth]{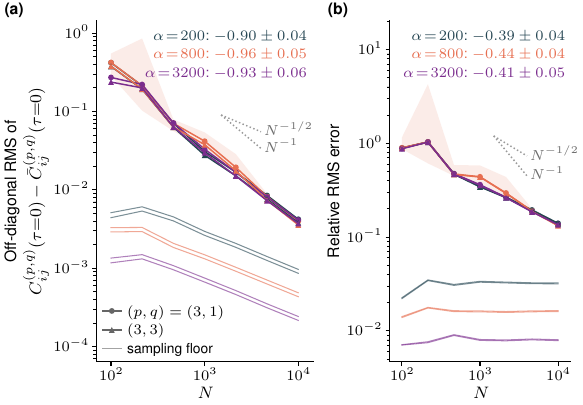}
    \caption{%
    Scaling of the error of Eq.~\eqref{eq:power_moment_ansatz} for two-site moments of odd powers with network size in the SCS model. Both panels show $g = 2.5$ at time lag $\tau = 0$, for three sampling ratios $\alpha$ and the moments $\tavgshort{\phi_i(t)^p\,\phi_j(t)^q}_t$ with $(p, q) = (3, 1)$ and $(3, 3)$. Lines show medians over ten realizations of $\bm{J}$ per $N$ (five for $\alpha = 3200$), and the shaded region shows the interquartile range for $(p, q) = (3, 1)$ at $\alpha = 800$. Keys give the fitted exponents for $(p, q) = (3, 1)$. Dotted gray lines show reference power laws.
    \textbf{(a)}~Off-diagonal RMS of the error of Eq.~\eqref{eq:power_moment_ansatz}, $C^{(p,q)}_{ij}(\tau = 0) - \bar{C}^{(p,q)}_{ij}(\tau = 0)$.
    \textbf{(b)}~Error of (a) relative to the off-diagonal RMS of $C^{(p,q)}_{ij}(\tau = 0)$.}
    \label{fig:moments_scaling}
\end{figure}

We test Eq.~\eqref{eq:power_moment_ansatz} in the simulations of the SCS model of Sec.~\ref{sec:numerics}, for the two-site moments $C^{(p,q)}_{ij}(\tau = 0)$ at $(p, q) = (3, 1)$ and $(3, 3)$. The single-site quantities of Eqs.~\eqref{eq:power_response} and~\eqref{eq:power_residual_dmft} are evaluated as Gaussian integrals over the DMFT solution. Figure~\ref{fig:moments_scaling}(a) shows the off-diagonal RMS of the error $C^{(p,q)}_{ij}(\tau = 0) - \bar{C}^{(p,q)}_{ij}(\tau = 0)$ as a function of $N$, and Fig.~\ref{fig:moments_scaling}(b) shows this error relative to the off-diagonal RMS of $C^{(p,q)}_{ij}(\tau = 0)$. The fitted exponents shown in Fig.~\ref{fig:moments_scaling} (fit protocol in Appendix~\ref{app:sim_floor}) are those for $(p, q) = (3, 1)$, and those for $(3, 3)$ agree with them within $0.01$. These exponents also agree, within $0.01$, with the exponents of the error of the linear-equivalent ansatz in the same simulations (Table~\ref{tab:exponents}).

\section{Sompolinsky--Crisanti--Sommers model and preactivation covariance}
\label{app:sompolinsky_specialization}

Here, we specialize the general results to the SCS model, Eq.~\eqref{eq:sompolinsky}, and derive the ansatz for the preactivation covariance matrix of this model.

\subsection{Single-site quantities, ansatz, and residual}

We first collect the formulas of the general theory in this model. Recall that, at site $i$, the dynamics of Eq.~\eqref{eq:sompolinsky} are
\begin{align}
    \phi_i(t) &= f\bigl(x_i(t)\bigr), \nonumber \\
    (1 + \partial_t) x_i(t) &= \eta_i(t) + \xi_i(t),
    \label{eq:sompolinsky_x_filter}
\end{align}
so the local response is
\begin{equation}
    R_i(t, t') = f'\bigl(x_i(t)\bigr)\,\Theta(t - t')\,e^{-(t-t')}.
    \label{eq:sompolinsky_local_response}
\end{equation}
At the level of the single-site DMFT, the cavity preactivation $x^\free_0(t)$ obeys $(1 + \partial_t)\,x^\free_0(t) = \etac_0(t) + \xi_0(t)$. In terms of $x^\free_0(t)$, the gain $\beta_\star$ is
\begin{equation}
    \beta_\star = \sstavg{f'\bigl(x^\free_0(t)\bigr)}.
    \label{eq:beta_star_def}
\end{equation}
Equation~\eqref{eq:dmft_S_sc} then gives the single-site response,
\begin{align}
    \Sstar(\tau) &= \beta_\star\,e^{-\tau}\,\Theta(\tau), \nonumber \\
    \Sstar(\omega) &= \frac{\beta_\star}{1 + i\omega}.
    \label{eq:Sstar_sompolinsky_freq}
\end{align}
Consequently, the residual power spectrum, Eq.~\eqref{eq:Cdelta_def}, is
\begin{equation}
    \CstarDelta(\omega) = \left(1 - \frac{g^2\,\beta_\star^2}{1 + \omega^2}\right)\Cstar(\omega),
    \label{eq:Cdelta_sompolinsky}
\end{equation}
and the linear-equivalent ansatz of \citet{shen2025covariance}, Eq.~\eqref{eq:ansatz}, becomes
\begin{equation}
    \bar{\bm{C}}^\phi(\omega) = \Cstar(\omega)\,\left(1 - \frac{g^2\,\beta_\star^2}{1 + \omega^2}\right)\,\left(\bm{I}_N - \frac{\beta_\star}{1 + i\omega}\,\bm{J}\right)^{-1}\,\left(\bm{I}_N - \frac{\beta_\star}{1 + i\omega}\,\bm{J}\right)^{-\dagger}.
    \label{eq:ansatz_scs}
\end{equation}
By Eq.~\eqref{eq:Sstar_sompolinsky_freq}, the residual of Eq.~\eqref{eq:residual_def} is
\begin{equation}
    \Delta_i(t) = \phi_i(t) - \beta_\star y_i(t),
    \label{eq:sompolinsky_residual}
\end{equation}
where $y_i(t)$ obeys the preactivation dynamics of Eq.~\eqref{eq:sompolinsky_x_filter} without the external drive, $(1 + \partial_t)\,y_i(t) = \eta_i(t)$. In the zero-drive case, $y_i(t) = x_i(t)$, and the residual becomes
\begin{equation}
    \Delta_i(t) = \phi_i(t) - \beta_\star x_i(t). \label{eq:sompolinsky_residual_zero_drive}
\end{equation}

\subsection{Preactivation covariance matrix}

We next derive the ansatz for the preactivation covariance matrix $\bm{C}^x(\omega)$, assuming Gaussian external drive with power spectrum $\sigma^2_\xi(\omega)$. Taking the outer product of the Fourier-space dynamics $(1 + i\omega)\,\bm{x}(\omega) = \bm{J}\,\bm{\phi}(\omega) + \bm{\xi}(\omega)$ with its conjugate transpose and averaging over time gives
\begin{equation}
    (1 + \omega^2)\,\bm{C}^x(\omega) = \bm{J}\,\bm{C}^\phi(\omega)\,\bm{J}^T + \bm{J}\,\bm{C}^{\phi\xi}(\omega) + \bm{C}^{\phi\xi}(\omega)^\dagger\,\bm{J}^T + \sigma^2_\xi(\omega)\,\bm{I}_N,
    \label{eq:cx_raw}
\end{equation}
where $\bm{C}^{\phi\xi}(\omega) = \tavg{\bm{\phi}(\omega)\, \bm{\xi}(\omega)^\dagger}_t$. Since $\bm{\xi}(\omega)$ is Gaussian, the Furutsu--Novikov theorem gives $\bm{C}^{\phi\xi}(\omega) = \sigma^2_\xi(\omega)\,\bm{S}^\phi(\omega)$. We substitute the response-matrix ansatz $\bar{\bm{S}}^\phi(\omega) = \Sstar(\omega)\,\bm{M}(\omega)$ (Eq.~\eqref{eq:response_ansatz}) for $\bm{S}^\phi(\omega)$. By the resolvent identity, Eq.~\eqref{eq:simpler_identity}, and its conjugate transpose, with $\bm{I}_N$ in place of $\bm{I}_{N+2}$, the last three terms in Eq.~\eqref{eq:cx_raw} combine to $\sigma^2_\xi(\omega)\,\bigl(\bm{M}(\omega) + \bm{M}(\omega)^\dagger - \bm{I}_N\bigr)$. Finally, we apply Eq.~\eqref{eq:M_identity} to this combination and substitute the ansatz $\bar{\bm{C}}^\phi(\omega) = \CstarDelta(\omega)\,\bm{M}(\omega)\,\bm{M}(\omega)^\dagger$ for $\bm{C}^\phi(\omega)$ in the first term. After division by $1 + \omega^2$, the result is the ansatz for the preactivation covariance matrix,
\begin{equation}
    \bar{\bm{C}}^x(\omega) = \frac{\CstarDelta(\omega) - \sigma^2_\xi(\omega)\,|\Sstar(\omega)|^2}{1 + \omega^2}\,\bm{J}\,\bm{M}(\omega)\,\bm{M}(\omega)^\dagger\,\bm{J}^T + \frac{\sigma^2_\xi(\omega)}{1 + \omega^2}\,\bm{M}(\omega)\,\bm{M}(\omega)^\dagger.
    \label{eq:cx_ansatz_main}
\end{equation}

The first prefactor in Eq.~\eqref{eq:cx_ansatz_main} can be re-expressed using the power spectrum of the cavity preactivation, $C^x_\star(\omega) = \sstavg{|x^\free_0(\omega)|^2}$, which satisfies
\begin{equation}
    (1 + \omega^2)\,C^x_\star(\omega) = g^2\,\Cstar(\omega) + \sigma^2_\xi(\omega).
    \label{eq:Cx_star_dmft}
\end{equation}
Combining Eq.~\eqref{eq:Cx_star_dmft} with Eqs.~\eqref{eq:Sstar_sompolinsky_freq} and~\eqref{eq:Cdelta_sompolinsky} gives $\CstarDelta(\omega) - \sigma^2_\xi(\omega)\,|\Sstar(\omega)|^2 = \Cstar(\omega) - \beta_\star^2\,C^x_\star(\omega)$, so Eq.~\eqref{eq:cx_ansatz_main} can be written as
\begin{equation}
    \bar{\bm{C}}^x(\omega) = \frac{\Cstar(\omega) - \beta_\star^2\,C^x_\star(\omega)}{1 + \omega^2}\,\bm{J}\,\bm{M}(\omega)\,\bm{M}(\omega)^\dagger\,\bm{J}^T + \frac{\sigma^2_\xi(\omega)}{1 + \omega^2}\,\bm{M}(\omega)\,\bm{M}(\omega)^\dagger,
    \label{eq:cx_ansatz_full}
\end{equation}
matching the form reported in \citet{shen2025covariance}.

\subsection{Precision of the preactivation ansatz}
The preactivation ansatz has the same precision as $\bar{\bm{C}}^\phi(\omega)$. Substituting $\bm{C}^\phi(\omega) = \bar{\bm{C}}^\phi(\omega) + \bm{\mathcal{E}}(\omega)$ (Sec.~\ref{sec:ansatz}) and $\bm{S}^\phi(\omega) = \bar{\bm{S}}^\phi(\omega) + \bm{M}(\omega)\,\bm{\mathcal{E}}^S(\omega)$ (Eq.~\eqref{eq:response_solution}) into Eq.~\eqref{eq:cx_raw} gives
\begin{equation}
    (1 + \omega^2)\,\bigl[\bm{C}^x(\omega) - \bar{\bm{C}}^x(\omega)\bigr] = \bm{J}\,\bm{\mathcal{E}}(\omega)\,\bm{J}^T + \sigma^2_\xi(\omega)\,\bm{J}\,\bm{M}(\omega)\,\bm{\mathcal{E}}^S(\omega) + \sigma^2_\xi(\omega)\,\bm{\mathcal{E}}^S(\omega)^\dagger\,\bm{M}(\omega)^\dagger\,\bm{J}^T.
\end{equation}
Because $\lVert \bm{J}\rVert_{\mathrm{op}}$, $\lVert \bm{M}(\omega)\rVert_{\mathrm{op}}$, $\lVert \bm{\mathcal{E}}(\omega)\rVert_F$, and $\lVert \bm{\mathcal{E}}^S(\omega)\rVert_F$ are all $\order{1}$, the right-hand side has Frobenius norm $\order{1}$, which gives an RMS error of $\order{N^{-1/2}}$ on the diagonal and of $\order{N^{-1}}$ off the diagonal.

\section{Derivation for nonzero mean activities}
\label{app:nonzero_mean}

Here, we derive the TAP-like equation, Eq.~\eqref{eq:tap_equation}, and the ansatz for the connected covariance matrix, Eq.~\eqref{eq:centered_ansatz}, for the dynamics of Eq.~\eqref{eq:biased_dynamics}, in which each unit receives a quenched bias and the activities have nonzero temporal means. The derivation parallels Secs.~\ref{sec:dmft} and~\ref{sec:method1}, with site-specific quantities in place of site-independent ones. Throughout, a bar and a breve denote the temporal mean and the centered part of a signal, as in Eq.~\eqref{eq:bar_breve_def}.

\subsection{Single-site dynamical mean-field theory and temporal means}
\label{app:nzm_single_site}

We first extend the single-site DMFT of Sec.~\ref{sec:dmft} to site-specific temporal means.

Because the temporal mean input differs across sites at $\order{1}$, the diagonal elements $C^\phi_{ii}(\tau)$ do not concentrate around a single value, unlike in Eq.~\eqref{eq:C_diagonal_concentration}. We therefore define the order parameters as averages over sites,
\begin{align}
    C^{\bar{\phi}}_\star &= \lim_{N \to \infty} \frac{1}{N}\sum_{i=1}^N \bar{\phi}_i^2, \label{eq:disconnected_order_param} \\
    C^{\breve{\phi}}_\star(\tau) &= \lim_{N \to \infty} \frac{1}{N}\sum_{i=1}^N C^{\breve{\phi}}_{ii}(\tau). \label{eq:centered_order_param}
\end{align}
When all sites are equivalent, Eq.~\eqref{eq:centered_order_param} reduces to the definition of $\Cstar(\tau)$ in Sec.~\ref{sec:dmft}.

As in Sec.~\ref{sec:dmft}, we introduce a cavity unit, labeled $0$, with cavity field $\etac_0(t)$ (Eq.~\eqref{eq:cavity_field_single_site}), which differs from the full local field $\eta_0(t)$ at $\order{N^{-1/2}}$ and can be used in place of $\eta_0(t)$ at the level of single-site DMFT. The temporal mean of the cavity field is
\begin{equation}
    \bar{\eta}^{\mathrm{c}}_0 = \sum_{i=1}^N J_{0i}\,\bar{\tilde{\phi}}_i. \label{eq:mean_cavity_field}
\end{equation}
Due to independence of the couplings $J_{0i}$ and unperturbed reservoir activities $\tilde{\phi}_i(t)$, the two parts of $\etac_0(t) = \bar{\eta}^{\mathrm{c}}_0 + \breve{\eta}^{\mathrm{c}}_0(t)$ have the following statistics at large $N$. The centered cavity field $\breve{\eta}^{\mathrm{c}}_0(t)$ is Gaussian under the time average, with mean zero and covariance $g^2\, C^{\breve{\phi}}_\star(\tau)$ (Property~1, Appendix~\ref{app:property_1}). The temporal mean $\bar{\eta}^{\mathrm{c}}_0$ is a constant under the time average. Across sites, it is Gaussian with mean zero and variance $g^2 C^{\bar{\phi}}_\star$, and independent of $b_0$.
The cavity unit thus receives a temporal mean input $\bar{\eta}^{\mathrm{c}}_0 + b_0$, which differs from $\bar{h}_0$ of Eq.~\eqref{eq:mean_input} at $\order{N^{-1/2}}$, plus the centered cavity field, plus the external drive. We denote by $\tavgshort{\cdots}_{\breve{\eta},\xi}$ the $N \to \infty$ measure on $(\breve{\eta}, \xi)$ under which $\breve{\eta} \sim \mathcal{GP}(0, g^2 C^{\breve{\phi}}_\star)$ and $\xi \sim \mathcal{P}_\xi$ independently. In place of Eqs.~\eqref{eq:dmft_C_sc} and~\eqref{eq:dmft_S_sc}, we define three functions of the temporal mean input $\bar{h}$,
\begin{align}
    \mathcal{M}_\star(\bar{h}) &= \tavg{\mathcal{T}[\bar{h} + \breve{\eta} + \xi](t)}_{\breve{\eta},\xi}, \label{eq:conditional_mean} \\
    C^{\breve{\phi}}_\star(\tau \,|\, \bar{h}) &= \tavg{\mathcal{T}[\bar{h} + \breve{\eta} + \xi](t+\tau)\,\mathcal{T}[\bar{h} + \breve{\eta} + \xi](t)}_{\breve{\eta},\xi} - \mathcal{M}_\star(\bar{h})^2, \label{eq:conditional_C} \\
    S_\star(\tau \,|\, \bar{h}) &= \tavg{\left. \frac{\delta \mathcal{T}[h](t+\tau)}{\delta h(t)} \right|_{h = \bar{h} + \breve{\eta} + \xi}}_{\breve{\eta},\xi}. \label{eq:conditional_S}
\end{align}
Single-site quantities are heterogeneous and concentrate only after conditioning on the temporal mean input of the site. In particular, treating each site $i$ as the cavity unit gives
\begin{align}
    \bar{\phi}_i &= \mathcal{M}_\star(\bar{h}_i) + \order{N^{-1/2}}, \label{eq:mean_concentration} \\
    C^{\breve{\phi}}_{ii}(\tau) &= C^{\breve{\phi}}_\star(\tau \,|\, \bar{h}_i) + \order{N^{-1/2}}, \label{eq:C_concentration} \\
    \tavg{R_i(t+\tau,t)}_t &= S_\star(\tau \,|\, \bar{h}_i) + \order{N^{-1/2}}, \label{eq:S_concentration}
\end{align}
where the local response $R_i(t,t')$ of Eq.~\eqref{eq:Ri_def} is now evaluated at $h = \eta_i + b_i + \xi_i$. Substituting Eqs.~\eqref{eq:mean_local_field} and~\eqref{eq:mean_input} into Eq.~\eqref{eq:mean_concentration} gives the TAP-like equation, Eq.~\eqref{eq:tap_equation}, with $\order{N^{-1/2}}$ error per element.

Equation~\eqref{eq:tap_equation} determines the disconnected part, provided we know $C^{\breve{\phi}}_\star(\tau)$, which enters $\mathcal{M}_\star(\bar{h})$ through the measure $\tavgshort{\cdots}_{\breve{\eta},\xi}$. To determine $C^{\breve{\phi}}_\star(\tau)$, we use the fact that $\bar{\eta}_i$, like $\bar{\eta}^{\mathrm{c}}_0$, is Gaussian across sites with mean zero and variance $g^2 C^{\bar{\phi}}_\star$, and independent of $b_i$. Squaring Eq.~\eqref{eq:mean_concentration} and averaging it and Eq.~\eqref{eq:C_concentration} over sites gives self-consistent equations for the order parameters,
\begin{align}
    C^{\bar{\phi}}_\star &= \tavg{\mathcal{M}_\star(\bar{\eta} + b)^2}_{\bar{\eta}, b}, \label{eq:closure_disconnected} \\
    C^{\breve{\phi}}_\star(\tau) &= \tavg{C^{\breve{\phi}}_\star(\tau \,|\, \bar{\eta} + b)}_{\bar{\eta}, b}, \label{eq:closure_connected}
\end{align}
with $\tavgshort{\cdots}_{\bar{\eta}, b}$ the $N \to \infty$ measure under which $\bar{\eta} \sim \mathcal{N}(0, g^2 C^{\bar{\phi}}_\star)$ and $b \sim \mathcal{P}_b$ independently. Due to the dependence of $\tavgshort{\cdots}_{\breve{\eta},\xi}$ on $C^{\breve{\phi}}_\star(\tau)$ and of $\tavgshort{\cdots}_{\bar{\eta}, b}$ on $C^{\bar{\phi}}_\star$, the two equations are coupled; once they are solved, $S_\star(\tau \,|\, \bar{h})$ is given by Eq.~\eqref{eq:conditional_S}.

If only the full autocovariance $\Cstar(\tau) = C^{\bar{\phi}}_\star + C^{\breve{\phi}}_\star(\tau)$ is desired, the conventional disorder-averaged treatment applies unchanged. Indeed, adding Eqs.~\eqref{eq:closure_disconnected} and~\eqref{eq:closure_connected}, and using the fact that $\etac_0(t) = \bar{\eta}^{\mathrm{c}}_0 + \breve{\eta}^{\mathrm{c}}_0(t)$ is, across sites, a Gaussian process with mean zero and covariance $g^2\,\Cstar(\tau)$, gives
\begin{equation}
    \Cstar(\tau)
    =
    \tavg{\mathcal{T}[\etac_0 + b_0 + \xi_0](t+\tau)\,
    \mathcal{T}[\etac_0 + b_0 + \xi_0](t)}_{\etac_0 \sim \mathcal{GP}(0,\, g^2 \Cstar),\; b_0 \sim \mathcal{P}_b,\; \xi_0 \sim \mathcal{P}_\xi},
    \label{eq:uncentered_dmft}
\end{equation}
whose solution decays to $C^{\bar{\phi}}_\star$ as $\tau \to \infty$ when the connected autocovariance decays. We determine the two parts separately because the site-specific functions of Eqs.~\eqref{eq:conditional_mean}--\eqref{eq:conditional_S} require them individually.

\subsection{Residuals for the connected part}
\label{app:nzm_residuals}

We now turn to linear equivalence for the connected part of the covariance matrix, reducing the ansatz of Eq.~\eqref{eq:centered_ansatz} to two conditions on the centered residuals, as in Sec.~\ref{sec:method1}.

We define the residual, generalizing Eq.~\eqref{eq:residual_def}, and its centered part,
\begin{align}
    \Delta_i(t) &= \phi_i(t)-[S_\star(\,\cdot \,|\, \bar{h}_i)\ast\eta_i](t), \label{eq:site_residual} \\
    \breve{\Delta}_i(t) &= \breve{\phi}_i(t)-[S_\star(\,\cdot \,|\, \bar{h}_i)\ast\breve{\eta}_i](t), \label{eq:centered_residual}
\end{align}
where $\bar{h}_i$ is the temporal mean input of the network, Eq.~\eqref{eq:mean_input} (the change in the ansatz from forming $\bar{h}_i$ instead from the solution of Eq.~\eqref{eq:tap_equation} is bounded in Appendix~\ref{app:nzm_error}). We collect the site-specific responses into the diagonal matrix
\begin{equation}
    \bm{D}^S(\omega) = \operatorname{diag}\bigl(S_\star(\omega \,|\, \bar{h}_1), \ldots, S_\star(\omega \,|\, \bar{h}_N)\bigr). \label{eq:DS_def}
\end{equation}
In Fourier space, Eq.~\eqref{eq:centered_residual} reads
\begin{equation}
    \breve{\bm{\phi}}(\omega)
    =
    \bm{D}^S(\omega)\,\bm{J}\,\breve{\bm{\phi}}(\omega)
    +
    \breve{\bm{\Delta}}(\omega),
    \label{eq:centered_residual_dynamics}
\end{equation}
with the closed-form solution $\breve{\bm{\phi}}(\omega) = \breve{\bm{M}}(\omega)\,\breve{\bm{\Delta}}(\omega)$, where the resolvent is
\begin{equation}
    \breve{\bm{M}}(\omega)
    =
    \bigl(\bm{I}_N-\bm{D}^S(\omega)\,\bm{J}\bigr)^{-1}.
    \label{eq:centered_resolvent}
\end{equation}
We assume that, at leading order, $\bm{D}^S(\omega)\,\bm{J}$ behaves as if $\bm{D}^S(\omega)$ were independent of $\bm{J}$. As $N \to \infty$, the eigenvalues of $\bm{D}^S(\omega)\,\bm{J}$ then fill the disk $|\lambda| \leq g\sqrt{\tavgshort{|S_\star(\omega \,|\, \bar{\eta} + b)|^2}_{\bar{\eta}, b}}$~\cite{ahmadian2015properties}, whose radius is less than one (Appendix~\ref{app:nzm_disk}). The connected covariance matrix takes the form
\begin{equation}
    \bm{C}^{\breve{\phi}}(\omega) = \breve{\bm{M}}(\omega)\,\bm{C}^{\breve{\Delta}}(\omega)\,\breve{\bm{M}}(\omega)^\dagger,
    \label{eq:centered_cov_from_residual}
\end{equation}
where $\bm{C}^{\breve{\Delta}}(\tau)$ is the covariance matrix of the centered residuals, with elements $C^{\breve{\Delta}}_{ij}(\tau) = \tavgshort{\breve{\Delta}_i(t+\tau)\,\breve{\Delta}_j(t)}_t$.

As in Sec.~\ref{sec:method1}, the ansatz holds given two conditions on $\bm{C}^{\breve{\Delta}}(\omega)$. First, the diagonal elements $C^{\breve{\Delta}}_{ii}(\omega)$ must concentrate around site-specific values $C^{\breve{\Delta}}_\star(\omega \,|\, \bar{h}_i)$ with $\order{N^{-1/2}}$ fluctuations (Appendix~\ref{app:nzm_diagonal}). Second, the off-diagonal elements $C^{\breve{\Delta}}_{ij}(\omega)$ must be $\order{N^{-1}}$ for $i \neq j$ (Appendix~\ref{app:nzm_offdiagonal}).

\subsection{Diagonal elements of \texorpdfstring{$\bm{C}^{\breve{\Delta}}(\omega)$}{C-Delta(omega)}}
\label{app:nzm_diagonal}

We now verify the first condition. The elements $C^{\breve{\Delta}}_{ii}(\omega)$ are computed within the single-site DMFT of Appendix~\ref{app:nzm_single_site}. Specializing the residual of Eq.~\eqref{eq:site_residual} to the cavity units of the two-site construction (Sec.~\ref{sec:cavity}), with the cavity field $\etac_\mu(t) = \bar{\eta}^{\mathrm{c}}_\mu + \breve{\eta}^{\mathrm{c}}_\mu(t)$ in place of the local field, we define
\begin{align}
    \phi^\free_\mu(t) &= \mathcal{T}[\etac_\mu + b_\mu + \xi_\mu](t), \nonumber \\
    R^\free_\mu(t,t') &= \left. \frac{\delta \mathcal{T}[h](t)}{\delta h(t')} \right|_{h = \etac_\mu + b_\mu + \xi_\mu}, \nonumber \\
    \Delta^\free_\mu(t) &= \phi^\free_\mu(t) - [S_\star(\,\cdot \,|\, \bar{h}_\mu) \ast \etac_\mu](t),
    \label{eq:site_free_residual} \\
    \breve{\Delta}^\free_\mu(t) &= \breve{\phi}^\free_\mu(t) - [S_\star(\,\cdot \,|\, \bar{h}_\mu) \ast \breve{\eta}^{\mathrm{c}}_\mu](t).
    \label{eq:centered_free_residual}
\end{align}
A diagonal element $C^{\breve{\Delta}}_{00}(\omega)$ concentrates, to $\order{N^{-1/2}}$, around $\tavgshort{|\breve{\Delta}^\free_0(\omega)|^2}_{\breve{\eta},\xi}$, which expands into four terms as in Eq.~\eqref{eq:diagonal_four_terms}. Using the Gaussianity of $\breve{\eta}^{\mathrm{c}}_0(t)$ under the time average, the Furutsu--Novikov theorem gives
\begin{equation}
    \tavg{\breve{\phi}^\free_0(\omega)\,\breve{\eta}^{\mathrm{c}}_0(\omega)^*}_{\breve{\eta},\xi} = g^2\, S_\star(\omega \,|\, \bar{h}_0)\, C^{\breve{\phi}}_\star(\omega),
    \label{eq:centered_FN}
\end{equation}
and the four terms combine to the residual power spectrum
\begin{equation}
    C^{\breve{\Delta}}_\star(\omega \,|\, \bar{h}_0)
    =
    C^{\breve{\phi}}_\star(\omega \,|\, \bar{h}_0)
    -
    g^2\,|S_\star(\omega \,|\, \bar{h}_0)|^2\,C^{\breve{\phi}}_\star(\omega),
    \label{eq:centered_residual_spectrum}
\end{equation}
which reduces to Eq.~\eqref{eq:Cdelta_def} when all sites are equivalent (in particular, when $\bar{h}_i = 0$ for all $i$). We collect the site-specific residual power spectra into the diagonal matrix
\begin{equation}
    \bm{D}^{\breve{\Delta}}(\omega) = \operatorname{diag}\bigl(C^{\breve{\Delta}}_\star(\omega \,|\, \bar{h}_1), \ldots, C^{\breve{\Delta}}_\star(\omega \,|\, \bar{h}_N)\bigr).
    \label{eq:DDelta_def}
\end{equation}

\subsection{Off-diagonal elements of \texorpdfstring{$\bm{C}^{\breve{\Delta}}(\omega)$}{C-Delta(omega)}}
\label{app:nzm_offdiagonal}

It remains to verify the second condition, namely, that $C^{\breve{\Delta}}_{ij}(\omega) = \order{N^{-1}}$ for $i \neq j$. The steps are those of Sec.~\ref{sec:method1}, with centered factors and site-specific responses. The cavity expansion of the residual, Eq.~\eqref{eq:Delta_cavity}, applies with $S_\star(s \,|\, \bar{h}_\mu)$ in place of $\Sstar(s)$,
\begin{align}
    \Delta_\mu(t) &= \Delta^\free_\mu(t) + \delta\Delta_\mu(t) + \order{N^{-1}}, \label{eq:site_Delta_cavity} \\
    \delta\Delta_\mu(t) &= \int_0^\infty \dd s\,
    \bigl(R^\free_\mu(t,t-s) - S_\star(s \,|\, \bar{h}_\mu)\bigr)\, \delta\eta_\mu(t-s), \label{eq:site_deltaDelta_def}
\end{align}
with $\delta\eta_\mu(t)$ the perturbation of Eq.~\eqref{eq:delta_eta_def}. Subtracting the time average of Eq.~\eqref{eq:site_Delta_cavity} from itself, and writing $\overline{\delta\Delta}_\mu = \tavgshort{\delta\Delta_\mu(t)}_t = \order{N^{-1/2}}$, gives the expansion of the centered residual,
\begin{equation}
    \breve{\Delta}_\mu(t) = \breve{\Delta}^\free_\mu(t) + \delta\Delta_\mu(t) - \overline{\delta\Delta}_\mu + \order{N^{-1}}.
    \label{eq:centered_Delta_cavity}
\end{equation}
When Eq.~\eqref{eq:centered_Delta_cavity} is substituted into $C^{\breve{\Delta}}_{00'}(\tau)$, each constant $\overline{\delta\Delta}_\mu$ either multiplies the temporal mean of the centered cavity residual at the other site, $\tavgshort{\breve{\Delta}^\free_\nu(t)}_t = 0$ with $\nu \neq \mu$, or appears in a product of two $\order{N^{-1/2}}$ quantities. Therefore,
\begin{equation}
    C^{\breve{\Delta}}_{00'}(\tau)
    =
    \tavg{\breve{\Delta}^\free_0(t+\tau)\,
    \breve{\Delta}^\free_{0'}(t)}_t
    +
    \tavg{\breve{\Delta}^\free_0(t+\tau)\,
    \delta\Delta_{0'}(t)}_t
    +
    \tavg{\delta\Delta_0(t+\tau)\,
    \breve{\Delta}^\free_{0'}(t)}_t
    +
    \order{N^{-1}}.
    \label{eq:centered_offdiag_expand}
\end{equation}

We first evaluate the cavity-field term in Eq.~\eqref{eq:centered_offdiag_expand}. Using Eq.~\eqref{eq:centered_free_residual} and the cross covariance of the centered cavity fields,
\begin{equation}
    C^{\breve{\eta}^{\mathrm{c}}}_{00'}(\tau) = \tavg{\breve{\eta}^{\mathrm{c}}_0(t+\tau)\,\breve{\eta}^{\mathrm{c}}_{0'}(t)}_t,
    \label{eq:centered_cavity_cross}
\end{equation}
we obtain
\begin{multline}
    \tavg{\breve{\Delta}^\free_0(\omega)\,
    \breve{\Delta}^\free_{0'}(\omega)^*}_t
    =
    \tavg{\breve{\phi}^\free_0(\omega)\,
    \breve{\phi}^\free_{0'}(\omega)^*}_t
    + S_\star(\omega \,|\, \bar{h}_0)\, S_\star(\omega \,|\, \bar{h}_{0'})^*\, C^{\breve{\eta}^{\mathrm{c}}}_{00'}(\omega) \\
    - S_\star(\omega \,|\, \bar{h}_{0'})^*
    \tavg{\breve{\phi}^\free_0(\omega)\,
    \breve{\eta}^{\mathrm{c}}_{0'}(\omega)^*}_t
    - S_\star(\omega \,|\, \bar{h}_0)\,
    \tavg{\breve{\eta}^{\mathrm{c}}_0(\omega)\,
    \breve{\phi}^\free_{0'}(\omega)^*}_t.
    \label{eq:centered_free_residual_four_terms}
\end{multline}
The centered cavity fields $\breve{\eta}^{\mathrm{c}}_0(t)$ and $\breve{\eta}^{\mathrm{c}}_{0'}(t)$ are jointly Gaussian under the time average at leading order, with autocovariance $g^2\, C^{\breve{\phi}}_\star(\tau)$ and cross covariance $C^{\breve{\eta}^{\mathrm{c}}}_{00'}(\tau) = \order{N^{-1/2}}$ (Property~1). Price's theorem, applied to $\breve{\phi}^\free_0(t)$ and $\breve{\phi}^\free_{0'}(t)$ as functionals of $\breve{\eta}^{\mathrm{c}}_0(t)$ and $\breve{\eta}^{\mathrm{c}}_{0'}(t)$, and the Furutsu--Novikov theorem, applied to $\breve{\phi}^\free_0(t)$ with respect to $\breve{\eta}^{\mathrm{c}}_{0'}(t)$ and to $\breve{\phi}^\free_{0'}(t)$ with respect to $\breve{\eta}^{\mathrm{c}}_0(t)$, then give
\begin{align}
    \tavg{\breve{\phi}^\free_0(\omega)\,
    \breve{\phi}^\free_{0'}(\omega)^*}_t
    &=
    S_\star(\omega \,|\, \bar{h}_0)\, S_\star(\omega \,|\, \bar{h}_{0'})^*\, C^{\breve{\eta}^{\mathrm{c}}}_{00'}(\omega)
    + \order{N^{-1}}, \nonumber \\
    \tavg{\breve{\phi}^\free_0(\omega)\,
    \breve{\eta}^{\mathrm{c}}_{0'}(\omega)^*}_t
    &=
    S_\star(\omega \,|\, \bar{h}_0)\, C^{\breve{\eta}^{\mathrm{c}}}_{00'}(\omega)
    + \order{N^{-1}}, \nonumber \\
    \tavg{\breve{\eta}^{\mathrm{c}}_0(\omega)\,
    \breve{\phi}^\free_{0'}(\omega)^*}_t
    &=
    S_\star(\omega \,|\, \bar{h}_{0'})^*\, C^{\breve{\eta}^{\mathrm{c}}}_{00'}(\omega)
    + \order{N^{-1}},
    \label{eq:centered_free_phi_phi_price}
\end{align}
where the marginal-product term of Price's theorem is absent from Eq.~\eqref{eq:centered_free_phi_phi_price} because both factors are centered. Substituting into Eq.~\eqref{eq:centered_free_residual_four_terms} produces a cancellation, $\tavgshort{\breve{\Delta}^\free_0(\omega)\,\breve{\Delta}^\free_{0'}(\omega)^*}_t = \order{N^{-1}}$.

We next evaluate the cross terms in Eq.~\eqref{eq:centered_offdiag_expand}. First,
\begin{multline}
    \tavg{\breve{\Delta}^\free_0(t+\tau)\,
    \delta\Delta_{0'}(t)}_t \\
    =
    \frac{1}{\sqrt N}
    \sum_{\nu\in\{0,0'\}}
    \int_0^\infty \dd s \int_0^\infty \dd s'\,
    \tavg{
    \breve{\Delta}^\free_0(t+\tau)
    \left(R^\free_{0'}(t,t-s)-S_\star(s \,|\, \bar{h}_{0'})\right)
    F_{0'\nu}(t-s,t-s-s')\,
    \phi^\free_\nu(t-s-s')
    }_t.
    \label{eq:centered_one_deltaDelta_right}
\end{multline}
As in Sec.~\ref{sec:method1}, Property~2 decouples $F_{0'\nu}(t-s,t-s-s')$ from the cavity-field-dependent factors, and Property~1 decouples cavity-field functionals at sites $0$ and $0'$, both at leading order. For $\nu = 0$, the site-$0'$ marginal, the time average of $R^\free_{0'}(t,t-s) - S_\star(s \,|\, \bar{h}_{0'})$, is $\order{N^{-1/2}}$ by Eq.~\eqref{eq:S_concentration}. Combined with the $N^{-1/2}$ prefactor in Eq.~\eqref{eq:centered_one_deltaDelta_right}, this contribution is $\order{N^{-1}}$. For $\nu = 0'$, the site-$0$ marginal is $\tavgshort{\breve{\Delta}^\free_0(t+\tau)}_t$, which vanishes because $\breve{\Delta}^\free_0(t)$ is centered. This centering replaces the zero-mean assumption of Sec.~\ref{sec:method1}.
The remaining term in Eq.~\eqref{eq:centered_offdiag_expand} follows by exchanging $0$ and $0'$. This completes the demonstration that the off-diagonal elements are suppressed.

\subsection{Radius of the eigenvalue disk}
\label{app:nzm_disk}

Treating $\bm{D}^S(\omega)$ as independent of $\bm{J}$ (Appendix~\ref{app:nzm_residuals}), the eigenvalue disk of $\bm{D}^S(\omega)\,\bm{J}$ has radius $g\sqrt{\tavgshort{|S_\star(\omega \,|\, \bar{\eta} + b)|^2}_{\bar{\eta}, b}}$~\cite{ahmadian2015properties}. We now show that this radius is less than one, generalizing $g\,|\Sstar(\omega)| < 1$ (Sec.~\ref{sec:ansatz}). As in Sec.~\ref{sec:ansatz}, we start from the Cauchy--Schwarz inequality,
\begin{equation}
    \bigl|\tavg{\breve{\phi}^\free_0(\omega)\,\breve{\eta}^{\mathrm{c}}_0(\omega)^*}_{\breve{\eta},\xi}\bigr|^2
    \leq
    g^2\, C^{\breve{\phi}}_\star(\omega)\; C^{\breve{\phi}}_\star(\omega \,|\, \bar{h}_0).
    \label{eq:centered_cauchy_schwarz}
\end{equation}
By Eq.~\eqref{eq:centered_FN}, the left-hand side is $g^4\,|S_\star(\omega \,|\, \bar{h}_0)|^2\, C^{\breve{\phi}}_\star(\omega)^2$, so
\begin{equation}
    g^2\,|S_\star(\omega \,|\, \bar{h}_0)|^2\, C^{\breve{\phi}}_\star(\omega)
    \leq
    C^{\breve{\phi}}_\star(\omega \,|\, \bar{h}_0),
    \label{eq:centered_site_stability}
\end{equation}
i.e., $C^{\breve{\Delta}}_\star(\omega \,|\, \bar{h}_0) \geq 0$ in Eq.~\eqref{eq:centered_residual_spectrum}. Averaging over $\bar{h}_0 = \bar{\eta} + b$ with $\tavgshort{\cdots}_{\bar{\eta}, b}$ and using Eq.~\eqref{eq:closure_connected} gives
\begin{equation}
    g^2\,\tavg{|S_\star(\omega \,|\, \bar{\eta} + b)|^2}_{\bar{\eta}, b} \leq 1,
    \label{eq:centered_stability}
\end{equation}
with strict inequality under the same nondegeneracy condition as in Sec.~\ref{sec:ansatz}.

\subsection{Error propagation}
\label{app:nzm_error}

We now propagate the two conditions to the error of the ansatz. These conditions give, in analogy with Eq.~\eqref{eq:controlled_err_expr},
\begin{equation}
    \bm{C}^{\breve{\Delta}}(\omega) = \bm{D}^{\breve{\Delta}}(\omega) + \bm{\mathcal{E}}^{\breve{\Delta}}(\omega),
\end{equation}
where $\bm{\mathcal{E}}^{\breve{\Delta}}(\omega)$ has diagonal elements of $\order{N^{-1/2}}$ and off-diagonal elements of $\order{N^{-1}}$, and thus $\lVert \bm{\mathcal{E}}^{\breve{\Delta}}(\omega)\rVert_F = \order{1}$. Substituting into Eq.~\eqref{eq:centered_cov_from_residual} gives
\begin{equation}
    \bm{C}^{\breve{\phi}}(\omega) = \bar{\bm{C}}^{\breve{\phi}}(\omega) + \breve{\bm{M}}(\omega)\,\bm{\mathcal{E}}^{\breve{\Delta}}(\omega)\,\breve{\bm{M}}(\omega)^\dagger,
    \label{eq:centered_error_sandwich}
\end{equation}
with $\bar{\bm{C}}^{\breve{\phi}}(\omega)$ the ansatz of Eq.~\eqref{eq:centered_ansatz}. We assume that $\lVert \breve{\bm{M}}(\omega)\rVert_{\mathrm{op}} = \order{1}$, the counterpart of the bound on $\bm{M}(\omega)$ in Sec.~\ref{sec:method1}, as expected when $\bm{D}^S(\omega)$ behaves as if independent of $\bm{J}$ and the eigenvalue disk has radius less than one (Appendix~\ref{app:nzm_disk}), so that
\begin{equation}
    \lVert \breve{\bm{M}}(\omega)\,\bm{\mathcal{E}}^{\breve{\Delta}}(\omega)\,\breve{\bm{M}}(\omega)^\dagger\rVert_F
    \leq
    \lVert \breve{\bm{M}}(\omega)\rVert_{\mathrm{op}}^2\,
    \lVert \bm{\mathcal{E}}^{\breve{\Delta}}(\omega)\rVert_F,
    \label{eq:centered_error_bound}
\end{equation}
which is $\order{1}$. The sum of the squared errors over all elements is thus $\order{1}$, giving an RMS error of $\order{N^{-1/2}}$ on the diagonal and, as in Eq.~\eqref{eq:ansatz_precision}, of $\order{N^{-1}}$ off the diagonal. Forming $\bar{h}_i$ from the solution of Eq.~\eqref{eq:tap_equation} near the temporal mean activities of the network, rather than from these activities, does not affect this precision. Since $\dd\mathcal{M}_\star(\bar{h})/\dd\bar{h} = S_\star(\omega = 0 \,|\, \bar{h})$, the Jacobian of Eq.~\eqref{eq:tap_equation} is $\bm{D}^S(\omega = 0)\,\bm{J}$, and $\bm{I}_N - \bm{D}^S(\omega = 0)\,\bm{J}$ has the bounded inverse $\breve{\bm{M}}(\omega = 0)$. Therefore, the $\order{N^{-1/2}}$ errors of Eq.~\eqref{eq:mean_concentration} change the solution of Eq.~\eqref{eq:tap_equation} by $\order{N^{-1/2}}$ per element in the RMS sense, and using this solution rather than the temporal mean activities of the network changes each element of $\bm{D}^S(\omega)$ and $\bm{D}^{\breve{\Delta}}(\omega)$ by $\order{N^{-1/2}}$ in the same sense. The changes $\delta\bm{D}^S(\omega)$ and $\delta\bm{D}^{\breve{\Delta}}(\omega)$ thus have Frobenius norm $\order{1}$. The induced change in the ansatz consists of terms such as $\breve{\bm{M}}(\omega)\,\delta\bm{D}^S(\omega)\,\bm{J}\,\breve{\bm{M}}(\omega)\,\bm{D}^{\breve{\Delta}}(\omega)\,\breve{\bm{M}}(\omega)^\dagger$ and $\breve{\bm{M}}(\omega)\,\delta\bm{D}^{\breve{\Delta}}(\omega)\,\breve{\bm{M}}(\omega)^\dagger$, each of Frobenius norm $\order{1}$ because the remaining factors have operator norm $\order{1}$ and the site-specific functions and their first derivatives are bounded uniformly in $\bar{h}$ (as for the models of Sec.~\ref{sec:other_models}). These changes are of the same order as $\breve{\bm{M}}(\omega)\,\bm{\mathcal{E}}^{\breve{\Delta}}(\omega)\,\breve{\bm{M}}(\omega)^\dagger$.

\subsection{Reduction to the zero-mean case under sign-flip symmetry}
\label{app:nzm_reduction}

When $\mathcal{T}[h](t)$ is odd, $\mathcal{P}_\xi(\xi) = \mathcal{P}_\xi(-\xi)$, and $b_i = 0$, the stationary state is sign-flip symmetric, and $\mathcal{M}_\star(0) = 0$. Thus, $C^{\bar{\phi}}_\star = 0$ and $\bar{h}_i = 0$ for all $i$, and the construction reduces to that of Sec.~\ref{sec:method1}.

\section{Numerical details}
\label{app:sim_details}

Here, we give the numerical details of the three families of simulations in this paper, namely, the simulations of the SCS model at $g = 2.5$ (Secs.~\ref{sec:introduction} and~\ref{sec:numerics}, Appendices~\ref{app:pairwise_moments}, \ref{app:sim_gains}, \ref{app:sim_lags}, and~\ref{app:sim_noise}), the cavity tests (Appendix~\ref{app:cavity_numerics}), and the simulations of the nonzero-mean models (Sec.~\ref{sec:other_models}). Their parameters are collected in Table~\ref{tab:sim_params} (Appendix~\ref{app:sim_params}).

\subsection{Simulations of the Sompolinsky--Crisanti--Sommers model}
\label{app:sim_scs}

All results for the SCS model at $g = 2.5$, namely, those of Figs.~\ref{fig:schematic}--\ref{fig:scaling} and~\ref{fig:moments_scaling}--\ref{fig:autocorrelation} and of the top of Table~\ref{tab:exponents}, come from one set of simulations. The two exceptions are the measurement of the site gains, defined in Appendix~\ref{app:sim_gains}, and the noise-driven networks of Fig.~\ref{fig:noise_drive} (Appendix~\ref{app:sim_noise}). The parameters of this set of simulations are as follows.
\begin{itemize}
    \item The nonlinearity is the error function $f(x) = \mathrm{erf}\!\left(\sqrt{\pi} x / 2 \right)$. It matches $\tanh(x)$, the original choice of \citet{sompolinsky1988chaos}, in slope at the origin ($f'(0) = 1$) and in saturation ($f(\pm \infty) = \pm 1$), and permits closed-form evaluation of the Gaussian averages of the single-site DMFT.
    \item The coupling strength is $g = 2.5$, and there is no external drive.
    \item The dynamics $(1 + \partial_t)\,x_i(t) = \eta_i(t)$ are integrated with a forward Euler scheme at step size $\Delta t = 0.025$.
    \item After a burn-in time $T_{\mathrm{burn}} = 300$, activity snapshots $\phi_i(t) = f(x_i(t))$ are recorded at intervals $T_{\mathrm{save}} = 0.1$ (four Euler steps).
    \item For each realization of $\bm{J}$, trajectories from $n_{\mathrm{ics}}$ independent initial conditions run in parallel, each recorded for a duration $T$ after the burn-in. Lagged statistics use a circular buffer of the snapshots from the most recent $\tau_{\mathrm{max}} = 40$ units of time, the largest recorded lag (Appendix~\ref{app:sim_recorded}). All statistics are accumulated only once this buffer is full, so the total simulation time is $T_{\mathrm{tot}} = n_{\mathrm{ics}}\,(T - \tau_{\mathrm{max}})$.
    \item As in Sec.~\ref{sec:numerics}, $T_{\mathrm{tot}} = \alpha N$, with sampling ratios $\alpha = 200$, $800$, and $3200$ and with ten, ten, and five realizations of $\bm{J}$ per $N$, respectively. To reach a given $\alpha$, we take $T = 2000$ for $N \leq 2000$, $T = 4000$ for $2000 < N \leq 4000$, and $T = 8000$ for larger $N$, and set $n_{\mathrm{ics}}$ accordingly. When $\alpha$ would require $n_{\mathrm{ics}} > 400$, we fix $n_{\mathrm{ics}} = 400$ and lengthen $T$ instead. The realized sampling ratios lie within $2\%$ of these values.
\end{itemize}

\subsection{Recorded quantities}
\label{app:sim_recorded}

Storing the full time series at large $N$ is not feasible, so the following statistics are accumulated across trajectories during the simulation.
\begin{itemize}
    \item The lagged covariance matrices $\bm{C}^\phi(\tau)$ are accumulated at the $21$ lags $\tau = 0, 0.5, \ldots, 10$, via the circular buffer of recent snapshots.
    \item The autocovariances $C^\phi_{ii}(\tau)$ are accumulated for all sites at all lags up to $\tau_{\mathrm{max}}$ with spacing $T_{\mathrm{save}} = 0.1$. Each lag uses reference times spaced by $5\,T_{\mathrm{save}}$.
    \item The equal-time residual covariance $\bm{C}^\Delta(\tau = 0)$ is accumulated with $\Delta_i(t) = \phi_i(t) - \beta_\star x_i(t)$ (Eq.~\eqref{eq:sompolinsky_residual_zero_drive}), and also with the measured site gains (Appendix~\ref{app:sim_gains}).
    \item The equal-time two-site moments of odd powers of the activities are accumulated for the test of Appendix~\ref{app:pairwise_moments}.
    \item To limit memory usage, each covariance matrix is accumulated only on the subblock of its first $\min(N, 1000)$ rows and columns, and off-diagonal RMS values are computed over this subblock. Since site labels are statistically exchangeable, this subblock is representative of the full matrix.
    \item The equal-time statistics are accumulated separately for two disjoint halves of the initial conditions, which give the sampling floor (Appendix~\ref{app:sim_floor}). At $\alpha = 800$, the lagged covariance matrices are also accumulated separately for the two halves, giving the sampling floors at nonzero lags in Fig.~\ref{fig:lags_diagonal}.
\end{itemize}

\subsection{Sampling floor and fitted exponents}
\label{app:sim_floor}

A covariance estimated from a finite total simulation time differs from its infinite-time value by a sampling error. To measure the size of this error, we use the two halves of the initial conditions. Up to corrections of order $1/T_{\mathrm{tot}}$, each covariance estimate is the mean of the estimates from the two halves. Since the two halves are independent, half the difference of their estimates cancels the infinite-time value and has the same variance as the sampling error of the full estimate. The off-diagonal RMS of this half-difference is the sampling floor of Sec.~\ref{sec:numerics}.

To see how the sampling floor enters the off-diagonal RMS, we write $\hat{A}_{ij}$ for an estimated off-diagonal element of a covariance matrix or of the error of an ansatz that does not use the recorded trajectories, and $A_{ij}$ for its infinite-time value. Because the sampling errors average to zero, up to corrections of order $1/T_{\mathrm{tot}}$ that are negligible here,
\begin{equation}
    \tavg{\hat{A}_{ij}^2}_{\bm{z}} = A_{ij}^2 + \tavg{(\hat{A}_{ij} - A_{ij})^2}_{\bm{z}},
    \label{eq:floor_squares}
\end{equation}
where $\tavgshort{\cdots}_{\bm{z}}$ denotes the average over the initial conditions $\bm{z}$ (and external drives, if any) at fixed $\bm{J}$. Averaged over the off-diagonal elements, the left-hand side of Eq.~\eqref{eq:floor_squares} is the mean squared off-diagonal RMS of the estimates, and the two terms on the right-hand side are the squared off-diagonal RMS of the infinite-time values and the mean squared sampling floor. Consequently, subtracting the sampling floor in quadrature gives an unbiased estimate of the squared off-diagonal RMS of the infinite-time values. This estimate still scatters from one set of initial conditions to the next, so it is reliable only while this scatter, which shrinks with the number of off-diagonal elements, is small compared with the squared off-diagonal RMS of the infinite-time values, as it is here.

We subtract the sampling floor, per realization, in the panels where it is comparable at large $N$ to most of the quantities shown, namely, for the residual cross covariances (Figs.~\ref{fig:scaling}(c) and~\ref{fig:nzm_scaling}(d)), for the cavity-test cumulants of Appendix~\ref{app:cavity_numerics} (Fig.~\ref{fig:appendix_numerics}(a) and (c)), and for the error of the ansatz in the noise-driven networks of Appendix~\ref{app:sim_noise} (Fig.~\ref{fig:noise_drive}(a) and (b)). For the quantities shown without subtraction, in Figs.~\ref{fig:scaling}(a) and (b), \ref{fig:nzm_scaling}(a)--(c), \ref{fig:moments_scaling}, and~\ref{fig:lags_diagonal}, the sampling floor lies well below the quantity shown, so subtracting it would change the fitted exponents by at most $0.02$. In these figures, the sampling floor is drawn as a thin line, and, in the SCS model, the fitted exponents do not depend systematically on $\alpha$ without subtraction. As further evidence that the subtraction removes the sampling contribution, the floor-subtracted medians of the residual cross covariance with $\beta_\star$ (Fig.~\ref{fig:scaling}(c)), which share several realizations of $\bm{J}$ across $\alpha$, agree closely at $N \geq 2154$.

We summarize each quantity as a function of $N$ as follows.
\begin{itemize}
    \item A fitted exponent is the least-squares slope of the logarithm of the median against $\log N$.
    \item The fit range is $N \geq 464$ for the SCS model and the cavity tests. Smaller networks ($N = 100$ and $215$) are not yet in the scaling regime, so they are shown but excluded from the fit. For the nonzero-mean models, the fit uses all sizes, $N = 500$ to $8000$.
    \item The standard error of an exponent is the standard deviation of the slope over $1000$ bootstrap resamples of the realizations, drawn with replacement within each $N$.
    \item We fit medians because a few realizations ($8$ of the $125$ in the SCS model at $N \geq 464$, from $4$ distinct $\bm{J}$) have errors of the ansatz larger than the off-diagonal RMS of $C^\phi_{ij}(\tau = 0)$ and would dominate a fit to the mean over realizations. In each of these realizations, the rightmost eigenvalue of $\beta_\star\bm{J}$ has real part closer to $1$ than in typical realizations, so the linear-equivalent network is closer to the edge of stability (Sec.~\ref{sec:numerics}). For one of these $\bm{J}$, this real part slightly exceeds $1$, a finite-size effect, since it approaches $g_{\mathrm{eff}} = 0.96$ as $N \to \infty$.
\end{itemize}
The fitted exponents are collected in Table~\ref{tab:exponents}.

\begin{table*}[t]
\centering
\caption{Fitted exponents, with standard errors. Top, SCS model at $g = 2.5$ (Fig.~\ref{fig:scaling}), by sampling ratio $\alpha$. Bottom, models with nonzero mean activities (Fig.~\ref{fig:nzm_scaling}) at $g = 20$ and $\alpha = 200$, with the ansatz evaluated in three ways. The first way uses the site-specific responses and residual power spectra measured in the network (measured), the second computes them from the single-site theory at the measured temporal mean inputs (single-site), and the third computes them from the single-site theory at the temporal mean inputs of the solved TAP-like equation (TAP-solved). For $C^\Delta_{ij}(\tau = 0)$ and $C^{\breve{\Delta}}_{ij}(\tau = 0)$, the sampling floor is subtracted in quadrature.}
\label{tab:exponents}
\begin{tabular}{@{}lccc@{}}
\toprule
$\alpha$ & error of the ansatz, $C^\phi_{ij}(\tau = 0) - \bar{C}^\phi_{ij}(\tau = 0)$ & relative error & $C^\Delta_{ij}(\tau = 0)$ with $\beta_\star$ \\
predicted & $-1$ & $-1/2$ & $\leq -1$ \\
\midrule
$200$ & $-0.90 \pm 0.04$ & $-0.39 \pm 0.04$ & $-1.28 \pm 0.04$ \\
$800$ & $-0.96 \pm 0.05$ & $-0.44 \pm 0.04$ & $-1.30 \pm 0.04$ \\
$3200$ & $-0.93 \pm 0.06$ & $-0.41 \pm 0.05$ & $-1.30 \pm 0.05$ \\
\bottomrule
\end{tabular}

\vspace{1em}

\medmuskip=1mu\begin{tabular}{@{}l*{8}{@{\hspace{6.5pt}}c}@{}}
\toprule
 & $\bar{\phi}_i - \mathcal{M}_\star(\bar{h}_i)$ & \multicolumn{3}{c}{$C^{\breve{\phi}}_{ij}(\tau = 0) - \bar{C}^{\breve{\phi}}_{ij}(\tau = 0)$} & \multicolumn{3}{c}{relative error} & $C^{\breve{\Delta}}_{ij}(\tau = 0)$ \\
\cmidrule(lr){3-5} \cmidrule(lr){6-8}
Model &  & measured & single-site & TAP-solved & measured & single-site & TAP-solved &  \\
predicted & $-1/2$ & $-1$ & $-1$ & $-1$ & $-1/2$ & $-1/2$ & $-1/2$ & $\leq -1$ \\
\midrule
Sigmoid & $-0.51 \pm 0.01$ & $-0.97 \pm 0.02$ & $-1.00 \pm 0.06$ & $-1.02 \pm 0.04$ & $-0.46 \pm 0.01$ & $-0.50 \pm 0.05$ & $-0.50 \pm 0.03$ & $-1.07 \pm 0.02$ \\
Sigmoid, biases & $-0.49 \pm 0.01$ & $-0.93 \pm 0.03$ & $-0.90 \pm 0.04$ & $-0.92 \pm 0.04$ & $-0.42 \pm 0.03$ & $-0.43 \pm 0.03$ & $-0.45 \pm 0.04$ & $-1.05 \pm 0.03$ \\
Hebbian, $k = -2$ & $-0.51 \pm 0.01$ & $-0.96 \pm 0.02$ & $-0.94 \pm 0.02$ & $-0.94 \pm 0.03$ & $-0.47 \pm 0.01$ & $-0.44 \pm 0.02$ & $-0.47 \pm 0.02$ & $-1.03 \pm 0.02$ \\
Hebbian, $k = +2$ & $-0.52 \pm 0.01$ & $-0.91 \pm 0.02$ & $-0.97 \pm 0.02$ & $-0.98 \pm 0.03$ & $-0.44 \pm 0.02$ & $-0.47 \pm 0.02$ & $-0.48 \pm 0.03$ & $-1.06 \pm 0.03$ \\
Lotka--Volterra & $-0.52 \pm 0.02$ & $-0.94 \pm 0.05$ & $-0.90 \pm 0.05$ & $-0.90 \pm 0.04$ & $-0.51 \pm 0.03$ & $-0.44 \pm 0.02$ & $-0.43 \pm 0.02$ & $-1.11 \pm 0.05$ \\
\bottomrule
\end{tabular}
\end{table*}

\subsection{Evaluating the ansatz in the Sompolinsky--Crisanti--Sommers model}
\label{app:sim_ansatz}

The ansatz $\bar{\bm{C}}^\phi(\omega)$ of Eq.~\eqref{eq:ansatz} is evaluated with the formulas of Appendix~\ref{app:sompolinsky_specialization}, in particular Eqs.~\eqref{eq:Sstar_sompolinsky_freq} and~\eqref{eq:Cdelta_sompolinsky}, in the following steps.
\begin{itemize}
    \item The single-site DMFT is solved for $\Cstar(\tau)$ by fixed-point iteration of the single-site Gaussian integrals for the Euler-integrated unit, on the recording grid in $\tau$. At $g = 2.5$, this gives $\beta_\star = 0.3847$ and $\Cstar(\tau = 0) = 0.649$, the same values as the ``particle in a potential'' method of \citet{sompolinsky1988chaos}.
    \item Since the covariances are estimated from the Euler-integrated dynamics recorded at interval $T_{\mathrm{save}}$, the ansatz uses the exact discrete-time transfer function of the Euler-integrated unit in place of $1/(1 + i\omega)$.
    \item The frequency grid extends to the Nyquist frequency $\pi/T_{\mathrm{save}}$ of the recorded process, with spacing $\Delta\omega = 0.05$ ($629$ bins). A frequency grid four times finer changes the fitted exponents of Fig.~\ref{fig:scaling}(a) and (b) by at most $0.01$.
    \item The resolvent $\bm{M}(\omega) = (\bm{I}_N - \Sstar(\omega)\,\bm{J})^{-1}$ is evaluated by an $N \times N$ matrix inverse at each frequency, and the inverse Fourier transform to $\bar{\bm{C}}^\phi(\tau)$ is a direct sum over frequency bins.
\end{itemize}

\subsection{Measured site gains}
\label{app:sim_gains}

The theory uses the gain $\beta_\star$ at every site, whereas each site of a finite network has its own \emph{site gain} $\beta_i$, the long-time average of $f'(x_i(t))$, measured in a separate simulation of the same network (eight trajectories of duration $1500$). The site gains differ from $\beta_\star$ by $\order{N^{-1/2}}$, and using them has opposite effects on the residual and on the ansatz. For the residual, $\Delta_i(t) = \phi_i(t) - \beta_i\,x_i(t)$ removes the site-to-site gain fluctuations, and the floor-subtracted residual cross covariance becomes smaller at $N \geq 464$, with fitted exponents between $-1.38$ and $-1.41$. Its decay flattens at the largest $N$, because the sampling error of the site gains is shared by both halves of the initial conditions and so is not removed with the sampling floor. For the ansatz, we replace the single-site response and residual power spectrum from DMFT by the site-specific responses and residual power spectra measured in the network, which tests the linear-equivalent structure independently of the single-site theory. With these measured inputs, the fitted exponents are between $-0.89$ and $-0.91$ for the error and between $-0.41$ and $-0.42$ for the relative error, but the error itself is larger than with the DMFT inputs at $N \geq 464$. This larger error comes from using measured site-specific quantities, not from the order parameter. Indeed, the response and residual power spectrum computed from the single-site theory with the measured mean autocovariance, in place of the DMFT autocovariance, give nearly the same error as the DMFT inputs. The DMFT inputs are therefore the better choice for this model. In the nonzero-mean models, by contrast, the responses differ across sites at $\order{1}$, since the temporal mean inputs do, so the theory itself is site-specific.

\subsection{Nonzero lags, diagonal elements, and autocovariances}
\label{app:sim_lags}

\begin{figure*}
    \centering
    \includegraphics[width=\textwidth]{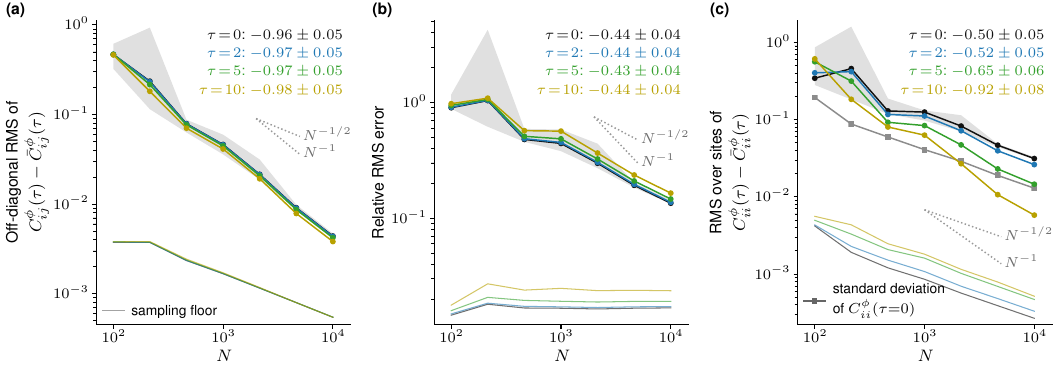}
    \caption{%
    Scaling of the error of the ansatz with network size at nonzero time lags and for the diagonal elements in the SCS model. All panels show $g = 2.5$ and sampling ratio $\alpha = 800$, for four time lags $\tau$. Lines show medians over ten realizations of $\bm{J}$ per $N$, and the shaded region shows the interquartile range at $\tau = 0$. Keys give the fitted exponents. Dotted gray lines show reference power laws.
    \textbf{(a)}~Off-diagonal RMS of the error of the ansatz, $C^\phi_{ij}(\tau) - \bar{C}^\phi_{ij}(\tau)$.
    \textbf{(b)}~Error of (a) relative to the off-diagonal RMS of $C^\phi_{ij}(\tau)$.
    \textbf{(c)}~RMS over sites of the diagonal error of the ansatz, $C^\phi_{ii}(\tau) - \bar{C}^\phi_{ii}(\tau)$, computed on the first $\min(N, 200)$ sites, together with the standard deviation of $C^\phi_{ii}(\tau = 0)$ across the first $\min(N, 1000)$ sites. Thin lines show the sampling floor of the diagonal elements.}
    \label{fig:lags_diagonal}
\end{figure*}

Figure~\ref{fig:scaling} shows off-diagonal elements at $\tau = 0$. We now turn to nonzero lags and diagonal elements (Fig.~\ref{fig:lags_diagonal}). At every lag up to $\tau = 10$, the off-diagonal fitted exponents are between $-0.96$ and $-0.98$ for the error of the ansatz (Fig.~\ref{fig:lags_diagonal}(a)) and between $-0.43$ and $-0.44$ for the relative error (Fig.~\ref{fig:lags_diagonal}(b)), with standard errors $0.04$--$0.05$.

The error of the diagonal elements (Fig.~\ref{fig:lags_diagonal}(c)) is bounded by $\order{N^{-1/2}}$ (Sec.~\ref{sec:ansatz}). Its fitted exponent is $-0.50 \pm 0.05$ at $\tau = 0$, consistent with this bound, and steeper at longer lags. At $\tau = 0$, the RMS over sites of the diagonal error is $2.2$--$2.9$ times the standard deviation of $C^\phi_{ii}(\tau = 0)$ across sites at $N \geq 464$ (median over realizations of this ratio), and most of this RMS comes from the spread of the diagonal error across sites. The diagonal elements of the ansatz therefore fluctuate across sites more than those of the network, which we attribute, as for the off-diagonal fitted exponents of Fig.~\ref{fig:scaling}(a) and (b) (Sec.~\ref{sec:numerics}), to the proximity of the linear-equivalent network to the edge of stability ($g_{\mathrm{eff}} = 0.96$). Consequently, for the diagonal elements alone at $\tau = 0$, the order parameter $\Cstar(\tau = 0)$ is the more accurate predictor at these sizes. At $\tau = 5$ and $10$, where the diagonal error decays faster with $N$, the ansatz is the more accurate predictor at $N \geq 4642$.

\begin{figure*}
    \centering
    \includegraphics[width=\textwidth]{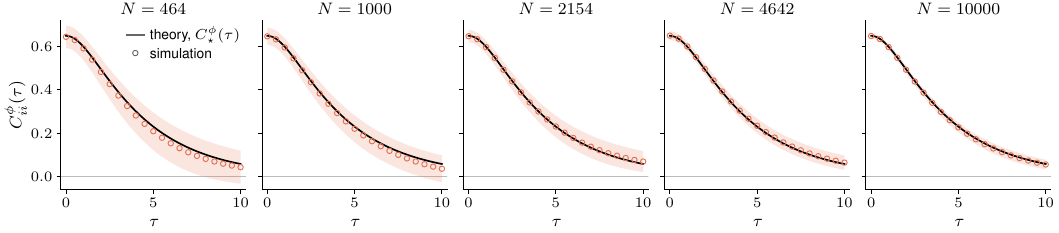}
    \caption{%
    Autocovariances of the SCS model in simulation and theory. All panels show $g = 2.5$ and sampling ratio $\alpha = 800$, with one realization of $\bm{J}$ per $N$. Circles show the mean of $C^\phi_{ii}(\tau)$ across the first $\min(N, 1000)$ sites, and shaded regions show $\pm 1$ standard deviation across these sites.}
    \label{fig:autocorrelation}
\end{figure*}

Finally, the single-site DMFT accurately predicts the empirical mean autocovariance across sites (Fig.~\ref{fig:autocorrelation}). For the realization shown, the largest deviations occur at long lags at $N = 464$ and $1000$. The spread across sites shrinks with $N$, consistent with the $\order{N^{-1/2}}$ diagonal concentration of Eq.~\eqref{eq:C_diagonal_concentration}.

\subsection{Nonzero-mean models}
\label{app:sim_nzm}

The models and results are given in Sec.~\ref{sec:other_models}. The simulations follow Appendices~\ref{app:sim_scs} and~\ref{app:sim_recorded} with the parameters of Table~\ref{tab:sim_params}. Site labels remain exchangeable here, even though each site has its own temporal mean. The remaining details are as follows.
\begin{itemize}
    \item The site-specific responses measured in the network are obtained as follows. For sigmoid units, the response of site $i$ is $\beta_i$ (Appendix~\ref{app:sim_gains}) times the transfer function of the Euler-integrated unit. For the network with Hebbian plasticity and the Lotka--Volterra model, we propagate a unit impulse in the local field through the linearized dynamics of the unit alone along network trajectories (from a separate simulation, as for $\beta_i$), without feedback through $\bm{J}$, and average the response over injection times and trajectories. These measured responses agree, site by site, with those of the single-site theory at the measured temporal mean inputs, with an RMS difference over sites that decays as $N^{-1/2}$.
    \item The measured residual power spectrum of each site is the Fourier transform of its residual autocovariance. Negative values of this spectrum, which arise from finite sampling, are set to zero. All Fourier transforms from lags to frequencies use a cosine taper over the last quarter of lags.
    \item The frequency grid extends to the Nyquist frequency $\pi/T_{\mathrm{save}}$ with spacing $\Delta\omega = 0.05$, as in the SCS model (Appendix~\ref{app:sim_ansatz}). For the Lotka--Volterra model, it instead extends to $\omega = 60$ with spacing $\Delta\omega = 0.1$, since nearly all of the residual power lies below this frequency.
    \item In the network with Hebbian plasticity, the single-site problem has the self-coupling kernel $(k/p)\,e^{-s/p}\,[C^{\bar{\phi}}_\star + C^{\breve{\phi}}_\star(s)]$, with $C^{\bar{\phi}}_\star$ and $C^{\breve{\phi}}_\star(s)$ measured in the network. In both the network and the single-site problem, the integrals over past times are truncated at a lag of $8p$ and evaluated by the rectangle rule on the Euler grid, including the current time step.
    \item The site-specific functions of Eqs.~\eqref{eq:conditional_mean}--\eqref{eq:conditional_S} are Gaussian integrals for the sigmoid units. For the other two models, they are obtained by sampling $512$ Gaussian-process inputs for the network with Hebbian plasticity and $1024$ for the Lotka--Volterra model, each of duration at least $2600$, with $S_\star(\tau \,|\, \bar{h})$ computed from the linearized dynamics along the sampled trajectories.
    \item For the third way of evaluating the ansatz (Sec.~\ref{sec:other_models}), we solve the TAP-like equation, Eq.~\eqref{eq:tap_equation}, by Newton's method from $\bar{\phi}_i = \mathcal{M}_\star(b_i)$, with the same function $\mathcal{M}_\star(\bar{h})$ as in the second way. All solves converge, and starting instead from the measured temporal mean activities gives the same solution.
    \item The residual cross covariance $C^{\breve{\Delta}}_{ij}(\tau = 0)$ is formed with the site-specific responses measured in the network, truncated at $\tau = 10$ for the Lotka--Volterra model.
    \item For the sigmoid units, the solutions of the self-consistent equations, Eqs.~\eqref{eq:closure_disconnected} and~\eqref{eq:closure_connected}, agree closely with the order parameters measured at $N = 8000$. In all of the models, the empirical distribution of the $\bar{h}_i$ is close to Gaussian.
    \item For the Lotka--Volterra model, the largest Lyapunov exponent is estimated from the growth rate of the separation between a trajectory and a nearby copy, renormalized every $20$ integration steps.
\end{itemize}

\subsection{Distance from the edge of stability}
\label{app:sim_noise}

In the undriven SCS model, the fitted exponents of the error of the ansatz are slightly smaller in magnitude than predicted, which we attribute to the proximity of the linear-equivalent network to the edge of stability, $g_{\mathrm{eff}} = 0.96$ (Sec.~\ref{sec:numerics}). Increasing $g$ would not help, since $g_{\mathrm{eff}}$ stays close to $1$ throughout the chaotic regime (Sec.~\ref{sec:discussion}). To test this attribution, we therefore drive the network with i.i.d.\ white noise $\xi_i(t)$ of power spectrum $\sigma^2_\xi(\omega) = \sigma_\xi^2$, with $\sigma_\xi = 1.5$, $2$, and $2.5$, which lowers $g_{\mathrm{eff}}$ to $0.92$, $0.86$, and $0.79$, respectively, according to the single-site DMFT with drive. All other parameters are those of Appendix~\ref{app:sim_scs} at $\alpha = 3200$, with five realizations of $\bm{J}$ per $N$. At $\sigma_\xi \geq 2$, the largest Lyapunov exponent, estimated as in Appendix~\ref{app:sim_nzm}, is negative, so these networks are not chaotic, a case the theory also covers.

\begin{figure*}
    \centering
    \includegraphics[width=\textwidth]{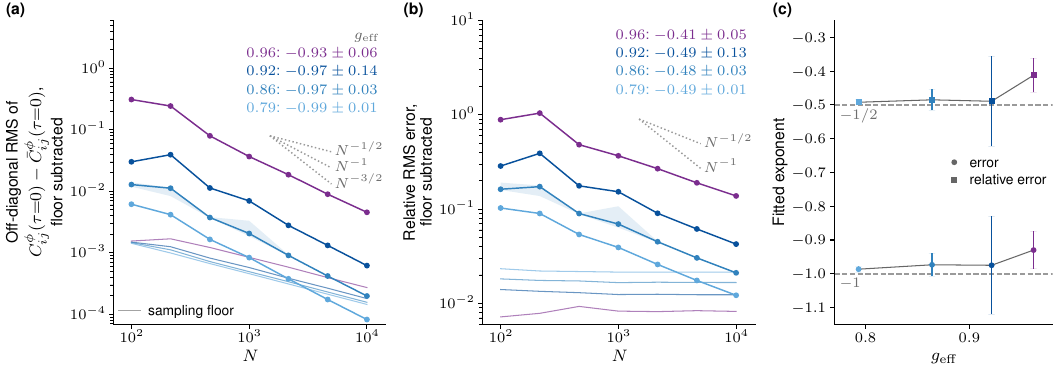}
    \caption{%
    Approach to the predicted exponents away from the edge of stability in the SCS model with white-noise drive. All panels show $g = 2.5$ at time lag $\tau = 0$ and sampling ratio $\alpha = 3200$, for drive strengths $\sigma_\xi = 0$, $1.5$, $2$, and $2.5$, which give $g_{\mathrm{eff}} = 0.96$, $0.92$, $0.86$, and $0.79$, respectively. Lines show medians over five realizations of $\bm{J}$ per $N$, and the shaded region shows the interquartile range at $g_{\mathrm{eff}} = 0.86$. Keys give the fitted exponents. Dotted gray lines show reference power laws.
    \textbf{(a)}~Off-diagonal RMS of the error of the ansatz, $C^\phi_{ij}(\tau = 0) - \bar{C}^\phi_{ij}(\tau = 0)$, after subtraction of the sampling floor.
    \textbf{(b)}~Error of (a) relative to the off-diagonal RMS of $C^\phi_{ij}(\tau = 0)$.
    \textbf{(c)}~Fitted exponents of (a) and (b) against $g_{\mathrm{eff}}$, with standard errors. Dashed lines show the predicted exponents.}
    \label{fig:noise_drive}
\end{figure*}

Further from the edge of stability, the error of the ansatz is smaller, and at the largest $N$ its sampling floor becomes comparable to it. We therefore subtract the sampling floor, as for the residual cross covariances (Appendix~\ref{app:sim_floor}). With a single $\alpha$, this subtraction is not checked by varying $\alpha$. Driving the network brings the fitted exponents to the predicted $-1$ and $-1/2$, from $-0.93 \pm 0.06$ and $-0.41 \pm 0.05$ without drive to $-0.99 \pm 0.01$ and $-0.49 \pm 0.01$ at $g_{\mathrm{eff}} = 0.79$ (Fig.~\ref{fig:noise_drive}). After subtraction of the sampling floor, the residual cross covariances of the driven networks, with the residual $\Delta_i(t) = \phi_i(t) - \beta_\star y_i(t)$ (Eq.~\eqref{eq:sompolinsky_residual}), decay as $N^{-1}$, with fitted exponents between $-0.99$ and $-1.02$, although at the largest $N$ the sampling floor approaches the measured value.

\subsection{Simulation parameters}
\label{app:sim_params}

The parameters of the three families of simulations are collected in Table~\ref{tab:sim_params}. The cavity tests record equal-time statistics only and do not evaluate the ansatz.

\begin{table*}[t]
\centering
\caption{Simulation parameters. In the nonzero-mean models, the sigmoid units with biases have $b_i \sim \mathcal{N}(0, 5^2)$, the network with Hebbian plasticity has $p = 2$ and $k = \pm 2$, and the Lotka--Volterra immigration rate is $\epsilon = 0.01$ (Sec.~\ref{sec:other_models}).}
\label{tab:sim_params}
\small
\begin{tabular}{@{}p{0.2\textwidth}p{0.25\textwidth}p{0.23\textwidth}p{0.25\textwidth}@{}}
\toprule
 & \raggedright SCS model & \raggedright Cavity tests & \raggedright Nonzero-mean models \tabularnewline
\midrule
\raggedright Figures & \raggedright \ref{fig:schematic}, \ref{fig:cij_tau}, \ref{fig:scaling}, \ref{fig:moments_scaling}, \ref{fig:lags_diagonal}, \ref{fig:autocorrelation}, \ref{fig:noise_drive} (with drive) & \raggedright \ref{fig:appendix_numerics} & \raggedright \ref{fig:nzm_scaling} \tabularnewline
\raggedright Units & \raggedright $f(x) = \mathrm{erf}(\sqrt{\pi} x / 2)$ & \raggedright $f(x) = \mathrm{erf}(\sqrt{\pi} x / 2)$ & \raggedright sigmoid without and with biases, Hebbian plasticity, saturating Lotka--Volterra \tabularnewline
\raggedright Coupling strength $g$ & \raggedright $2.5$ & \raggedright $20$ & \raggedright $20$ \tabularnewline
\raggedright Network size $N$ & \raggedright 100, 215, 464, 1000, 2154, 4642, 10000 & \raggedright 100, 215, 464, 1000, 2154, 4642, 10000 & \raggedright 500, 1000, 2000, 4000, 8000 \tabularnewline
\raggedright Fit range & \raggedright $N \geq 464$ & \raggedright $N \geq 464$ & \raggedright all $N$ \tabularnewline
\raggedright Sampling ratio $\alpha$ & \raggedright $200,\, 800,\, 3200$ & \raggedright $200,\, 800,\, 3200$ & \raggedright $200$ \tabularnewline
\raggedright Realizations of $\bm{J}$ per $N$ & \raggedright $10,\, 10,\, 5$ (by $\alpha$) & \raggedright $3$ & \raggedright $8$ (per model) \tabularnewline
\raggedright Euler step $\Delta t$ & \raggedright $0.025$ & \raggedright $0.025$ & \raggedright $0.05$ ($0.01$ for Lotka--Volterra) \tabularnewline
\raggedright Save interval $T_{\mathrm{save}}$ & \raggedright $0.1$ & \raggedright $0.5$ & \raggedright equal to $\Delta t$ \tabularnewline
\raggedright Burn-in time $T_{\mathrm{burn}}$ & \raggedright $300$ & \raggedright $500$ & \raggedright $300$ \tabularnewline
\raggedright Recording time $T$ & \raggedright $2000$--$8000$, longer when $n_{\mathrm{ics}} = 400$ & \raggedright about $5000$, longer when $n_{\mathrm{ics}} = 400$ & \raggedright $2000$--$8000$ \tabularnewline
\raggedright Cross-covariance sites & \raggedright $\min(N, 1000)$ & \raggedright $64$ & \raggedright $\min(N, 1000)$ \tabularnewline
\raggedright Cross-covariance lags & \raggedright $\tau = 0, 0.5, \ldots, 10$ & \raggedright $\tau = 0$ & \raggedright $\tau = 0$ shown \tabularnewline
\raggedright Autocovariance lags & \raggedright spacing $T_{\mathrm{save}}$, up to $\tau_{\mathrm{max}} = 40$ & \raggedright none & \raggedright spacing $T_{\mathrm{save}}$, up to $\tau_{\mathrm{max}} = 20$ \tabularnewline
\bottomrule
\end{tabular}
\end{table*}

\bibliography{refs}

\end{document}